\documentstyle{CUPconf}

\input psfig.sty    

\ifoldfss
\else
  \ifnfssone
    \newmathalphabet{\mathit}
      \addtoversion{normal}{\mathit}{cmr}{m}{it}
      \addtoversion{bold}{\mathit}{cmr}{bx}{it}
    \newmathalphabet{\mathcal}
      \addtoversion{normal}{\mathcal}{cmsy}{m}{n}
    \else
    \ifnfsstwo
    \fi
  \fi
\fi

\def\hexnumber#1{\ifcase#1 0\or1\or2\or3\or4\or5\or6\or7\or8\or9\or
 A\or B\or C\or D\or E\or F\fi }

\makeatletter
\ifx\CUP@mtlplain@loaded\undefined
\else
\fi
\makeatother
 \makeatletter
 \ifx\CUP@mtlplain@loaded\undefined
   \font\tenbmi=cmmib10 at 10pt
   \font\sevenbmi=cmmib10 at 7pt
   \font\fivebmi=cmmib10 at 5pt

   \newfam\bmifam
   \textfont\bmifam=\tenbmi
   \scriptfont\bmifam=\sevenbmi
   \scriptscriptfont\bmifam=\fivebmi
   
 \fi
 \makeatother
\ifnfsstwo

\fi
\ifnfssone

\fi
\ifoldfss

\fi

\mathchardef\varLambda="0103

\makeatletter
\ifx\CUP@mtlplain@loaded\undefined
\else
\fi
\makeatother
\makeatletter
\ifx\CUP@mtlplain@loaded\undefined
  \font\tenbms=cmbsy10
  \font\sevenbms=cmbsy10 at 7pt
  \font\fivebms=cmbsy10 at 5pt
  \newfam\bmsfam
  \textfont\bmsfam=\tenbms
  \scriptfont\bmsfam=\sevenbms
  \scriptscriptfont\bmsfam=\fivebms

  \edef\bsy@{\hexnumber\bmsfam}
  \mathchardef\bnabla="0\bsy@72
\fi
\makeatother
\begin{document}
\ifnfssone
\else
  \ifnfsstwo
  \else
    \ifoldfss
      \let\mathcal\cal
      \let\mathrm\rm
      \let\mathsf\sf
    \fi
  \fi
\fi

  \title[High-Redshift Galaxies: The IR and Sub-Mm View]{High-Redshift Galaxies:
\\ The Far-Infrared and Sub-Millimeter View
  }

  \author[A. Franceschini]{%
  A\ls L\ls B\ls E\ls R\ls T\ls O\ns 
  F\ls R\ls A\ls N\ls C\ls E\ls S\ls C\ls H\ls I\ls N\ls I$^1$,\ns 
%  M\ls A\ls E\ls V\ls E\ns 
%  O'\ls F\ls L\ls Y\ls N\ls N\ls $^1$\\
%  \and\ns
%  J\ls O\ls S\ls E\ls P\ls H\ls I\ls N\ls E\ns 
%  G\ls R\ls I\ls F\ls F\ls I\ls N$^2$
}
  \affiliation{$^1$Dipartimento di Astronomia, 
    University of Padova, I-35122 Padova, IT\\[\affilskip]
%    $^2$Department of Mechanical and Materials 
%    Engineering, Washington State University, 
%    Pullman, WA 99164-2920, USA
}
  \maketitle

\begin{abstract}
Observations at long wavelengths, in the wide interval from a few to 1000 $\mu m$,
are essential to study diffuse media in galaxies, including all kinds of atomic,
ionic and molecular gases and dust grains. Hence they are particularly suited
to investigate the early phases in galaxy evolution, when a very rich ISM
is present in the forming systems.

During the last few years a variety of observational campaigns in the far-IR/sub-mm,
exploiting both ground-based and space instrumentation,
have started to provide results of relevant cosmological impact. 
Most crucial among these have been the
discovery of an intense diffuse background in the far-IR/sub-mm of extragalactic
origin, and the deep explorations from space in the far-IR and with large millimetric
telescopes on ground. These results challenge those obtained
from optical-UV observations, by revealing luminous to very luminous phases
in galaxy evolution at substantial redshifts, likely corresponding to violent
events of star-formation in massive systems. This is bringing to significant
refinements of the present schemes of galaxy formation, as far as the history of
baryon transformations is concerned.

\end{abstract}

\firstsection % if your document starts with a section,

\vspace*{-9cm}\vspace*{-2cm}\noindent
{\sl
To appear in the Proceedings of the XI CANARY ISLANDS WINTER SCHOOL OF ASTROPHYSICS on 
"Galaxies at High Redshift", 
I. Perez-Fournon, M. Balcells, F. Moreno-Insertis and F. Sanchez Editors,
Cambridge University Press. }
\vspace*{9.6cm}

\section{INTRODUCTION}
 
\subsection{The history of baryon transformations}

Although baryons contribute a negligible fraction of the global mass density
of the universe,
their transformations and the associated energy releases are key elements
of the complex, puzzling history bringing from the 
primeval undifferentiated plasma to the highly structured present-day universe.

Two main driving mechanisms are able to circulate and transform 
baryons in astrophysical systems:
one is related with stars and thermonuclear processes occurring therein, the other 
with gravitational contraction of gas -- an important aspect of which, able to
generate vast amounts of energy and producing spectacular effects in Active
Galactic Nuclei and quasars, is gravitational accretion onto supermassive
black holes.

Obviously, these two fundamental motors of the baryon cycle produce very
different outcomes. While gravitational BH accretion irreversibly destroys
baryons to produce energy, gas cycling into stars has (more beneficial)
effects originating beautiful stellar systems, 
producing soft-energy photons, heavy elements, dust, and planetary systems
in the proper amounts to bring eventually to the life.

A basic aim of the present studies of the distant universe, 
exploiting the current most powerful astronomical instrumentation, 
is indeed to clarify the history of baryon circulation, and in particular
the paths through which the various different galaxy populations,
which we observe in the local universe, have built their stellar content,
created their hosted nuclear BH's and accumulated material in them.

While the overall story is driven by the evolving background of dark matter 
distribution, baryons are the observable traces of the evolving
large scale structure.

The history of star formation, in particular, is a fundamental descriptor
of cosmic evolution.
Different cosmogonic scenarios predict very different timetables for the 
formation of stars and structures.
For example, some models predict substantially different formation epochs
for stars among the various morphological classes of galaxies, 
in particular between early-type and late-type galaxy systems.
Some others, notably some specializations of the Cold Dark Matter-dominated 
models, do not.

\subsection{Long-wavelength observations of galaxies: a view on the diffuse media
and on the "active" phases in galaxy evolution}

The build up of stellar populations in high-redshift galaxies is most usually
investigated by looking at the optical/UV/near-IR emission from already formed stars
in distant galaxies.
The complementary approach, less frequently used, is to look at the
diffuse media -- atomic and molecular gas and dust -- in high-z systems, and
their progressive transformation into stars. 

While observations of the redshifted starlight emission in the optical/near-IR can 
exploit large telescopes on ground and very efficient photon detectors, reliable
probes of the diffuse media require longer-wavelength observations in the far-IR
and sub-millimeter: a large variety of lines from 
atomic species and molecules in the Inter-Stellar Medium (ISM) at all ionization levels 
are observable there. Another fundamental
component of the ISM, dust grains present in all astrophysical settings ranging from 
planetary disks to nuclear accretion torii around quasars, have the property
to emit at these wavelengths, typically between a few $\mu m$ to 1000 $\mu m$.

Observations at long$-\lambda$ are then essential to study diffuse media in galaxies
and {\sl are particularly suited [and needed] to study the early
phases in galaxy evolution, when a very rich ISM is present in the forming system}.

Under the generic definition of {\sl galaxy activity} we indicate transient phases 
in the secular evolution of a galaxy during which the various transformations
of the baryons undergo a significant enhancement with respect to the average rate,
for reasons to be ascertained. These phenomena concern both enhanced rates
of conversion of the ISM gas into stars (the {\sl starburst}  phenomenon),
and phases of increased activity of the nuclear emission following an event of
fast accretion of gas into the super-massive BH (the so-called AGN phase,
reaching parossistic levels of photon production of up to $\sim 10^{50}erg/s$ in 
some high-z quasars).

As we will describe in this paper, IR and sub-mm wavelengths provide a privileged
viewpoint to investigate galaxy "activity" in general, for two main reasons:
(a) in many cases this $\lambda$-interval includes a dominant fraction of the
whole bolometric output of active objects; (b) at long wavelengths the screening 
effect of diffuse dust, present in large amounts in "active" galaxies, 
is no more effective and an impeded access to even the most extreme
column-density regions is possible.

\subsection{Observational issues}

Unfortunately, the IR and sub-millimeter constitute a very difficult domain to access
for astronomy:
from ground this is possible only in a few narrow bands from 2.5 to 30 $\mu m$ and at 
$\lambda>300\ \mu m$.   From 30 to 300 $\mu m$ observations are only possible from 
space platforms, the atmosphere being completely opaque.

In any case, however, infrared observations even from space are seriously limited by
several factors. The most fundamental limitation is intrinsic in the energies $\epsilon$ 
of photons we are looking at:  the quantum-mechanics uncertainty principle
sets a boundary to the best achievable angular resolution $\theta$ due to diffraction of 
photons in the primary mirror of a telescope of size $D$: 
$\theta [FWHM] \geq  1.4*57.3*3600 \lambda / D\ [arcsec]$, ($\lambda=c h /\epsilon$). 
For a typical cooled space telescope of 1 meter diameter  working at  $\lambda=100 \ \mu m$ 
this corresponds to $\theta \sim 30\ arcsec$.  For deep surveys of high-redshift 
IR galaxies this limited spatial resolution implies a limiting flux detectable above the
noise due to confusion of several faint sources in the same elementary sky pixel. This
confusion limit sets in at flux levels corresponding to $\sim 0.04\ sources/area\ 
element$, or 0.16 sources/$arcmin^2=570\ sources/degree^2$ in the above example 
(see eq. [\ref{conf}] and further details below).
On this regard, recent surveys (see Sects. 10 and 11) have revealed that the far-IR sky 
is very much populated by luminous extragalactic sources, which implies that confusion 
starts to manifest already at relatively bright fluxes for even large space observatories.

Other limiting factors for IR observations come from the difficulty to reduce the 
instrumental background of  (even space)  telescopes due to photons generated by
the optics.
This adds to the ambient photon backgrounds, due to Zodiacal light from interplanetary dust,
dust emission from the Milky Way, and the terrestrial atmospheric emission.
 
The instrumental backgrounds are reduced by cooling the instrumentation, in particular
for space IR observatories,  but this requires either inserting the whole telescope in
large dewars (ISO, SIRTF), or by passively cooling the telescope with a very
efficient Sun-shielding (FIRST, NGST). All this is technologically very much demanding
and tends to limit the duration of space IR missions (because of the
finite reservoir of coolant) and the size of the primary photon collector.

Finally, photon detection is not as easy in the IR as it is in the optical, and 
limited performances are offered by bolometers
in the sub-mm and by photo-conductors in the mid- and far-IR. Furthermore, 
the need to cool detectors to fundamental temperatures entails problems of
response hysteresis and detector instabilities due to slow reaction of the electrons
to the incoming signal.

\subsection{These lectures}     %Recent discoveries for cosmology in the IR/sub-mm

In spite of the mentioned difficulties to observe at long wavelengths, it was clear 
since the IRAS survey in 1984 that very important phenomena can be investigated here.
Only recently, however, pioneering explorations of the high-redshift universe at
these long-wavelengths have been made possible by new space and ground-based 
facilities, and a new important chapter of observational cosmology has been opened.

These lectures are dedicated to a preliminary assessment of some results in the
field. Because of the very complex, often still elusive, nature of many of the
discovered sources, and because of the complicated astrophysical processes
involved, we dedicate a significant fraction of this paper to
review properties of diffuse media (particularly dust) in local galaxies, and
of their relation with stars (Sects. 2, 3, 4 and 5). We also devote a substantial
chapter (Sect. 6) to the description of local IR starbursts and ultra-luminous
IR galaxies, to improve our chances of understanding their high-redshift counterparts.

Then after a brief mention of historical (IRAS) results in the field (Sect. 7), 
we come to discuss in Sect. 8 the discovery and recent findings about the Cosmic Infrared
Background (CIRB), in Sect. 9 the deep IR surveys by the Infrared Space 
Observatory (ISO), and in Sect. 10 the pioneering observations by millimetric
telescopes (SCUBA, IRAM). Interpretations of the deep counts are given in Sect. 11, and 
the question of the nature of the fast-evolving IR source populations is addressed in 
Sect. 12. Sect. 13 is dedicated to discuss the global properties of the population and
some constraints set by the CIRB observations. A concise summary is given in
Sect. 14.   A Hubble constant $H_0=50\ Km/s/Mpc$ will be adopted unless otherwise stated.

%Three particularly relevant developments in the field of long wavelength astrophysics
%and observational cosmology:
% 
%Exploitation of the all-sky background maps by the COsmic Background Explorer
%{(COBE)}  discovery of the CIRB
%
%Operation of sensititive detector assemblies on large mm/sub-mm telescopes 
%{(JCMT, IRAM)} and centimeter/millimeter aperture-synthesis arrays (VLA, IRAM)
%
%Mid- and far-IR studies by the {Infrared Space Observatory (ISO), }following up with
%substantially improved capabilities the IRAS exploratory surveys.
%
%important phases in galaxy evolution -- namely the formation
%of spheroids -- are hidden in the optical but very clearly revealed by
%long wavelength (IR and mm) observations.
%   
%
%\subsection{These lectures}
%The properties of galaxies at long wavelengths:
%diffuse media, dust, ions, molecules (1.5 h)
%
%Quiescent and violent star formation in the local universe: luminous
%and ultra-luminous galaxies, interactions and mergers, dynamical modelling and simulations
%(1 h)
%
%Observations of high-z galaxies selected in the IR and
%sub-mm: confrontation with spectrophotometric models (1 h)
%
%The distant and very distant universe as viewed 
%in the IR: the history of SF and metal production, constraints from the background
%radiation (1 h)
%
%Relationship between galaxy and quasar formation (0.5 h)
 
\section{DUST IN GALAXIES}

\subsection{ Generalities}

Dust is one of the most important components of the ISM, including
roughly half of heavy elements synthesized by stars.
The presence of dust is relevant in many astrophysical environments
and has a crucial role in shaping the spectra of many cosmic bodies.
However, its existence has been inferred from very indirect evidences 
up untill recently.
The first evidence came from the discovery of a tenuous screen of small particles
around the Earth producing the {\sl zodiacal light}. 
Other evidences came from observations of 
dust trails of comets, circumstellar dust envelopes around evolved stars,
diffuse dust in the MW producing the interstellar extinction,
the discovery of IR emission by galaxies and ultra-luminous IR galaxies 
in the IRAS era, circumnuclear dust in AGNs (essential ingredient of the unified model for
AGNs), the cosmological IR background (COBE, 1996-1998),  and eventually
the discovery of sites of extremely active star formation at high redshifts 
(SCUBA and ISO, 1998-2000).

Accounting for the effects of dust is essential not only to
understand the {erosion of optical light}, but, even more importantly,
to evaluate the {energy re-emitted by dust at longer wavelengths,}
typically at $\lambda \sim 5$ to 1000 $\mu$.
This is crucial for estimating all basic properties of distant galaxies:
the {\sl Star Formation Rate} (SFR) from various optical and IR indicators, 
the {\sl ages of stellar populations}, which, based as they are on optical colours, have to
distinguish the reddening of the spectrum due to aging from that due to dust extinction, 
and finally to constrain the stellar {\sl initial mass function} (IMF).

\subsection{Dust grains in the ISM}

Rather than by stars, the available volume in a galaxy is occupied by the
ISM, which in local late-type
systems amounts to $\sim 10\%$ of the baryonic mass. The ISM includes gas mixed
with tiny solid particles, the {\sl dust}, with sizes ranging from a few 
\AA\ (the PAH molecules) up to $\sim
10 \, \mu$m. The mass in dust is typically 0.5 to 1\% of the ISM mass.

\subsubsection{Grain production}

The mechanisms of birth, growth and destruction of grains are very complex and  
poorly understood. It is believed that condensation nuclei for dust grains 
mostly form in dense regions of the ISM, which are better shielded from UV photons.
Main dust production sites are hereby listed.

\noindent
{\sl Envelopes of protostars:} during the process leading to the birth of a star
a solar nebula is produced, where silicate grains can be formed
and then blown away by a Pre-Main Sequence wind (T Tauri phase).

\noindent
{\sl Cold evolved stars:} in the cold atmospheres of evolved giants, dust grains can form
and drive a strong stellar wind, in particular graphite grains
from carbon stars and silicate grains in OH-IR stars. Stars with $M<8M_\odot$ 
are important dust producers; higher-$M$ stars, like Wolf-Rayets with high
mass-loss rates, are too rare.

\noindent
{\sl Type-II supernovae} are probably the most important contributors, as
revealed by a variety of tests, like those provided by the IR excesses in the light-curve 
and the extinction of background stars in SN ejecta.
Direct evidences of dust production came from the case-study of SN 1987a 
(CO and SiO molecules found in the ejecta),
the dark spots observed in the synchrotron nebula of Crab, 
the IR mapping by ISO of Cas-A which resolved
clumpy emission associated with the fast moving knots (Lagage et al. 1996).

\noindent
{\sl Type-I supernovae} have an uncertain role, with no evidence yet for dust formation 
(which would be otherwise relevant to solve the problem of the Fe depletion).

\noindent
{\sl The general interstellar medium} is also the site of a slow growth around 
pre-existent condensation nuclei (refractory cores); it is in this way that dirty icy grains
are produced.

\subsubsection{Grain destruction}

Grain survival is another, uncertain, chapter of the complex story of dust enrichment 
of the ISM.  Grain destruction is not likely a problem in stellar winds, the grain should
survive the injection into the ISM, while 
it is more a problem for SN ejecta (which have typical velocities in excess of 
$1000\ Km/s$).

Even after the ejection phase, the ISM is in any case a difficult
environment for grain survival: grains can be destroyed there by
evaporation, thermal sublimation in intense radiation fields,
evaporation in grain-grain collisions, and by radiative SN shocks.

\subsubsection{The evolution of the dust content in a galaxy}

Modelling the complex balance between grain production and destruction is also guided by
observations of isotopic anomalies in meteorites and of the elemental depletion pattern.
A detailed account of most plausible intervening processes in the dust life cycle can be found 
in Dwek (1998). The author also discusses evolution paths of the elemental abundances
in the gas and dust phases in a typical spiral galaxy, based on standard assumptions for
the infall of primordial gas and chemical evolution.
Type-II SN are found to be the main producers of silicate dust in a galaxy, while carbon
dust is due to lower mass (2-5 $M_\odot$) stars. The different lifetimes of the two imply
likely anomalous abundance ratios between the various dust grain types during the course of 
galaxy evolution, naturally evolving from an excess of silicate to an excess of carbon grains
with galactic time.

Altogether, the dust mass is found to be linearly proportional to the gas metallicity
and equal to 40\% of the total mass in heavy elements in a present-day galaxy.
Although the details can depend to some extent on the evolution of the SFR with
time (e.g. in the case of elliptical galaxies this evolution could have been more rapid,
see Mazzei, De Zotti \& Xu 1994), these general results are not believed to be much affected.

\subsection{Interactions between dust and radiation}

Dust particles interact with photons emitted by astrophysical sources by absorbing, 
scattering, and polarizing the light (the combined effect of absorption and
scattering takes the name of {\sl extinction}). They also emit photons at
wavelengths typically much greater than those of the absorbed photons.
The total intensity radiation field $I_\nu(\vec r, t)$ (defined as usual by
$dE \equiv I_\nu d\nu d\Omega dA dt$, $dE$ being the differential amount of 
radiant energy) is related to the field sources by the {\sl transfer equation}:
\begin{equation}
\frac{dI_\nu}{d\tau} = -I_\nu + S_\nu, 
\label{transfer}
\end{equation}
where $d\tau_\nu\equiv \alpha_\nu ds$ is the differential optical depth
corresponding to a spatial path $ds$,
$S_\nu\equiv j_\nu/\alpha_\nu$ is the  {\sl source function},
$\alpha_\nu$ and $j_\nu$ being the {\sl extinction} (true absorption
+ scattering) and emission (true emission + scattering) coefficients.
A medium is said {\it optically thin} or {\it thick} if
$\tau_\nu$ along a typical path trough the medium is $<<1$ or
$>> 1$.
{Absorption} includes those processes in which the energy of photons
is turned into other forms (may be internal energy of matter or fields),
{true emission} is the opposite processe, whilst in scattering the energy
of photons is simply {\sl deviated} into other directions.
Dust scattering is usually elastic.
A formal solution to eq.(1) [e.g. Rybicki \& Lightman 1979] is given by:
\begin{equation}
I_\nu(\tau_\nu)=I_\nu(0) \exp(-\tau_\nu) +\int_0^{\tau_\nu}
\exp(-\tau_\nu+\tau'_\nu) \, S_\nu(\tau'_\nu) \, d\tau'_\nu
\label{sol}
\end{equation}
If each dust grain has a $\lambda-$dependent effective {\it cross section} 
$\sigma_\nu$ and spatial density $n$, then  
$\alpha_\nu=n \, \sigma_\nu$ or $\tau_\nu=N \, \sigma_\nu$, where $N$ is
the {\it column density}. For dust grains it is common to write
$$
\sigma_\nu=Q_{\nu,e} \, \sigma_g=(Q_{\nu,a}+Q_{\nu,s}) \, \sigma_g
$$
where $\sigma_g$ is the geometrical cross section ($\pi a^2$ for spheres) 
and $Q_{\nu,e}$ is the {extinction efficiency} (true absorption + scattering).
At short$-\lambda$ (UV), diffraction effects in the photon-grain interaction become
negligible, and the effective cross-section coincides with the geometric one,
$Q_{\nu,e}\sim 1$.  Altogether: $\alpha_\nu=Q_{\nu,e} \sigma_g \, n$.

The {albedo} $a_\nu=Q_{\nu,s}/Q_{\nu,e}$ is the fraction
of extinguished light being scattered by the grain rather than absorbed.

The emission coefficient $j_\nu$ includes a {\sl true} emission
$j_{\nu,e}$ and an elastic scattering component, $j_{\nu,s}$, given by:
$$
j_{\nu,s}(\hat \omega)=Q_{\nu,s} \sigma_g \, n_d \, \frac{1}{4 \pi}
\int_{4 \pi} I_{\nu}(\hat \omega') f_\nu (\hat \omega- \hat \omega') d\Omega
$$
where $f_\nu$ is the {\it phase function}, depending on the
incidence--scattering angle.

The true emission of dust grains is thermal.
From Kirchoff's law [$j_\nu=\alpha_\nu \, B_\nu(T)$]:
\begin{equation}
j_{\nu,e}=n_d \, Q_{\nu,a} \, \sigma_g \, B_\nu(T_d) .
\label{eq1}
\end{equation}

It is clear that both terms of the emission coefficient depend on
the radiation field $I_\nu$. In particular $j_{\nu,e}$ depends on it trough the dust grain
temperature $T$: grain heating is almost always dominated by the radiation field.
Thus a primary task is to compute $T$. Two situations apply.

\noindent
{\sl (a) Grains sufficiently large and massive} don't cool in the time interval
between absorption of two photons: they are in {thermal
equilibrium} with the radiation field. Their temperature can
be determined by solving for $T$ an energy conservation equation 
{\it absorbed energy = emitted energy}:
\begin{equation}
\int Q_{\nu,a} \, J_{\nu} \, d\nu = \int Q_{\nu,a} \, B_\nu(T) \, d\nu
\label{balance}
\end{equation}
where $J_{\nu}=1/4\pi \int I(\nu,\omega) d\Omega$ is the angle-averaged $I_{\nu}$.

\noindent
{\sl (b) Small grains fluctuate in temperature} at any acquired photon. 
They never reach thermodynamic equilibrium (the cooling time is shorter than
that between two photons arrivals).
A probability distribution $P(T)dT$ to find a grain
between $T$ and $T+dT$ can then be computed based on a statistical approach
(Puget et al. 1985; Guhathakurta \& Draine 1989, Siebenmorgen \& Kruegel 1992).
Basic ingredients for this computation are:

- the specific heat $C(T,a)$ per C-atom of PAH's of size $a$ and the number $N_c$ 
of C atoms in the grain;

- the maximum T a PAH can attain after absorption of a photon $h\nu$, and
given by the relation:
$$
h\nu' = \int _{T_{min}}^{T_{max}} N_c(\nu',a) C(T,a) dT ;
$$

- the cooling rate of a PAH after being heated to $T_{max}$ is
$$
dT/dt = {4\pi a^2 F(T,a) \over C(T,a)}
$$
where $F(T,a)=\int Q_{abs}(\nu,a)\pi B(\nu,T)d\nu$ 
is the power radiated per unit grain surface.
The total IR spectrum radiated during the cooling down is:
$$
S(\nu',\nu,a) = \int_0^t dt\ \pi B(\nu,T) Q_{a}(\nu,a) 4\pi a^2 =
\int_{T_{min}}^{T_{max}} dT\ \pi B(\nu,T)Q_{a}(\nu,a) {N_c\ C(T,a) \over F(T,a)} .
$$
In any case, dust grains are destroyed by radiation-induced temperatures above
$\sim 1000 \div 2000$ K (depending mainly on composition). This is
the reason why their emission is relevant only longwards a few $\mu$m.

For mixtures of different species of particles
the equations must be summed over all the species. For spherical
grains of different compositions and sizes $a$ and density $n_i(a)$:
$$
\alpha_\nu=\sum_i \int n_i(a) \, Q_{i,\nu,e} \, \pi a^2 \, da.
$$

The interaction of a dusty medium with the radiation field then requires the 
knowledge of the quantities $Q_{\nu,a}$, $Q_{\nu,s}$ and $f_\nu$. 
The Mie (1908) theory provides analytic solutions for homogeneous spheres and infinite
cilinders. Otherwise, for irregularly shaped and inhomogeneous grains good 
approximations can be obtained by simple generalizations of the exact solutions
for spheres and cylinders (e.g. Hoyle \& Wickramasinghe 1991; Bohren and Huffman 1983).

As a source of scattering (like the $e^-$), another important effect of dust is 
to induce polarization in the emitted light. Two ways for dust to produce this
are through {\sl (a)} light transmission in a dusty medium including oriented 
bipolar components; or {\sl (b)} dust reflection (e.g. in AGNs).
Should we be interested in modelling these effects of dust on polarization, then
solutions of the transfer equation (\ref{transfer}) for all four Stokes parameters 
would be required.

\subsection{Alternative heating mechanisms for dust}

Two other heating mechanisms for dust grains can operate (Xu 1997).
  
{\sl (a) Collisional heating} for dust mixed with thermal gases.
In the HI component of the solar neighbourhood the ratio of collisional 
heating $G_{coll}$ to radiative heating $G_{rad}$ turns out to be
$$G_{coll}/G_{rad}|_ISRF \simeq v_{HI,thermal}/c ;$$
i.e. the collisional is 5 orders of magnitude less than radiative heating! Only in very hot 
plasmas (IC plasmas at $T>10^7$) the two can get comparable.

{\sl (b) Chemical heating}, a process occurring typically in the cold gas component of the ISM,
e.g. when an $H_2$ molecule is formed on the
surface of a grain from the combination of 2 H atoms:
$$  H+H \rightarrow H_2 + 4.48 eV  .  $$
Most of this chemical energy is absorbed by the grain (the rest is taken by the 
molecule).
The released energy turns out comparable with the collisional one (hence negligible).

\subsection{The interstellar extinction curve}  

Before IRAS, the properties of interstellar dust were mainly inferred
from the dimming of optical light of stars inside the Galaxy.
If we observe the light from a source through a dust screen, dust emission is 
negligible in the optical (dust emits significantly only in the IR), 
offline scattering is unimportant, and the formal solution (eq. \ref{transfer}) simplifies to
${I_\nu}={I_\nu(0)}e^{-\tau_\nu}$. Given a source with unextinguished flux 
$m_\lambda(0)$, the extinction in magnitudes is:
$$
A_\lambda \equiv m_\lambda-m_\lambda(0) = \frac{2.5}{\ln 10} \tau_\nu
\simeq 1.08 \, \tau_\nu.
$$
%
%Interstellar extinction can be recognized and measured
%through its , which modifies
%the apparent color of sources.
%
%Astronomers define the {color index}
%$C_{ij}\equiv m_{\lambda_i}-m_{\lambda_j}$. The
%observed c.i.\ is linked to the intrinsic one by
%\[
%C_{ij} = C_{ij}(0)+E_{ij}
%\]
%
%where  $E_{ij} \equiv A_{\lambda_i}-A_{\lambda_j}$ is the
%{color excess}.
%
%To study the wavelength dependence of extinction several
%equivalent quantities,
%depending only on the optical properties of the medium and not from its
%column density, are used in the literature.
%
%The `classical' ones are
%\[
%R_\lambda \equiv \frac{A_\lambda}{E_{ij}}\; \; {\rm and} \;\;
%e_\lambda \equiv \frac{A_\lambda-A_{\lambda_j}}{E_{ij}}
%\]
%
%were $\lambda_i$ and $\lambda_j$ are fixed wavelengths (mostly
%those of the B and V bands). For a Galactic extinction, $R_\lambda=A_\lambda/E(B-V)$ 
%is 3 for the V band and 4 for the B.
%
%Currently the quantity {$\tau_\lambda/N_H$} is more frequently used.
%
%
%\subsection{The interstellar extinction curve}
%
%We give emphasis here to the best-known extinction function, the one measured
%
%
The knowledge of the intrinsic colors for a source population allows to 
determine the wavelength dependence of the extinction curve.
The mean extinction curve along most line-of-sights in the Milky Way has been studied 
by many authors (see references in Hoyle \& Wickramasinghe 1991; and see Figure 1).
Its main properties are:
{\sl (a)} a growth in the optical--near UV, more than linear with frequency, 
$\tau \propto \nu^{1.6} \propto \lambda^{-1.6}$ ($0.6-5\ \mu m$);
{\sl (b)} a bump around $2175$ \AA;
{\sl (c)} a steeper rise in the far--UV;
{\sl (d)} two features in the mid--IR at 9.7 and 18$\mu$m.

The extinction curve is not universal: in the Milky Way it depends on the line of sight.
Data on other stellar systems (LMC and SMC for example) suggest a variable behaviour,
in particular in UV. 

More recently it has been possible to evaluate indirectly the extinction curve
is distant galaxies, by means of accurate photometric observations in narrow-band
filters. Gordon et al. (1997) (see also Calzetti 1997) 
analyze colour-colour plots for 30 starburst galaxies, 
inferring starburst ages and extinction properties. The 2175 $\AA$ 
bump is absent and the rise in the far-UV slower than observed for the Milky Way.
The authors suggest that the starburst has modified the grain distribution, in particular 
suppressing the 2175\AA ~feature observed in the MW.
Alternatively, Granato et al. (2000) reproduce the observed extinction law
in starbursts as a purely geometrical effect, by using the same dust grain mixture than 
for the MW and accounting for differential extinction for young and old stars (see Sect. 4
below).

\subsection{Models of the interstellar dust}

\begin{figure}
\vspace{1cm}
\psfig{figure=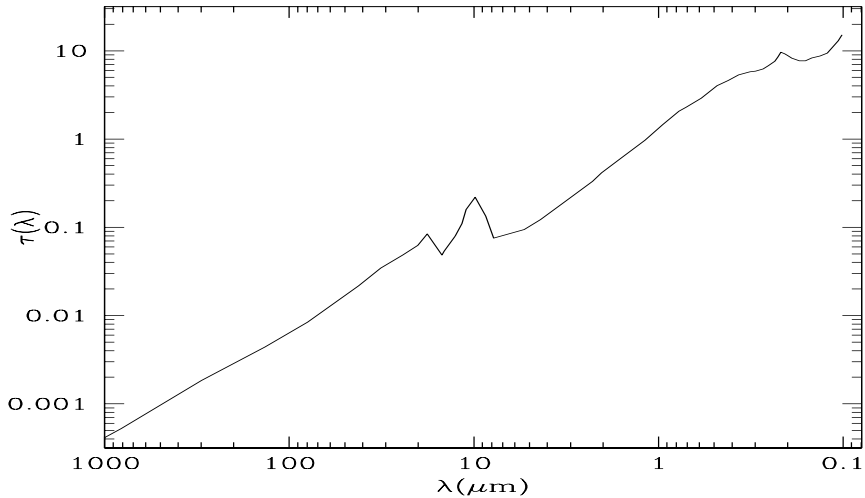,height=70mm,width=120mm}
\caption{The galactic extinction curve, in optical depth per unit value
of E(B-V). The two silicate features at 10 and 18 $\mu m$ and that of
carbonaceous grains at 2175 $\AA$ can be recognised.
}
\label{ext}
\end{figure}

The extinction curve, whose main features are reported in Fig.\ref{ext},
can be explained by a mixture of grains with different sizes and compositions.
The curve in the optical is reproduced by grains with $a \sim 0.1 \mu$m,
while the fast growth of the extinction curve in UV requires smaller
particles with $a \sim 0.01 \mu$m. Silicate grains explain the 9.7 $\mu$m
and 18 $\mu$m emission features, whose large widths 
suggest the presence of many impurities ({\it dirty} or
{\it astronomical} silicates).

On the contrary, silicates cannot explain the optical extinction, because of
their excessive albedo. Here carbonaceous grains (graphite or amorphous carbon)
are proposed as main absorbers, their 
resonance at $2175\ \AA$ nicely fitting the observed UV bump.
The non-linear growth in the FUV is probably due to very small grains
and PAH molecules (required also to explain the interstellar IR emission bands,
e.g. Puget \& Leger 1989).

Unfortunately, the extinction curve does not constrain enough the properties 
of interstellar dust. For this reason, a variety of models, all with the above 
basic ingredients, have been proposed to reproduce it.

% for example including or not 
%populations of very big grains assumed 

Draine and Lee (1984) adopt a power law size distribution of silicate and graphite
grains $dn/da \propto n_H a^{-3.5}$ for 0.005 $\mu m < a < 0.25 \mu m$.
A quite more complex model by Siebenmorgen \& Krugel (1992) includes five classes of
grains (amorphous carbon, silicates, very small grains, small PAH
and clusters PAH), providing an impressive fit to the extinction
curve. The one by Rowan--Robinson (1992) with a discrete set of nine kinds
of grain (amorphous carbon size $a= 30 \mu m$ and
$a= 0.1 \mu m$; graphites with $a=$0.03, 0.01, 0.002 and 0.0005 $\mu m$;
amorphous silicate $a= 0.1 \mu m$ and silicates with $a=$ 0.03 and 0.01 $\mu m$)
explains also the FIR emission from circumstellar envelopes.
The population of very big grains is assumed here to explain the
sub-mm emission of carbon stars.

The most relevant recent improvement with respect to the classical models by
Drain \& Lee is the addition to the grain mixture of
very small particles and macro-molecules reaching temperatures higher than 
equilibrium because of their small size, as described above.
Two regions of the extinction curve are particularly sensitive to the
presence of these small particles: the mid-IR spectrum (including the emission bands 
at 3.28, 6.2, 7.7, 8.6 and 11.3 $\mu m$, and appreciable continuum), 
and the fast far-UV rise.

The mid-IR emission bands, in particular, are
most commonly interpreted as due to a family of very stable planar molecules,
the PAH's (polyciclyc aromatic hydrocarbons), whose vibrational spectra closely
resemble, according to laboratory tests, those of emission bands.
PAH emission features originate mainly in the so-called {\it photo-dissociation
regions}, i.e. in the interfaces between molecular clouds and the HII
regions, where the cloud surfaces are illuminated by the 
high energy field of the young stars.
There are evidences that in denser environments and stronger UV 
field intensities the PAHs (and the associated mid-IR bands) could be depleted.
In the circum-nuclear dusty regions around AGNs PAH emission is not observed.

PAH emission features have been observed by ISO to display Lorentian profiles,
whose broad overlapping wings may mimic a kind of continuum (Boulanger et al. 1998). 
This may possibly explain the observed underlying mid-IR continuum in many astrophysical
objects.

%Current models (most recent: Desert et al. 1990; Dwek et al. 1997; Silva et al. 
%1998) reproduce fairly well the IS extinction curve from the UV to the IR.
%In the sub-mm ($\lambda>300\mu$) a few observations seem to display excess
%extinction, possibly requiring the presence of very large grains 
%($a= 30 \mu$m) in the dust mixture (Rowan-Robinson 1992).
%In detail, the match of the extinction curve and dust emission from our Galaxy
%obtained by current models is not perfect: in particular the {galactic emission
%might require slightly larger fractions of very small grains, which would imply
%a larger UV extinction than observed}.

\section{EVALUATING THE DUST EMISSION SPECTRA}

Knowing, or guessing, the optical properties of dust, one can predict
the spectra of dusty systems. From a computational point of view,
we have to distinguish two cases.

\begin{itemize}

\item
If the IR dust emission is not self-absorbed ($\tau_{IR} <<1$),
the emitted spectrum is simply the volume integral of the local
emissivity. 
An ambient of this kind is the diffuse dust in the IR galactic {\it cirrus}.
Solution of the energy balance equation (\ref{balance}) provides the $T$ distribution 
for the various grain species. Since, in particular, $Q_{\nu,(a,s)} \sim 1$ in UV
and $Q_{\nu,(a,s)} \propto a/\lambda^ {1.5-2}$ in the far-IR,
and considering that the left-hand side is dominated 
by absorption of UV photons while the right hand by emission at long wavelengths,
eq. (\ref{balance}) can be re-written for a given grain specie as:
\begin{equation}
I_{bol}=\int J_{\nu}d\nu \simeq  \int B_\nu[T_g(a)]Q_{\nu,(a)} d\nu.
\label{T} 
\end{equation}  
Since $\int B_\nu[T_g(a)] d\nu = aT_g^4$ and because of the additional dependence implied
by $Q_{\nu,(a)}\propto \nu^{1.5-2}\propto T^{1.5-2}$,  the grain
equilibrium temperature $T_g$ is found to depend very weakly on the intensity of the local
radiation field:
\begin{equation}
T_g \propto (I_{bol}/a)^{1/6}.
\label{T}
\end{equation}
This implies that dust emission spectra in a variety of galactic environments 
(from quiescent to actively starbursting galaxies and AGNs) 
are quite stable and robust, with peak emission mostly confined to the 
wavelength interval $\lambda_{peak} \simeq 100$ to 30 $\mu m$.
Longward of $\lambda_{peak}$ and after eq. (\ref{eq1}), dust spectra converge according to
the RJ law as 
$$I_{\nu} \propto  B_\nu[T_g(a)]Q_{\nu,(a)} \propto \nu ^{3.5-4},$$
in agreement with mm observations of local IRAS galaxies by Andreani \& Franceschini (1996) 
and Chini et al. (1995).

\item
Otherwise, in the presence of IR-thick media (e.g. dense molecular clouds and dusty torii 
in AGNs), one is faced by the difficult task to solve the transfer equation. 
We expect  that in thick media the IR spectrum will be erased at the
short wavelengths (typically in the near- and mid-IR, but sometimes even in the
far-IR) by self-absorption.

%{\it lambda-iteration method}, a straightforward application of
%the formal solution.
%
\end{itemize}

\subsection{Radiative transfer in thick dusty media}

In most practical cases, the radiative transfer equation can be solved
only with numerical techniques. We mention in this Section a couple of such
approaches quite often used.

\subsubsection{Numerical solutions based on iterative schemes}

A first class of solutions adopt an iterative numerical scheme based on applications
of the formal solution of the transfer equation (eq.[2]).
This was originally developed for interpreting AGN spectra
(Granato \& Danese 1994; Pier \& Krolick 1992; Granato, Danese \& Franceschini 1997),
but is useful to treat more generally radiative transfer in thick media.
Although the source function can be any kind in principle, we discuss here an application
by Granato \& Danese for a central point-source and for a planar and azimuthal symmetry 
of the dust distribution within a minimum $r_m$ and maximum $r_M$ radii. 
A condition is set on $r_m$ because of dust sublimation: it cannot be lower than
$r_m= L_{46}^{0.5}\ T_{1500}^{-2.8} \ (pc)$ to avoid exceeding an equilibrium grain 
temperature of $T_{gr}=1500$ for graphite and $T_s=1000$ for silicates. 

The two fields to solve for are the radiation field intensity $I_\nu(r,\Theta,\theta ,\phi)$
and the grain temperature distribution $T(r,\Theta)$.
%
%{Setting parameter grids}: 3 polar coordinates (r, $\Theta$, $\Phi$) set the position %within the dust distribution, while another 2-coordinate set ($\theta$, $\phi$) sets the %direction of light emission from the point
%${\omega}$ being the direction specified by $\theta ,\phi$, $\alpha$ and $j$ the
%absorption and emission coefficients. Assuming LTE and writing as $B_\nu(T)$ the
%Planck function:
%$$ \alpha_\nu = \sum_l Q_{\nu, e} \sigma n     $$
%$$  j_\nu = \sum_l Q_{\nu, a} \sigma n B_\nu(T)  + \\
%\sum_l   Q_{\nu, a} \sigma n {1\over 4\pi} \int I_\nu (\omega) f(\theta) d\Omega  , $$
%$Q_{\nu, e}$, $Q_{\nu, a}$, $Q_{\nu, e}$, $Q_{\nu, s}$ are the extinction, absorption and
%scattering efficiencies of each species of grains, $f$ the scattering phase functions.
%
The solution is found by representing the field intensity as the contribution of two terms
\begin{equation}
 I_\nu = I_\nu ^1 + I_\nu^2  ,
\label{two}
\end{equation}
the first term being the radiation field emitted by the central source
and estinguished by the dust, with trivial solution from eq.(\ref{transfer}):
\begin{equation}
I_\nu ^1 ={1\over 4\pi} {L_\nu(\Theta) \over 4\pi r^2} 
exp[-\tau_\nu(r,\Theta)] , 
\label{three}
\end{equation}
$L_\nu(\Theta)$ becoming dependent on direction because of differential 
extinction, $\tau$ being the optical depth to the point (r, $\Theta$, $\Phi$).
The second term originates from thermal emission by dust, and may be expressed at the
zero-th order as the formal solution (eq. \ref{sol}) of the transfer equation: 
\begin{equation}
I_\nu ^2(r,\Theta, \theta,\phi) = \int _{(r,\Theta)}^{(\infty, \infty)} 
S_\nu(r^\prime,\Theta^\prime) exp[-\tau_\nu(r^\prime,\Theta^\prime)] 
d\tau_\nu  
\label{four}
\end{equation}
The quantity $S_\nu$ is the source function $j_\nu/\alpha_\nu$ which,
if the scattering is isotropic, can be expressed as a weigthed average of the 
scattering and absorption (Rybicki and Lightman 1979) summed over all grain species:
\begin{equation}
S_\nu={ \sum_i \sigma n [Q_{\nu, a} B_\nu(T) + Q_{\nu, s} J_\nu(r,\Theta)]
\over  \sum_i \sigma n Q_{\nu, s}}
\label{five}
\end{equation}
The function $J_\nu$ is the direction-averaged radiation field intensity
$\int I_\nu d\Omega$: this integral obviously includes both contributions
to the total intensity in eq.(\ref{two}).
Finally, assuming radiative equilibrium for the dust grains, the grain temperature 
distribution is found from eq.(\ref{balance}). The following {iterative scheme} is used 
to obtain a solution for $I_\nu$:

(1) the zero-th order approximation for $I_\nu^1$ in eq.(\ref{two}) is obtained from
 eq.(\ref{three}) given $L_\nu$ and the adopted dust distribution;

(2) then a zero-th value for the $T$-field is found from eq.(\ref{balance});

(3) the source function $S_\nu$ is then computed from eq.(\ref{five}) including the
contribution from thermal dust emission;

(4) after eq.(\ref{four}) the second term $I_\nu^2$ of the radiation field is computed
and the total field intensity in eq.(\ref{two}) is updated;

(5) convergence is achieved when e.g. $dT$ from one step to the other
is less than a small fixed amount.
Suitable scaling rules are usually adopted to accelerate the convergence.

%Iterative solutions like this typically require
%some checks: constancy of L from one shell to the other in the 
%dust distribution, and satistaction of the condition for radiative
%equilibrium eq.(6) have to be controlled in the approximate solution.

\subsubsection{Monte Carlo solutions}

The advantage of brute-force solutions like a Monte Carlo simulation is that it is better suited
to treat complex situations for the geometries of the source function and of the spatial 
distribution of the absorber.
Also velocity fields can be naturally considered in the code to map the kinematical structure
of the emission lines (e.g. Jimenez et al. 1999).

The usual approach is to assume a given geometrical distribution for the absorber,
possibly including a velocity field, and to generate inside (or outside) it 
photons according to a given source function (plus a background photon distribution).
All these fields are usually discretized into appropriate spatial grids.
Each photons are then followed through the distribution of the absorber, their
interaction being ruled by the optical depth, albedo and scattering phase functions
at that point.
The simplest geometrical distributions adopted are (e.g. Disney et al. 1989; Gordon et al. 1997):
{\sl the mixed}, in which the source and absorber are homogeneously distributed;
{\sl  the shell}, where the source and absorber are separated, typically the former inside
and the latter outside acting as a screen.
However, much more complex situations can be described this way, up to fully 3D
distributions without any symmetries (Jimenez et al. 1999).

\section{GENERALIZED SPECTRO-PHOTOMETRIC MODELS OF GALAXIES}

Twenty years after the first serious models of stellar population synthesis
(Tinsley 1977; Bruzual 1983), the most relevant recent progresses have been
the attempts to provide a self-consistent description of the effects of 
dust (and gas) in galaxy spectra and spectral evolution.  We review in this Section
some recent efforts
of generalized spectral synthesis of galaxies from the UV to the sub-mm,
including dust effects (as for both the extinction of the primary optical spectrum, 
and dust re-radiation at longer $\lambda$)  in the various galactic environments.

Dust plays an important role in all relevant galactic sites:
{\sl (1)} the neutral interstellar medium, whose associated dust is heated by the general
radiation field ({infrared cirrus}, prominent in the $100 \mu$m IRAS band);
{\sl (2)} the dense cores of molecular clouds, where dust optical-depth is very high
and prevents light from very young stars to be observed;
{\sl (3)} dust in the external layers of molecular clouds (PRD regions), heated by the
interstellar radiation field and OB associations formed in the clouds;
{\sl (4)} dust around protostars;
{\sl (5)} dust around evolved giants and young planetary nebulae;
{\sl (6)} hot dust associated with HII regions.

%The mid- and far-IR observations on galaxies can be explained by the presence of
%three dust components: a { cold}
%($T\sim 20$ K) cirrus component, a {warm} ($T\sim 40\div 100$ K)
%component associated with star birth and death, and a {hot} component,
%including the very small grains and PAHs, and associated with the previous
%sites in different proportions.

{\sl The inclusion of dust means a dramatic complication of  spectro-photometric
models: the usual assumption of population-synthesis codes -- 
that the global emission of a whichever complex stellar system is simply the
addition of the integrated flux of all components independently on the geometry
of the system -- is no more valid:
not only the extinction process depends in a complex way on the relative
distributions of stars and dust, but also 
dust emission itself, at high dust column densities and according to 
geometry, may be optically thick.}

In principle, accounting for dust effects in detail may require a very complex 
description of: (1) the physical-chemical-geometrical properties of grains,
determining their interactions with the radiation field (e.g. 
amorphous, porous low-albedo grains vs. highly reflective grains);
(2) the chemical composition of the ISM where grains have condensed (which affects
the dust composition), given by the integrated contribution of all previously
active stellar populations in the galaxy;
(3) the modifications that grains and molecules undergo during the course of
evolution, i.e. sublimation in strong UV radiation fields, sputtering, etc.
(see Sect. 2.2).

{\sl These complications of the classical purely stellar evolutionary
codes cannot be avoided, if we want a complete and reliable description of physical processes
inside galaxies}. As we will discuss in later Sections, this turns out to be
particularly critical when describing what we called the {\sl active phases}
during galaxy evolution: {\sl neglecting dust effects in such cases would bring to
entirely wrong conclusions}.

On the other hand, the uncertainties introduced by the large number of new parameters 
are largely reduced by adopting a multi-wavelength (UV through mm)
approach, which balances the unknowns with the number of 
constraints coming from a wide-band observed spectrum.

\subsection{Semi-empirical approaches}

A phenomenological approach to a global spectrophotometric description of
galaxy evolution was recently discussed by Devriend, Guiderdoni \& Sadat (1999).
This paper elaborates separately the code for stellar population synthesis from 
that of dust emission. The former is treated with the most recent prescriptions.
The dust emission is schematically represented
as the contribution of four different components: 
the PAH emission features, very small grains, big grains illuminated by the 
general galactic radiation field (cold dust), and big grains illuminated by young 
stars in starburst regions.
These four components are modelled using typical parameter values for the dust
composition, radiation field intensity, mass, etc.
Relative normalizations of the four components are finally calibrated using the 
observed relationship between the IRAS colours of
galaxies and the bolometric luminosity.

This approach is quite fast as for computation time (in particular it
overcomes the problem of solving the radiative transfer equation),
and is particularly useful for statistical analyses of large galaxy databases.

\subsection{Detailed self-consistent spectro-photometric models}
  
More physically detailed descriptions of the galactic dust emission are
discussed by several teams. These models interface two logical procedures:

\begin{itemize}
\item    (1)
the first is to describe, given a prescription for the IMF, the history of 
star-formation in the galaxy as a function $\psi(t)$ detailing the mass in stars formed
per unit time $t$, the actual gas metallicity
$Z(t)$, the abundances of various elements produced by stars during evolution,
and the residual gas fraction $g(t)$ as a function of time;

\item  (2)
the second step is to sum up, at any galactic age $t$, the fluxes from all populations 
of stars, by solving the radiative transfer equation taking into account how stars 
and the residual gas and dust are geometrically distributed.

\end{itemize}

\subsubsection{Chemical evolution of the ISM}

While point (1) above is addressed in detail by other contributions to these 
Book (Bruzual), we remind here a few basic concepts. 

A galaxy is usually modelled from the chemical point of view as a single environment 
where primordial gas flows in according to an exponential law
\begin{equation}
\dot M(t)\propto exp(-t/t_{inf}).
\label{inf}
\end{equation}
The SFR follows a general Schmidt law
\begin{equation}
\psi(t)=\nu M_g(t)^k  \propto g(t)^k
\label{schmidt}
\end{equation}
with the addition of one or more bursts of star-formation to describe starburst
episodes possibly triggered by galaxy interactions or mergers. 
The typically adopted value for $k$ is 1.
For the initial mass function (IMF) the usual assumption is a Salpeter law
\begin{equation}
d \phi(M) \propto M^{-x} dM, \ \ \ x=2.35, \  M_{min}<M<M_{max}=100\ M_\odot  
\label{salp}
\end{equation}
with typically $M_{min}=0.1\ M_\odot$ (but higher values may apply for example
in the case of starbursts). The observed photometric properties of
galaxies of various types and morphologies are reproduced by varying in particular
$t_{inf}$ and $\nu$.

Given the above parameters, the solution of the equations of chemical evolution allow to 
compute at any given galactic time all basic quantities, in particular the functions 
$g(t)$ and $Z(t)$, and then, after eq.(\ref{schmidt}), the number of 
stars generated at that time with metallicity $Z(t)$.
The integrated spectrum of each stellar generation 
(Single Stellar Population, SSP) then evolves according to the 
prescriptions of stellar evolution, defining a 2D sequence (spectral intensity
$L[\nu,t]$ vs. frequency $\nu$ as a function of time, $t$).

\subsubsection{Geometrical distributions of gas and stars}

In the model by Silva et al. (1998) three different stellar and ISM components are 
considered in the generic galaxy:
{\sl (a)} star-forming regions, comprising molecular clouds (MC), with young stars, gas 
and dust in a dense phase, and HII regions;
{\sl (b)} young stars escaped from the MC complexes;
{\sl (c)} diffuse dust ("cirrus") illuminated by the general interstellar radiation
field.

For disk galaxies the adopted geometry is a flattened system with azimuthal
simmetry and a density distribution for the 3 above components described by {\sl
double exponentials}: $\rho=\rho_0 exp(-r/r_d) exp(-|z|/z_d)$.

For spheroidal galaxies, spherical symmetry is adopted with {King profiles}
$\rho=\rho_0 ( (1+[r/r_c]^2)^{-\gamma} - (1+[r_t/r_c]^2)^{-\gamma})$,
with $\gamma=3/2, [r_t/r_c]\sim 200, r_c\sim 300 pc $ as typical values.

\subsubsection{Models of the molecular clouds (MC)}

High-resolution CO and radio observations show that MCs are highly structured objects 
containing very dense cores where stars are actually formed. 
Typical values for the MCs are: size$\sim 10$ pc, mass$\sim 10^6\ M_\odot$.

All star-formation in the Galaxy 
happens in dusty MCs, the early evolution phases of young star clusters occurr inside
such dusty regions, hence are optically hidden.
Later, on the lifetime of OB stars ($10^6-10^7\ yrs$), the radiation power of 
young stars, stellar winds and the first SNs destroy the parent MCs
and allow the young stellar population to appear in the optical.

Note that, bacause of the clumpiness of MCs, this is in any case a {\sl statistical 
process}: in some clouds even the emission of the youngest OB stars is 
already visible, while in others all young stars are completely embedded
in dust. Silva et al. (1998) describe schematically this transition of the MC from a 
dust-embedded phase to the
optically dominated phase, as a process in which the fraction $f$ of the light
from the SSP generated within the cloud still embedded into
dust decreases linearly with time as $f(t) = 2-t/t_0$,
$t_0$ being the time interval during which the SSP is entirely extinguished.

\begin{figure}
\vspace{1cm}
\psfig{figure=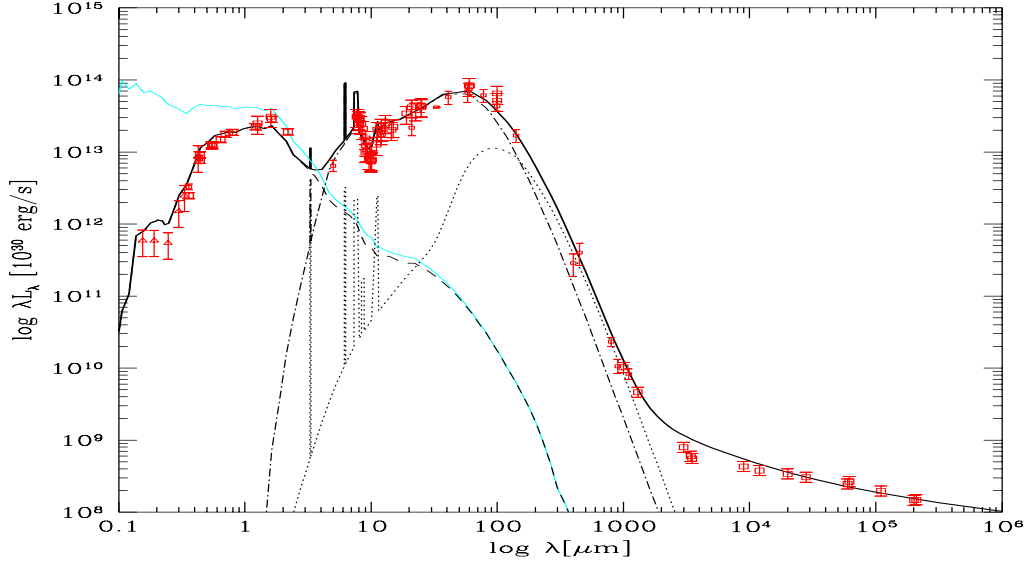,height=80mm,width=140mm}
\caption{
The broad-band (UV through radio) data on the prototype nearby starburst galaxy M82. 
The ordinate axis is normalized to $10^{30}\ erg/sec$.
[Courtesy of G.L. Granato].
}
\label{M82}
\end{figure}

The spectrum emitted by the MC and filtered by dust is computed by solving the 
{\sl transfer equation}, e.g. by assuming that the primary SSP spectrum comes from
a point source in the center of the cloud (this rather crude assumption allows 
substantial semplifications in the numerical code, see above).

%This geometric semplification implies an overestimate of the amount of very hot
%dust with respect to the more realistic case of a random distribution of 
%lower intensity centers of illumination (but this small bias can be easily
%corrected "a posteriori" by cutting off the T-distribution)

A more detailed description of molecular cloud structure and emission is provided by
Jimenez et al. (1999).
Their model is based on fully three-dimensional simulations of the {\sl density}
and {\sl velocity fields} obtained by solving 3D compressible magneto-hydrodynamical
(MHD) equations in supersonic turbolent flows, as typical of the motions in
Galactic molecular clouds (Padoan et al. 1998).
The MHD turbolence generates a large density contrast, with the density field spanning
a range of 4 to 5 orders of magnitude. This brings to a {\sl highly filamentary and 
clumpy morphology}.
All this is consistent with observed properties of the clouds.

Young stars with $M>15-20\ M_\odot$ in this model are heavily extinguished for 
virtually all their live.
A detailed Monte Carlo approach is required to solve the radiative
transfer equation.
The simultaneous knowledge of the density and velocity fields allows also
to estimate in great detail the molecular emission lines (CO).

\subsubsection{Models of diffuse dust (cirrus)}

Diffuse dust in the galaxy is responsible for a general attenuation of the light
emitted by all stars and MC complexes.
In this case the dust column density is not so high to require a detailed
solution of the transfer eq. ($\tau_\nu$ for IR photons is small).
One can express an effective optical depth to account for combined absorption
and scattering (Rybicky and Lightman 1979):  $\tau_{eff} = \tau_a (\tau_a+\tau_s)$.
The galaxy is divided into small volume elements $V_i$, such that the local
radiation field in this elementary volume is
$$ J(\lambda)_i = \sum_k {V_k [j(\lambda)_k^{mc}+j(\lambda)_k^{star}]
e^{-\tau_{eff}(i,k)}\over r^2_{i,k} } , $$
$r^2_{i,k}$ being the distance between the i-th and k-th volumes.
This determines the temperature of the local diffuse dust, whose integrated flux 
seen by an observer in a direction $\theta$ is a 
simple sum over all volume elements of the diffuse dust emissivity:
$$ S(\lambda, \theta) = 4\pi \sum_k V_k j(\lambda)_k e^{-\tau_{eff}(k,\theta)} $$
$\tau$  being the optical depth from the V-element to the outskirts in that direction
and $j(\lambda)_k = j(\lambda)_k^{mc}+j(\lambda)_k^{star}+j(\lambda)_k^{cirrus}$.

\subsubsection{Modelling the SEDs of normal and starburst galaxies }

Figure \ref{M82} shows the broad-band (UV through radio) spectrum of the prototype
starburst galaxy M82, a closeby well studied object at 3.2 Mpc. 
The lines in the figure come from the fit obtained by Silva et al. (1998).
The thin (cyan colored) continuous line peaking at 0.1 $\mu m$ corresponds to the
unextinguished integrated spectrum of all stellar population, while the long-dashed line
is the reddened stellar continuum. The dot-dashed line is the contribution of dust in
molecular clouds, while the dotted line comes from diffuse dust in the "cirrus".
In this model, the optical-NIR spectrum of the galaxy is contributed mostly
by old stellar populations unrelated to the ongoing starburst, whereas the starburst
emission is mostly observable at $\lambda>4\ \mu m$ in the form of dust re-radiation,
radio SN and free-free emissions.

Equal areas in the $\lambda L(\lambda)$ plot of Fig. (\ref{M82}) subtend equal amounts of
radiant energy: it is then clear from the figure that in this moderate starburst 
$\sim 80\%$ of the bolometric flux emerges as dust re-radiation above 5 $\mu m$.
In higher luminosity starbursts and in Ultra-Luminous IR Galaxies (ULIRGs, e.g. Arp 220) 
this fraction gets close to 100\%.
On the contrary, for local normal galaxies the average fraction is only $\sim 30\%$,
as found from comparison of the far-IR with the optical luminosity functions of galaxies 
(Saunders et al. 1990).

\section{INFRARED AND SUB-MM LINE SPECTRA}

IR/sub-mm spectroscopy offers unique opportunities to probe the physical conditions
($n[atoms]$, $P$, $T$, extinction, ionization state) in the various components
of the ISM, because:

\begin{itemize}
\item
the widespread presence of dust makes optical-UV-NIR line diagnostics
completely unreliable;
\item
 most of the line emission by MCs is extinguished and does not appear in the optical;
\item
 lines from molecular phases (including most of the ISM mass)
appear in the FIR-mm;
\item
 several fundamental cooling lines of gas happen in the FIR;
\item
 lines from an extremely wide range of ionization states are observable in the IR.
\end{itemize}

Table 1 summarizes IR tracers of the various ISM components.
Clearly, IR spectroscopy is essential for studies of galaxy activity, though
it requires a continuous coverage of the IR spectrum, possible only from space. 
While ISO allowed to invertigate spectroscopically nearby IR active galaxies, 
future missions (SIRTF, NGST, FIRST) will make possible similar studies 
for galaxies at any redshifts.

\nobreak\begin{table}
{{Table\enskip 1.}\enskip{Relevant components and line tracers of the ISM} \ \ \ \ \ \ \ \ \ \ \ \ \ \ \ \ \ }
\begin{tabular}{llll}\hline
%\multicolumn{1}{c}{
Component & Temperature & Density & Tracers and IR lines\\
%\multicolumn{1}{c}{($\mu$m)} &\multicolumn{1}{c}
%{$\hbox{erg}\,\hbox{s}^{-1}\,\hbox{cm}^{-2}\,\hbox{sr}^{-1}$} 
%&\multicolumn{1}{c}{$\hbox{erg}
%\,\hbox{s}^{-1}\,\hbox{cm}^{-2}\,\hbox{sr}^{-1}$} 
\hline
 Cold gas   & 10--100 K     & 1--1000 $cm^{-3}$  &  $H_2$, CO, PAH's \\
 Diffuse HI & 100--1000 K   & 1 $cm^{-3}$        &  HI 21cm, [CII], [OI] \\
 HII regions& 1000--10000 K & 3-300 $cm^{-3}$    &  $H\alpha$, [OII], [OIII] \\
\end{tabular}
%{$^{(a)}$~Total observed sky brightness in a dark direction
%$^{(b)}$~Extragalactic component}
\end{table}

\subsection{The cold molecular gas}

Looking at the mm/sub-mm spectral lines is the usual way to study the cold
molecular gas, which typically includes the largest mass fraction of the ISM.
The lines come from {\sl rotational} and {\sl vibrational} transitions of
diatomic and polyatomic molecules. 

The very many molecules observable allow to accurately sample the various regimes of
$\rho$, $T$ and elemental abundance.
Unfortunately, the most abundant molecule ($H_2$) is not easily observed directly.
It is seen in absorption in UV, or in the NIR roto-vibrational transitions at 
2.121 and 2.247 $\mu m$. 
Only with mid-IR spectroscopy by ISO it was possible to observe the
fundamental rotational lines at 17 $\mu m$ (S[1]), 28.2 $\mu m$ (S[0]),
and 12.3 $\mu m$ (S[2]) in NGC6946, Arp220, Circinus, NGC3256, NGC4038/39).
These observations indicate very cool gas to be present with very high 
column densities (the transition probabilities of the lines are very low).

Because of the difficulty of a direct measure, the amount of molecular gas ($H_2$)
is often inferred from easier measurement of CO emission lines, assumed an $H_2/CO$
conversion.
CO rotational transitions allow excellent probes of cold ISM in galaxies: the CO brightness
temperature ($\propto$ line intensity) is almost independent on $z$ at $z=1$ to 5,
due to the additional $(1+z)^2$ factor with respect to the usual scaling with the luminosity
distance (Scoville et al. 1996).
CO line measurements have been performed for all IRAS sources in the Bright Galaxy Sample, 
the majority have been detected with single-dish telescopes. 
In the most luminous objects the molecular mass is $0.2-5\ 10^{10}\ M_\odot$,
i.e. 1 to 20 times the content of Milky Way.
Typically 50\% or more of this mass is found within the inner kpc from the nucleus,
the molecular mass substantially contributing to the total dynamical mass
($>50\%$ of $M_{dyn}$).
Unfortunately, detecting CO emission by high-z galaxies has proven to be
difficult (see below).

\subsection{The cold neutral gas}

The diffuse neutral ISM is commonly traced by the HI 21 cm line from ground-based
observations.
HI cooling, which is essential to achieve temperatures and densities needed to trigger
SF, depends mainly on emission by the 158$\mu m$ [CII] line, the 21 cm line and the 
63$\mu m$ [OI] line.

The 158 $\mu m$ [CII] line is a major coolant for the diffuse neutral gas and a fundamental 
cooling channel for the photo-dissociation regions (PDR's), the dense
phase interfacing cold molecular clouds with the HII or HI lower-density gas.
Carbon is the most abundant element with ionization potential (11.3 eV) below the H limit
(13.6 eV): CII atoms are then present in massive amounts in neutral atomic clouds.
The two levels in the ground state of CII responsible for the $\lambda=158\ \mu m$
transition correspond to a
relatively low critical density $n_{crit}\simeq 300\ cm^{-3}$ [the density
at which collisional excitation balances radiative de-excitation]: 
CII is excited by electrons and protons and cools down by emitting a FIR photon.
The CII line intensity is also weakly dependent on $T$, hence a good measure for $P$.
The [OI]145$\mu m$ and 63$\mu m$ lines are also coolants, though less efficient.

\nobreak\begin{table}
{{Table\enskip 2.}\enskip{ The most important IR fine-structure lines.
{\small $^{(a)}$~Line intensity compared with the observed [CII]158$\mu m$ for the
prototypical starburst M82, when available, or predicted by Spinoglio \& Malkan (1992)
from a model reproducing the physical conditions in M82.
}}
}
\begin{tabular}{lllll}\hline
Species & Excitation & $\lambda$ & $n_{crit}$ & F/F[CII]$^{(a)}$ \\
       & potential  & $(\mu m)$ & $cm^{-3}$  &                \\
\hline
 OI   & -     & 63.18  &  5 10$^5$ & 1.4 \\
 OI   & -     & 145.5  &  5 10$^5$ & 0.06\\
 FeII &  7.87 & 25.99  &  2 10$^6$ &     \\
 SiII &  8.15 & 34.81  &  3 10$^5$ & 2.6 \\
  CII &  11.26& 157.7  &  3 10$^2$ &  1  \\
  NII &  14.53& 121.9  &  3 10$^2$ & 0.37\\
  NII &  14.53& 203.5  &  5 10$^1$ & 0.11\\
 ArII &  15.76&   6.99 &  2 10$^5$ & 0.11\\
 NeII &  21.56& 12.81  &  5 10$^5$ & 2.1 \\
 SIII &  23.33& 18.71  &  2 10$^4$ & 0.68\\
 SIII &  23.33& 33.48  &  2 10$^3$ & 1.1 \\
 ArIII&  27.63&  8.99  &  3 10$^5$ & 0.23\\
 NIII &  29.60& 57.32  &  3 10$^3$ & 0.31\\
 OIII &  35.12& 51.82  &  5 10$^2$ & 0.74\\
 OIII &  35.12& 88.36  &  4 10$^3$ & 0.66\\
 NeIII&  40.96& 15.55  &  3 10$^5$ & 0.16\\
 OIV  &  54.93& 25.87  &    10$^4$ & --  \\
\end{tabular}
\end{table}

%\begin{figure}
%\vspace{1cm}
%\psfig{figure=mmastro.eps1,height=110mm,width=140mm}
%\caption{Ionization potenzial for ions in the far-IR.}
%\label{pot}
%\end{figure}

\subsection{The ionized component of the ISM}

Again, a number of lines from atomic species, covering an extremely wide range 
of ionization conditions, %(see Fig. \ref{pot}), 
are observable in the far-IR. Their observations allow
extensive analyses of the physical state of the gas. This,
coupled with the modest sensitivity to dust extinction, provides the
ideal tool to probe even the most compact, extinguished sites, 
e.g. in the inner galactic nuclei.

For a detailed physical investigation, line ratios sensitive to either gas temperature $T$
or density $n$ are used. To estimate electron density $n$ one can use the strong dependence of 
the fine-structure line intensities for doublets of the same ion on $n$:
one example are the [OIII] lines at 5007 \AA, 52 $\mu m$ and 88 $\mu m$.
Similarly one can estimate $T$ and the shape of the ionizing continuum.

Particularly relevant to test the spectral shape of the ionizing continuum are
the {\sl fine-structure lines from photo-ionized gas}, which allow to discriminate
spectra of stellar and quasar origin.
Low-ionization transitions typically strong in starbursts are [OIII]52 and 88, [SiII]34, 
[NeII]12.8, [NeIII]15.6, [SIII]18.7 and 33.4, while higher ionization lines in AGNs
are  [OIV]25.9 and [NeV]24. Table 2 reports a few of the most important IR ionic lines.

One important application of IR spectroscopy was by Genzel et al. (1998),
to investigate the nature of the primary energy source in IR luminous galaxies
(see Sect. 6.8).

\section{IR STARBURST AND ULTRA-LUMINOUS GALAXIES IN THE LOCAL UNIVERSE}

For a variety of reasons it is unlikely that star-formation (SF) in galaxies 
has proceeded quietly during the Hubble time. 'A posteriori' evidence 
has accumulated that a fraction of stars in stellar systems was produced
during short-lived events (see e.g. the excellent review in Moorwood, 1996). 
These SF events are expected to be very luminous, either in the optical or in the
IR, and are expected to contribute substantially to the 
global energetics from baryon thermonuclear reactions, to the
synthesis of metals, and the generation of background radiations in the optical,
IR and sub-mm. Also the origin of Es, S0s and of the bulges of spirals may
have some relationship with luminous and  ultra-luminous starburst events at high-z.

If the study of star-formation in high-redshift sources is a primary task for modern 
cosmology, it is obvious that relevant information for the interpretation of
distant objects comes from a close up on local galaxies with enhanced SF.
For this reason we consider in this Section a sub-class of local galaxies, {\sl the starburst
galaxies and the IR luminous and ultra-luminous galaxies}, including a small fraction 
(few \%) of all local objects, but accounting for
a large percentage of the present-day star formation in galaxies.

The discovery of the starburst phenomenon dates back to the 1970's and came
almost simultaneously from two quite independent lines of investigation:
from objective-plate (Markarian) surveys of UV-excess galaxies, and from the first 
pioneering IR observations of galaxies in the local universe.
IR observations, in particular, revealed the existence of galaxies with IR luminosities
and L/M ratios appearently too high to be sustained over their lifetimes (Harwit and Pacini
1975). This brought to the idea that some galaxies undergo a 
sudden burst of massive star formation, with dust reprocessing of UV photons emitted 
by the young stars interpreted as the source of the IR light.

From these observations it was clear that SF has a twofold appearence, a UV excess and an 
IR excess, which may be explained by the 
{\sl stocastical nature of the interaction between photons and dust} in
star-forming regions of galaxies (see above).

However, the abilities of UV and IR surveys to sample the starburst phenomenon are
very different: while at low bolometric luminosities UV and IR surveys sample 
roughly the same kind of objects, at high luminosities the UV flux is no more a 
good tracer of the SF, which is better sampled by the IR emission.
This effect is due to dust extinction of the UV-light by young stars becoming more
and more relevant at the higher bolometric luminosities ($L_{bol}>10^{11}\ L_\odot$, 
Sanders and Mirabel 1996). At the highest values of $L_{bol}$ ($>10^{12}\ L_\odot$) 
most ($>95\%$) of the flux comes out in the IR.

%This difference between IR and optical-UV emission in starbursts entails a remarkable 
%difference in the {luminosity functions}:
%while the optical one is described by the well known Schechter (1986) 
%{exponential function},
%the FIR is entirely inconsistent with this, and fitted rather by a 
%{2-power-law function}.
%The difference is particularly relevant at the highest L, where the IR LF shows no
%characteristic break.

$L_{bol}$ is also tightly correlated with the optical morphology:
while at low-L there is a "natural" mix of various (mostly late) types, 
at the higher-L nearly all objects appear to be interacting galaxies, and at the 
highest-L they look as advanced mergers.
Also, the correlation is in the sense that while in low-L objects the SF activity is
spread over the galactic disk (enhanced in the spiral arms), at increasing luminosity 
the SF gets more and more concentrated in the nuclear regions.

In the higher-L objects in particular, it is often observed a concomitant stellar 
and nuclear non-thermal (AGN) activity, usually the latter occurring in the dynamical 
center of the galaxy and the former in a circum-nuclear ring (at $\sim 1\ kpc$).

A basic difficulty encountered in studies of active galaxies is to
disentangle between starburst-dominated and AGN-dominated energy sources of the
IR-luminosity. In fact, the two astrophysical processes are quite
often associated in the same object.
Optical line ratios (high vs. low excitation, e.g. [OIII]5007/H$\beta$ vs. 
[NII]6583/H$\alpha$, the Osterbrock diagram)  and line widths (few
hundreds Km/s for starbursts, larger for AGNs) are sometimes useful indicators, 
even in the presence of dust.

Useful near-IR lines, accessible from ground, are the Hydrogen Br$\gamma$2.166$\mu$,
HeI2.058$\mu$, H$_2$, but also higher atomic number species, [FeII] among others.
The Br$\gamma$2.166$\mu$ and HeI2.058$\mu$, in particular, so
close in $\lambda$ that differential extinction is negligible, constrain 
the underline ionization spectrum.

However, the most reliable information is provided by mid- and far-IR spectroscopy
by space observatories. Extremely promising in this field, in addition to ISO and SIRTF in 
the next few years, are the planned large space telescopes: NGST in the mid-IR and FIRST in
the far-IR.
%
%It is however with mid- and fir-IR spectroscopy, now possible with space equipments 
%like ISO and SIRTF in the next years, that one achieves the most reliable information,
%by exploiting the variety of IR fine-structure lines.
%There is an extremely promising perspective in this field, if we consider the power
%of the planned large space observatories, particularly NGST in the mid-IR and FIRST
%in the far-IR.

%Other {fine-structure lines excited in photo-dissociation regions}
%which are believed to arise predominantly in UV-excited regions at the boundaries 
%between the HI and HII gas phases: [OI]63, [CII]158

\subsection{The infrared-radio correlation}

While there is no direct proof for the basic interpretation of the IR starburst 
phenomenon (i.e. being due to UV light from newly formed stars absorbed by dust and re-emitted 
in the IR), an indirect support comes from the well-known radio to far-IR luminosity 
correlation (de Jong et al. 1985, Helou et al. 1985).
%relation (see below), proving
%that SNe following SF emit non-thermal electrons almost simultaneously to the activity
%of young OB stars illuminating the dust.
%
This, which is the tightest correlation involving global properties of galaxies,
provides an important constraint on the physics ruling starbursts of any luminosities.
It not only involves luminous active starbursting galaxies, but also many other
galaxies, like quiescent spirals.

The correlation is parametrized by the ratio of the bolometric far-IR flux 
$F_{FIR}$ (in $erg/s/cm^{2}$) to the radio flux $S_\nu$ (in $erg/s/cm^2/Hz$):
\begin{equation}
 q= log[F_{FIR}/3.75\ 10^{12}\ Hz/S_\nu(1.4\ GHz)] \simeq 2.35,\ \ \ \ \ 
\sigma(q)\simeq 0.2
\label{FIRrad}
\end{equation}
which is observed to keep remarkably constant 
with $L_{bol}$ ranging over many orders of magnitude, from low-luminosity spirals up to
ultraluminous objects (Arp 220)
[small departures from linearity appearing at the low- and high-luminosity ends].

The relation is interpreted as an effect of the ongoing star formation:
the far-IR emission comes from dust heated by UV photons by young stars, which also heat
the ISM producing free-free emission and generate SN originating high-energy e$^-$ 
and synchrotron flux mostly by interaction with the general galactic magnetic field.
This same scheme explains the departures from linearity: e.g. $q$ slightly increases at the 
low-luminosity end because $L_{FIR}$ is also contributed by the flux by old stars heating 
the dust.
The radio emission tends to be less concentrated than the far-IR, because of
fast e$^-$ diffusion.

\subsection{Estimates of the star formation rate (SFR)}

As the bolometric luminosity increases, the optical indicators of the
SFR (e.g. the UV flux, or the EW of H$\alpha$) become increasingly uncertain, as a larger
and larger fraction of short-$\lambda$ photons are extinguished.
In such a situation, the IR luminosity (proportional to the luminosity by young stars) 
becomes the most reliable indicator of the SFR.
A slight complication here is that older stars illuminating the diffuse cirrus dust
in galaxies also contribute to the far-IR flux, particularly in 
low-luminosity inactive systems.

The SFR is estimated by Telesco (1988) from the energy released by the CNO cycle and assuming 
a Salpeter IMF (eq. \ref{salp}):
$$ SFR (OBA) = 2.1\ 10^{-10}\ L_{FIR}/L_\odot   \ \  [M_\odot/yr], \ \ \   
SFR (All) = 6.5\ 10^{-10}\ L_{FIR}/L_\odot    \ \  [M_\odot/yr]  $$
the former relation referring to the OBA star formation.
A refined calibration is given by Rowan-Robinson et al. (1997):
$${SFR (All) = 2.6\ 10^{-10}\ \phi\ \epsilon\ L_{60\mu}/L_\odot   \ \  [M_\odot/yr] }$$
where $\phi$ incorporates the correction from a Salpeter IMF to the true IMF
($\phi \sim 3.3$ going to a Miller-Scalo) and includes corrections for the cut in the 
IMF (e.g. $\phi \sim 1/3$ if only OBA stars are formed),
$\epsilon$ being the fraction of photons re-radiated in the IR.

Another mean of estimating the ongoing SFR exploits the radio flux
(Condon 1992), by relating the SN rate to the rate of SF and using observations of 
the radio luminosity of the Milky Way to calibrate the relation.
Since the synchrotron emission (proportional to the rate of SN remnant production)
and thermal radiation (from HII regions heated by young OB stars) dissipate in 
$10^7-10^8$ yrs, the radio flux provides a good measure of the instantaneous
SFR. Operatively, one needs to estimate the fraction of stars with masses $M>8\ M_\odot$, 
progenitors of type-II SN, formed per unit time.
The problem with faint radio-source observations is that the radio emission
of stellar origin gets easily confused with non-thermal emission by a radio-loud AGN.

Finally, ISO observations indicate that also the mid-IR flux [dominated by hot dust and
PDR emission] traces very well the SFR (see Sect. 12.1.2 below).

All these long-wavelength methods provide obvious advantages, in terms of
robustness with respect to dust-extinction, compared with the optical ones,
namely the relation of SFR with the UV continuum flux by Madau et al. (1996):
$SFR (all\ stars) = 5.3\ 10^{-10}\ L_{2800\AA}/L_\odot \ \  [M_\odot/yr]$;
$SFR (metals) =1/42\ SFR(stars)$, and that between the $H\alpha$ line flux
and the SFR (Kennicut 1998): 
$$ SFR(all)= 7\ 10^{-42}\ L_{H\alpha}\ [erg/s].   $$
Poggianti, Bressan \& Franceschini (2000) and Franceschini et al. (2000) 
have shown that even after correcting for extinction the $H\alpha$ flux
using measurements of the Balmer decrement, the $H\alpha$-based SFR is typically
a factor $\sim 3$ lower than the appropriate value inferred from the bolometric flux
in IR-luminous galaxies.

Altogether, with these calibrations, moderate luminosity IR starbursts have 
SFR$\sim$3--30 $[M_\odot/yr]$,
(corresponding to $\sim 10^5$ O stars present during a typical burst).
The most luminous objects, if indeed powered by SF, have
SFR up to 1000$[M_\odot/yr] $.
Bolometric flux and SFR are correlated with the broad-band IR to optical
luminosity ratio:  $L_{IR}/L_B \sim 0.1$ in inactive galaxies (M31, M33),
 $L_{IR}/L_B \sim 3-10$ in luminous ($L\sim 10^{11}\ L_\odot$) SBs,
 $L_{IR}/L_B \sim 100$ in ultra-luminous objects ($L>10^{12}\ L_\odot$,
e.g. Arp 220).

\subsection{Gas reservoirs, depletion times, starburst duration}

%We consider here the {duration of the SB phenomenon}, within 2 extremes of an
%instantaneous burst observed at somewhat later time, and a long duration of 
%$\sim 10^9$ yrs as considered in the first modellistic attempts.

The duration of the starburst is critically related with the mass fraction of stars
produced during the event and to the available gas reservoir.
Assuming that the SB dominates the spectrum on top of the old stellar population 
emission, an estimator of the SB duration is the EW 
of the Br$\gamma$ line, which is a measure of the ratio of the OB stellar flux 
(the excitation flux) to the red supergiant star flux (evolved OBA stars). 
The EW is then expected to evolve monotonically with time. 
Also, the comparison of the  Br$\gamma$ line with the CO NIR absorption lines is
an age indicator (Rieke et al. 1988). Moorwood et al. (1996) find
in this way ages of $10^7$ to $\sim 10^8$  yrs.

However, 
{\sl the most direct way to estimate at least an upper limit to the burst duration
is the comparison of the total mass of molecular material in the galaxy nucleus with
the estimated SFR}, which is also a measure of the efficiency of SF.
The gas mass is usually estimated from mm-wave CO line emission and from mm continuum 
observations of dust emission (assuming suitable conversion factors for $H_2/CO$ and
dust/gas).
Chini et al. (1995) have found that the two independent evaluations of the molecular mass
provide consistent results, showing that luminous IR galaxies are very rich in gas ($2\ 
10^9$ to $2\ 10^{10}\ M_{\odot}$).
The ratio $L_{IR}/M_{gas}$ assumes enormously different values in different stages
of galaxy activity: in normal inactive spirals $L_{IR}/M_{gas}\sim 5\ (L_\odot/M_\odot)$ 
(e.g. M31), in moderate starbursts $L_{IR}/M_{gas}\sim 20$ (M82, NGC253), in
ultra-luminous IR galaxies $L_{IR}/M_{gas}\sim 200$ (Arp 220),
in quasars $L_{IR}/M_{gas}\sim 500$.

A limit to the SB duration is then given by
$ t_{depletion}  =  10^{10}\ M_{gas} /L_{IR}\ yrs$, ranging from typically several Gyrs for 
inactive spirals down to a few $10^7$ yrs for the more active SBs.

\subsection{Starburst-driven super-winds }
 
There are several evidences that extremely energetic outfows of gas are taking
place in starbursts:
{\sl (a)} from optical spectroscopy, evidence for Wolf-Rayet lines indicative
 of very young SBs ($<10^7$ yrs) and outflow of ionized gas, with velocities
 up to 1000 Km/s (Heckman, Armus \& Miley 1990; Lehnert \& Heckman 1996);
{\sl (b)} from optical imaging there are evidences of bubbles and cavities left over
by large, galactic-scale explosions;
{\sl (c)} from X-ray spectroscopy, evidence for plasmas at very high temperatures 
(up to few KeV), far in excess
of what the gravitational field could explain (e.g. Cappi et al. 1999).

These highly energetic processes are interpreted as due to {\sl radiative
pressure by massive stars, stellar winds and supernovae explosions}
occurring in a small volume in the galaxy core, able to efficiently energize the gas and
to produce a dynamical unbalance followed by a large scale outflow of the remaining gas.

This phenomenon has relevant implications. 
It is likely at the origin of the huge amounts of metals observed
in the Intracluster Plasma (ICP) in local galaxy clusters and groups.
It should be noted that the estimated Fe metallicity of the outflowing plasma 
($\sim 0.2-0.3$ solar, Cappi et al. 1999) is similar to the one observed in the ICP.
Also the higher abundances (1.5-2 solar) observed in the 2 archetypal starbursts
for $\alpha$ elements (Si, O, Mg) may indicate that type-II SN 
(those produced by very massive stars, M$>8\ M_\odot$) are mostly responsible 
(Gibson, Loewenstein \& Mushotzky 1997).
Similar properties are observed in the hot halo plasmas around elliptical galaxies,
also rising the question of a possible relationship of the hyper-luminous IR
galaxy phenomenon with the formation of early-type galaxies.
Therefore, {\sl the enriched plasmas found in local clusters and groups may represent the 
fossile records of ancient starbursts of the kind we see in local luminous IR starbursts}.

%More trivially, it could explain the LINERs (low ionisation narrow emission line region galaxy)
%phenomenon, i.e. in terms of a starburst activity rather than AGN activity, as it was commonly
%believed.

\subsection{Starburst models}
 
More precise quantifications of the basic parameters describing the SB phenomenon
require detailed modelling.
The first successful attempt accounting in some detail for the observed IR and radio data 
was by Rieke et al. (1980), who demonstrated that the remarkable properties of M82 and 
N253 are consistent with SB activity.

Since then, a number of groups elaborated sophisticated models of SBs. These successfully
reproduce SB properties assuming exponentially decaying SFRs with burst durations of  
10$^7$ to 10$^8$ yrs, whereas both instantaneous and long duration bursts are excluded.

An important issue addressed by these models is about the stellar IMF during the burst:
Rieke et al. found that assuming for M82 a Salpeter IMF with standard low-M cutoff 
at 0.1 $M_\odot$ resulted in a stellar mass exceeding the limit implied by dynamical
mass evaluations.
The problem was resolved by assuming that formation of stars with masses less than a few 
$M_\odot$ is strongly suppressed. This result however is not univocally supported by more
recent studies of M82: e.g. Leitherer and Heckman (1995) solution is for a 1 to 30 $M_\odot$ 
IMF.

Interesting constraints on the IMF come in particular from the analysis of CO line 
kinematics in Arp 220: Scoville et al. (1996) indicate that the dynamical mass,
the Lyman continuum, the SFR and the burst timescale can be reconciled by assuming
a IMF truncated outside 5 to 23 $M_\odot$, with a SFR$\sim 90\ M_\odot/yr$ for stars
within this mass range.
Altogether, there seem to be fairly clear indications for a "top-heavy" mass
function in the more luminous SBs, as compared with quiescent SF in the Milky 
Way and in spirals. This has relevant implications for the SFR history in galaxies,
the cosmic production of light and of heavy elements.

%
%\subsubsection{Modelling high-resolution SB spectra}

A very detailed modellistic study of starbursts was given by Leitherer \& Heckman (1995)
and Leitherer et al. (1999), incorporating all up-to-date improvements in the treatment of stellar evolution and
non-LTE stellar atmospheric models.
The model successfully explains most basic properties of starbursts, as observed
in the optical. Model predictions for a continuous SF over $10^8$ yrs and a 1-30 $M_\odot$ 
Salpeter IMF, normalized to a SFR=1 $M_\odot/yr$ are: 
bolometric luminosity = $1.3\ 10^{10}\ L_\odot$; number of O stars = $2\ 10^4$;
ionizing photon flux ($\lambda<912 \AA$) = $1.5\ 10^{53}\ photons/sec$; 
SN rate = 0.02 $yr^{-1}$; K=--20.5 mag; mass deposition rate = 0.25 $M_\odot/yr$;
mechanical energy deposition rate = $6\ 10^{41}\ erg/sec$.
Important outcomes of these papers are predictions for the EW
of most important line tracers of the SF ($H\alpha$, Pa$\beta$, 
Br$\gamma$), as a function of the time after the onset of SF and of IMF shape.
%These calibrations are currently used for analyses of distant high-z galaxies.

Models of dusty starbursts have been discussed by Silva et al. (1998), Jimenez et al. (1999), 
Siebenmorgen, Rowan-Robinson \& Efstathiou (2000), Poggianti \& Wu (1999),
Poggianti, Bressan \& Franceschini (2000). The latter two, in particular, address
the question of the classification and interpretation of optical spectra of
luminous IR starbursts: they find that the elusive class of {\sl e(a)} (emission+absorption)
spectra, representative of a large fraction ($>50\%$) of all IR SBs, are better understandable 
as ongoing
active and dusty starbursts, in which the amount of extinction is anti-correlated with
the age of the population (the youngest stars are the more extinguished, see
also Sect. 4.2.3), rather than post-starburst galaxies as sometimes have been interpreted.

\subsection{Statistical properties of active galaxy populations}
 
Statistical properties of SB galaxies provide guidelines to understand the origin
and triggering mechanisms of the phenomenon.
A fundamental descriptor of the population properties is provided by the Local 
Luminosity Function (LLF), detailing the distribution of space density as a function 
of galaxy luminosity in a given waveband.

While the faint luminosity end is important for cosmogonic purposes (providing constraints
on the formation models, its flattish shape being roughly similar at all wavelengths),
SBs and their complex physics dominate at the bright end of the LLF.
Indeed, the latter is observed to undergo substantial changes as a function
of $\lambda$: if the optical/near-IR LLF's display the classical "Schechter" exponential
convergence at high-luminosities (essentially tracing the galaxy mass function),
LLF's for galaxies selected at longer wavelengths show flatter and flatter slopes
(see Fig. \ref{llf12} below).
This flattening is progressive with $\lambda$ going from the optical up to
60 $\mu m$. When expressed in differential units ($Mpc^{-3}L^{-1}$), 
the bright-end slope of the 60 $\mu m$ LLF 
is $\propto L_{60\mu}^{-2}$, according to the extensive sampling by IRAS
(Saunders et al. 1990). Note that this flattening is not due to the contribution
of AGNs at 60 $\mu m$, which is modest here and quite more important instead at 12 $\mu m$.
What is progressively increasing with $\lambda$ up to $\lambda=60\ \mu$ is
the incidence of the starburst contribution to the luminosity: it is the
starbursting nature of 60$\mu$ selected galaxies responsible for
the shape of the LLF.

%At still longer-$\lambda$ the galaxy LLF gets still steep as for the optical/NIR

It is interesting to consider that almost the same slopes $\propto L^{-2}$
are found for all known
classes of AGNs, from the luminous radio-galaxies (Auriemma et al. 1977; Toffolatti
et al. 1987), to the optical 
and X-ray quasars (Miyaji, Hasinger \& Schmidt 2000; Franceschini et al. 1994b).
Also to note is the evidence that the $L^{-2}$ slope 
for AGNs keeps almost exactly the same at any redshifts, in spite of the drastic
increase of the source number-density and luminosity with $z$ due to evolution.

%Clearly the flat behaviour of the 60 $\mu m$ LLF is related to the starbursting 
%nature of high-L galaxies 
%selected at 60 $\mu m$:} note that this is not due to the presence of AGNs, their
%contribution is maximum at 12 $\mu m$

There should be a ruling process originating the same functional law in
a wide variety of categories of active galaxies and remarkably invariant with cosmic 
time, in spite of the dramatic differences in the environmental and physical 
conditions of the sources.
This remarkable behaviour may be simply understood as an effect of the triggering
mechanism for galaxy (AGN and starburst) activity: 
{\sl the galaxy-galaxy interactions (either violent mergers between gas rich 
objects or encounters triggering a slight increase of the activity).}

The physical mechanism ruling the process is the variation in the angular momentum
$\Delta J/J$ of the gas induced by the interaction, and the consequent 
gas accretion $\Delta m/m$ in the inner galaxy regions (Cavaliere \&
Vittorini 2000).  Starting for example from a $\delta-$function-shaped LLF, the 
starburst triggered by the interaction produces a transient increase of L which 
translates into a distortion of the LF towards the high-L's.
%this distortion is ruled by the factor $\Delta m/m$ multiplied by a suitable mass-energy
%convertion efficiency.
%The low-L end is instead unaffected and keep the shape of the parent (inactive)
%population
% 
Assumed $\Delta m/m$ is ruled by the probability distribution of the impact
parameter $b$, it is simple to reproduce in this way the LLF's observed
asymptotic shape at the high luminosities.

{\sl All this points at the interactions as ruling the
probability to observe a galaxy during the active phase}.

\subsection{Starburst triggering}

In normal inactive spirals the disk SFR is enhanced in spiral arms in 
correspondence with density waves compressing the gas. This favours the growth 
and collapse of  molecular clouds and eventually the formation of stars.
This process is, however, slow and inefficient
in making stars (also because of the feedback reaction to gas compression produced
by young stars). This implies that very 
long timescales (several Gyrs) are needed to convert the ISM into a significant stellar 
component.

On the contrary, because of the extremely high compression of molecular gas 
inferred from CO observations in the central regions of luminous
starburst galaxies, SF can proceed there much more efficiently.
Both on theoretical and observational grounds, it is now well established that the 
trigger
of  a powerful nuclear starburst is due to a galaxy-galaxy interaction or merger, driving
a sustained inflow of gas in the nuclear region.
This gas has a completely different behaviour with respect to stars: it is extremely
dissipative (gas clouds have a much larger cross-section and in cloud collisions
gas efficiently radiates thermal energy generated by shocks).
A strong dynamical interaction breaks the rotational symmetry and centrifugal support
for gas, induces violent tydal forces producing very extended tails and bridges
and triggers central bars, which produce shocks in the leading front,
and efficiently disperse the ordered motions and the gas angular momentum.
{\sl The gas is then efficiently compressed in the nuclear region and allowed to
form stars}.

These concepts are confirmed by numerical simulations of galaxy encounters.
Toomre (1977) was the first to suggest
that ellipticals may be formed by the interaction and merging of spirals. This suggestion 
is supported by various kinds of morphological features (e.g. tidal tails, rings)
observed in the real objects and predicted by his pioneering numerical simulations.

Much more physically and numerically detailed elaborations have more recently
been published by
Barnes and Hernquist (1992), who model the dynamics of  the encounters 
between 2 gas-rich spirals including disk/halo components, using a combined N-body
and gas-dynamical code based on the Smooth Particle Hydrodynamics (SPH).
Violent tidal forces act on the disk producing extended tails and triggering central bars,
who sweep the inner half of each disk and concentrates the gas into a single giant cloud.
The final half-mass radii of gas are much less than those of stars: for an $M^\ast$
galaxy of 10$^{11}\ M_\odot$, $\sim 10^{9}\ M_\odot$ of gas are compressed
within 100-200 pc, with a density of $10^3\ M_\odot/pc^3$ (Barnes \& Hernquist 1996).

Various other simulations confirm these finding. 
SPH/N-body codes show in particular that the dynamical interaction
in a merger has effects not only on the gas component, but also on the stellar one,
where the stars re-distribute following the merging and violent relaxation
of the potential.

\subsection{Ultra-luminous IR galaxies (ULIRGs)}

Defined as objects with bolometric luminosity $L_{bol}\simeq L_{IR}>10^{12}\ L_\odot$, 
they are at the upper rank of the galaxy luminosity function.
A fundamental interpretative problem for this population is to understand
the primary energy source, either an extinguished massive nuclear starburst, 
or a deeply buried AGN.

A systematic study of this class of sources was published by 
Genzel et al. (1998), based on ISO spectroscopy of low-excitation and 
high-excitation IR lines, as well as of the general shape of the mid-IR SED
(the intensity of PAH features vs. continuum emission; see also Lutz et al. 1998).
While the general conclusion of these analyses is that star-formation is
the process dominating the energetics in the majority of ultraluminous IR galaxies,
they have also proven that AGN and starburst activity are often concomitant
in the same source. This fact is also proven by the evidence (e.g. Risaliti et al. 2000;
Bassani et al. 2000) that many of the ULIRGs classified by Genzel et al. as 
starburst-dominated also show an hidden, strongly photoelectrically absorbed,
hard X-ray spectrum of AGN origin. Soifer et al. (2000) have also found that 
several ULIRGs show very compact (100-300 pc) structures dominating the mid-IR flux,
a fact they interprete as favouring AGN-dominated emission.
The relative role of SF and AGN in ULIRGs is still to be 
quantified, hard X-ray (spectroscopic and imaging) observations 
by CHANDRA and XMM, as well as IR spectroscopy by space observatories 
(SIRTF, FIRST) will provide further crucial information.
 
%
%\nobreak\begin{table}
%{{ Table\enskip 4.}\enskip{ Comparison of wide-band SED's of
%Active Galaxies.}}
%\begin{tabular}{lllll}\hline
%Name     & $L_X/L_{FIR}$      &  $L_X/L_{[OIII]}$   &$L_{[OIII]}/L_{FIR}$   & \\
%         &        ($10^{-4}$)  &           & ($10^{-5}$) &  \\
%\hline
% M82     &        3.2     &      700    &   0.05   &  \\
% NGC 253 &        1.5     &      114    &   0.1    &  \\
% M83     &        1.2     &      4.6    &   2.5    &      \\
%Arp 220  &        0.36    &      6      &   0.6    &          \\
%NGC 4038 &          3     &      30     &    1     &  (Antennae) \\
%NGC 3690 &        2.5     &      2.9    &   10     &  \\
%NGC 3256 &        1.4     &      1.1    &   12     &   \\
%NGC 4945 &        2.4     &     $<8.7$  &  $>3$    &                   \\
%IRAS 19254&      3.5      &      0.06   &   646    &  (Super-antennae) \\
%NGC 6240 &        14      &      1.7    &   90     &                   \\
%UGC23060 &        80      &      1      &   938    &                   \\
%         &        5       &      0.08   &   640    &  (Circinus) \\
%NGC 1068 &        1       &      0.02   &   443    &                   \\
%NGC 4151 &       1830     &      17     &   1100   &                   \\
%\end{tabular}
%\end{table}
% 

\subsection{Origin of elliptical galaxies and galaxy spheroids}

As pointed out for the first time by Kormendy \& Sanders (1992),
the typical gas densities found by interferometric imaging of CO emission in ultra-
luminous IR galaxies turn out to be very close to the high values of stellar densities in the 
cores of E/S0 galaxies.
This is suggestive of the fact that ULIRG's have some relationships with the 
long-standing problem of the origin of early-type galaxies and spheroids.

Originally suggested by Toomre (1977), the concept that E/S0 could form in mergers of 
disk galaxies immediately faced the problem to explain the dramatic difference in
phase-space densities between the cores of E/S0 and those of spirals.
Some efficient dissipation is required during the merger, which can be
provided by the gas. Indeed,
the CO line observations in ULIRG's, also combined with those of the stellar nuclear velocity
dispersions and effective radii, show them to share the same region of the "cooling diagram" 
occupied by ellipticals.

%The two basic observables to check this hypothesis (eg van der Werf 1996):
%the "Kormendy diagram" central surface brightness vs. effective radius
%the stellar velocity dispersion vs. core radius
%The two are "manifestations" of the fundamental plane of E/S0's:
%{ultra-luminous IR galaxies occupy the same region of the "cooling diagram" 
%occupied by ellipticals}.

Detailed analyses of the $H_2$ NIR vibrational lines in NGC 6240 and Arp 220 
(van der Werf 1996) have provided interesting information about the
mass, kinematics, and thermodynamics of the molecular gas.
The conclusion is that shocks, the fundamental drivers for dissipation, can fully explain 
the origin of the $H_2$ excitation.
The evidence that the $H_2$ emission is more peaked than stars, and located
in between the two merging nuclei, is consistent with the fact that
gas dissipates and concentrates more rapidly, while stars are expected to relax violently 
and follow on a longer timescale the new gravitational potential ensuing the merger.

A detailed study of Arp 220 by van der Werf (1996) has shown that 
most of the $H_2$ line emission, corresponding to $\sim 2\ 10^{10} M_\odot$
of molecular gas, comes from a region of 460 pc diameter,
the gas mass is shocked at a rate of $\sim 40\ M_\odot/yr$, not 
inconsistent with a SFR$\sim 50-100\ M_\odot/yr$ as discussed in Sect. 6.5.
Compared with the bolometric luminosity of Arp 200, this requires
a IMF during this bursting phase strongly at variance with respect to the Salpeter's
one (eq. \ref{salp}) and either cut at $M_{min}>>0.1\ M_\odot$ or
displaying a much flatter shape.

In support of the idea that ellipticals may form through merging processes there is 
evidence coming from high-resolution K-band imaging that the starlight distribution
in hyper-luminous IR galaxies follows a de Vaucouleurs $r^{-1/4}$ law typical of E/S0
(Clements \& Baker 1997).

Also proven by simulations, after the formation of massive nuclear star clusters 
from the amount of gas (up to 10$^{10}\ M_\odot$) collapsed in the inner Kpc, 
part of the stellar recycled gas has low momentum
and further contracts into the dynamical center, eventually producing a super-massive
Black Hole with the associated AGN or quasar activity 
(Norman and Scoville 1988, Sanders et al. 1988).

\section{IR GALAXIES IN THE DISTANT UNIVERSE: PRE-ISO/SCUBA RESULTS}

We have summarized in previous paragraphs 
the main properties of local galaxies when observed at long wavelengths,
and emphasized the unique capability of these observations to unveil classes of sources,
unnoticeable at other wavelengths, but extremely luminous in the IR.
It was clear from this that the most luminous objects in the universe
and the most violent starbursters can be reliably studied only at these
wavelengths.

Our previous discussion has also illustrated the complexity and
difficulty of modelling the long-wavelength spectra of galaxies, 
heavily dependent on the relative geometries of stars and dust.
%In any case, a delicate modellistic effort is required by the recent relevant results 
%emerging from IR/sub-mm observations of the high-redshift universe.

Now, assumed we have a decent understanding of the local universe and its 
IR galaxy populations, we dedicate the next Sections to illustrate and
discuss new emerging facts about their distant counterparts, which
entail important discoveries for cosmology.

%%%%%%%%%%%%%%%%%%%%%%%%\section{DISTANT IR SOURCES: PRE-ISO/SCUBA RESULTS}

The IRAS survey in 1983, allowing the first sensitive all-sky view
of the universe at long wavelengths, is considered as the birth date of IR astronomy.
Most of our knowledge
about local IR galaxies, as previously discussed, comes from the IRAS database.
The fair sensitivity of the IRAS surveys, coupled with the prominent emission
of IR galaxies at 60-100 $\mu m$, have also allowed to sample and study galaxies
at cosmological distances and to derive first tentative indications for
evolution.

Counts of IRAS galaxies (mostly at 60 $\mu m$, where the S/N was optimum, S including 
the source signal and N the instrumental and sky [cirrus] noise) have been obtained
by Hacking \& Houck (1987), Rowan-Robinson et al. (1991), Gregorich et al. (1995),
Bertin, Dennefeld, Moshir (1997).
Samples at 60 $\mu m$ with optical identifications and radial velocities 
have been published by Saunders et al. (1990, 1997), Lonsdale et al. (1990),
and Oliver et al. (1996).

Early evidence in favour of evolution for IRAS-selected galaxies have been
discussed by Hacking et al. (1987), Franceschini et al. (1988) and
Lonsdale et al. (1990), among others.
In the models by Franceschini et al. and Pearson \& Rowan-Robinson (1996),
a sub-population of starburst galaxies including a substantial fraction 
(30\%) of all galaxies in the local universe evolves as
$ L(z) = L(0) (1+z)^{3.1}$ (Pearson \& Rowan-Robinson)
or
$ L(z) = L(0) e^{\ 4.3 \tau(z)}$ (Franceschini et al. ), roughly
reproducing counts and redshift distributions.

However, the IRAS sensitivity was not enough to detect galaxies
at substantial redshifts, apart from a handful of exceptions (essentially due 
to gravitational lensing amplifying the flux): the most distant were found at 
$z\simeq 0.2-0.3$. Any conclusions based on IRAS data are to be considered as 
preliminary, large-scale inhomogeneities badly affecting these shallow samples.

Another problem for the IRAS surveys was the uncertain identification 
with faint optical counterparts, because of the large 
[$\sim 1\ arcmin^2$] IRAS error-box: this implied a systematic bias towards
associating IRAS sources with the brightest galaxy falling inside it, which
may systematically miss the fainter higher-redshift correct identification.

\section{THE BREAKTHROUGHS: DISCOVERY OF THE CIRB}

Cosmological background radiations are a fundamental channel of information
about cosmic high-redshift sources, particularly if, for technological limitations,
observations of faint sources in a given waveband are not possible.
This was clearly the case for the IR/sub-mm domain. The present Section
is dedicated to a review on a recently discovered new cosmic component, the cosmological 
background at IR and sub-millimetric wavelengths (CIRB), an important achievement
made possible by the NASA's Cosmic Background Explorer (COBE) mission.

To appreciate the relevance of this discovery (anticipated by a detailed modellistic
prediction by Franceschini et al. 1994), consider that
extragalactic backgrounds at other wavelengths contain only modest (undiscernible)
contributions by distant galaxies.
The Radio background is clearly dominated by radio-loud AGNs; 
the Cosmic Microwave Background includes photons generated at $z\sim 1500$;
the X-ray and $\gamma$-ray backgrounds are dominated by distant quasars and AGNs.
Also, diffuse light in the optical-UV (and partly the near-IR) will be hardly depurated 
of the foreground contaminations (in particular, Galactic starlight reflected by high
latitude "cirrus" dust, and Zodiacal-reflected Sun-light).

On the other hand, the recently
completed third experiment (DIRBE) of the COBE mission has brought to the 
{\sl first detection ever (with surprisingly small uncertainties) 
of the integrated emission of distant galaxies} in the form of an isotropic signal 
in the far-IR and sub-mm (Puget et al. 1996, Guiderdoni et al. 1997, Hauser et al. 1998, 
Fixsen et al. 1998).

\subsection{Observational status about the CIRB}

In spite of the presence of very bright foregrounds (Zodiacal and Interplanetary
dust emission, Galactic Starlight, high-latitude "cirrus" emission), relatively
clean spectral windows exist in the IR suitable for extragalactic research:
the near-IR cosmological window (2-4 $\mu m$) and the sub-mm window (100-500 $\mu m$).
At these wavelengths the Zodiacal, Starlight, and emission by high galactic 
latitude dust produce two minima in the total foreground intensity, which is 
much lower here than it is in the optical-UV.

These spectral windows occur where we would expect to observe the redshifted
photons from the two most prominent galaxy emission features:
the stellar photospheric peak at $\lambda \sim 1\ \mu m$ and the one at
$\lambda \sim 100\ \mu m$ due to dust re-radiation.
The best chances to detect the integrated emission of distant and primeval galaxies
are here.
 
For a curious coincidence, the (expected) integrated emission
of distant galaxies turns out to be comparable by orders of magnitude to the Galaxy emission 
at the Pole and to the Zodiacal light in the near-IR window.
This implies that a delicate subtraction of the foreground emissions 
is required to access the extragalactic domain.

Three main observational routes have been followed to measure the CIRB: 

\begin{itemize}
\item rocket flights of dedicated instrumentation (1970-1990), 
now only of historical interest;

\item all-sky surveys by space telescopes (IRAS and COBE, 1984-1996);

\item indirect estimates based on very high-energy spectral observations of 
extragalactic $\gamma$-ray sources
(Stecker, de Jager \& Salomon 1992; Stanev \& Franceschini  1998).

\end{itemize}

In sky directions outside obvious Galactic sources, like star-forming and 
low-galactic latitude regions,
the total far-IR background is due to the contribution of various dust
components in the ISM: galactic dust associated with neutral and ionized hydrogen, 
the interplanetary dust emission, all adding to the isotropic diffuse flux, the CIRB.
The way to subtract these various foregrounds when estimating
the CIRB intensity is to exploit the different
spatial dependencies of the various components, using the correlations with
appropriate dust tracers like the HI 21 cm or H$_\alpha$ lines.

\begin{figure}
\vspace{1cm}
\psfig{figure=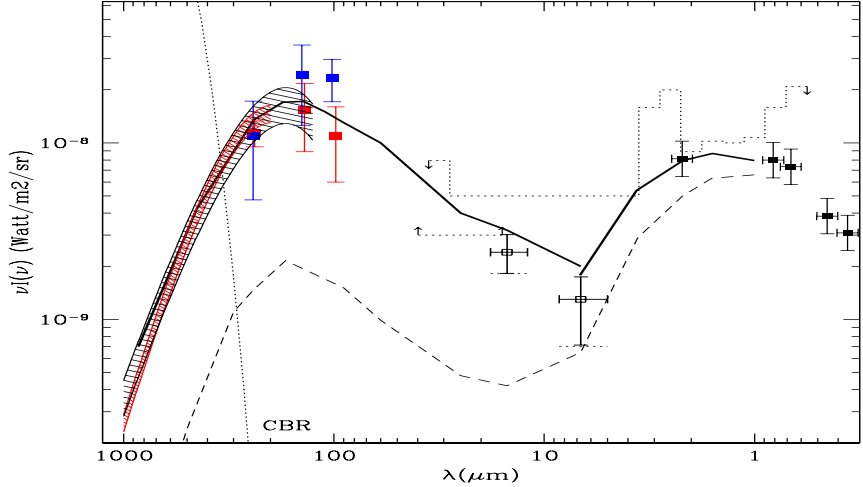,height=70mm,width=120mm}       %@ rifare figura
\caption{The Cosmic Infrared Background (CIRB) as measured by independent
groups in the all-sky COBE maps (e.g. Hauser et al.1998), compared 
with the optical extragalactic background estimated from ultradeep
optical integrations by the HST in the HDF (Madau \& Pozzetti 2000). 
Three datapoints in the far-IR are from a re-analysis
of the DIRBE data by Lagache et al. (1999), the shaded area from Fixsen 
et al. (1998) and Lagache et al. The two mid-IR points are the resolved fraction of the
CIRB by the deep ISO surveys GITES, while dashed lines are limits set by TeV cosmic
opacity (Sect. 8.2). The dotted line marks the expectation
based on the assumption that the IR emissivity of galaxies does not change
with cosmic time. The thick line is the predicted CIRB
intensity by the model discussed in Sect. 11.
}
\label{bkg}
\end{figure}

To subtract the most important foreground in the far-IR, the galactic dust
emission, the simplest procedure is to determine the parameters of the correlation
between the background intensity $I_\nu$ and the dust tracers expressed in terms
of equivalent hydrogen column density $N_H$, 
and then to evaluate the CIRB as the intercept of the total flux at $N_H=0$.

Another method is to perform an all-sky best-fit analysis of a relation like
$I_\nu = C_1 N_H(HI) +C_2 N_H(II) + CIRB$,  $N_H$ being the column densities
of the dust components associated with the neutral and ionized H,
$CIRB$ being the extragalactic background intensity at the working wavelength
(e.g, Lagache et al. 1999).
The best-fitting determines the constant $C_1$and $C_2$ and allows
to estimate a value for the parameter $CIRB$.

Puget {\it et al. } (1996) first recognized in the
all-sky FIRAS/COBE maps an isotropic signal (independent of Galactic coordinates)
with an intensity that can be represented by the law
$\nu B_{\nu}\simeq 3.4\times 10^{-9} (\lambda/400 \mu m)^{-3} \ 
W\ m^{-2} \ sr^{-1}$ in the 400--1000 $\mu m$ interval.  
 
This tentative detection has been later confirmed with independent
analyses by various other groups (e.g. by Fixsen et al. 1998, who find
significant isotropic signal from 200 and 1000 $\mu m$), as well as
by analyses of data from the DIRBE experiment on COBE in two broad-band
channels at $\lambda = 140$ and $240 \mu$ (Hauser et al. 1998).
Finkbeiner, Davies \& Schlegel (2000), 
after a very delicate subtraction of the far dominant Galactic and IPD foregrounds,  
found an isotropic signal at 60 and 100 $\mu m$ with intensity at the
level of $\sim 30\ 10^{-9}\ W\ m^{-2} \ sr^{-1}$. This latter result is presently under 
discussion, but appears to conflict with independent estimates (see Sect. 8.2).

Recent analyses by Dwek \& Arendt (1998) and Gorjian, Wright \& Chary (2000) 
have tentatively found also a 
signal in the near-IR cosmological window at $3.5\ \mu m$ and in the J, H and K DIRBE 
bands, however with 
large uncertainties because of the problematic evaluation of the Zodiacal
(scattered) light. Because of this, CIRB estimates particularly in J, H and K
are to be taken more reliably as upper limits.

To avoid overcrowding, we report in Figure \ref{bkg} only the most
recent results from DIRBE (Lagache et al. 1999; Finkbeiner, et al. 2000) 
and FIRAS (Fixsen et al. 1998).

No isotropic signals are significantly detected at 
$4 \mu <\lambda < 60 \mu$, any cosmological flux being far dominated here by the 
Zodiacal light, the Interplanetary dust (IPD) emission and by Galactic dust emission
(only missions to the outer Solar System would have chances to reduce the 
dominant IPD flux to achieve detection of the CIRB here).
The constraints we report at these wavelengths come from indirect estimates 
based on the cosmic high-energy opacity (Sect. 8.2 below).

Altogether, after four years of very active debate among various teams working 
on the COBE data, first about the existence and later on the intensity and spectral
shape of CIRB, there is now ample consensus even on details of CIRB's
spectral intensity, at least from 140 to 300 $\mu m$ where it is most reliably
measured and where two completely independent datasets (FIRAS and DIRBE, with independent
absolute calibrations) are available.
The CIRB flux has in particular stabilized at values $\nu I_\nu \simeq 24 \pm 5$
and $\nu I_\nu \simeq 15 \pm 5\ 10^{-9}\ Watt/m^2/sr$ at $\lambda=140$ and
240 $\mu m$.
Modest differences in the absolute calibration of 
FIRAS and DIRBE around 100 $\mu$ have been reported (Hauser et al. 1998),
but these do not seem to affect the overall result.

This was a fundamental achievement for observational cosmology, providing the
global energy density radiated by cosmic sources at any redshifts.
Two concomitant facts, the very strong K-correction for galaxies in the far-IR/sub-mm
implied by the very steep and featureless dust spectra, and their relative
robustness due to the modest dependence of dust equilibrium temperature $T$ on
the field intensity (eq.[\ref{T}]) have suggested to use the CIRB spectrum to infer
the evolution of the galaxy long-wavelength emissivity as a function of redshift
(Gisper, Lagache \& Puget 2000). 
Indeed, while the peak intensity at $\lambda =100$ to 200 $\mu m$ constrains the
galaxy emissivity at  $z=0$ to $z=1$, the quality of the FIRAS intensity maps
and the low foreground contamination at $\lambda> 200\ \mu m$ allow to
set important constraints on the universal emissivity at $z>1$. 

Between 100 and 1000 $\mu m$ the integrated CIRB intensity turns out to be 
$\sim 30\pm 5\ 10^{-9}\ Watt/m^2/sr$. In addition to this measured part of the CIRB, 
one has to consider the
presently un-measurable fraction resident in the frequency decade between 100 and 10 $\mu m$.
This flux is larger than the integrated "optical background" 
($\sim 17\ nWatt/m^2/sr$, see Fig.\ref{bkg}), 
obtained by counting all galaxies detected between 0.3 and 3 $\mu m$ by HST down to the 
faintest detectable sources. This procedure to estimate the
"optical background" relies on the fact that optical counts show a clear convergence
at magnitudes $m_{AB}\geq  22$ (Madau \& Pozzetti 2000), such that the expected 
contribution by sources fainter than HST limiting fluxes appears negligible
(a significant upwards revision of this optical background suggested by Bernstein et
al. [1998] to account for low surface brigtness emission by galaxies is not
confirmed).

Already the directly measured part of the CIRB sets a relevant constraint on
the evolution of cosmic sources, when compared with the fact mentioned in Sect.
4.2.5 that for local galaxies only 30\% of the bolometric flux is absorbed
by dust and re-emitted in the far-IR. 
{\sl The CIRB's intensity matching or even exceeding the optical
background  tells unequivocally that galaxies in the past should have been much more
"active" in the far-IR than in the optical, and very luminous in an absolute sense.
A substantial fraction of the whole energy emitted by high-redshift galaxies should 
have been reprocessed by dust at long wavelengths.}

\subsection{Constraints from observations of the cosmic high-energy opacity}

As originally suggested by F. Stecker soon after the discovery of high-energy
photon emissions from distant blazars, high-energy spectral observations may provide
a suitable alternative to the direct detection of the CIRB at wavelengths where
it is currently impossible.
The idea is to infer the CIRB intensity from combined GeV and TeV observations of a set
of Blazars by exploiting the $\gamma \rightarrow \gamma$ interaction of their emitted high
energy photons with those of the CIRB.

The absorption cross-section of $\gamma$--rays of energy $E_\gamma$ [TeV] 
has a maximum for IR photons with energies obeing the condition (Stecker, de Jager \& 
Salomon 1992):
$$\epsilon_{max} = 2 (m_e c^2)^2/E_\gamma , $$
which implies
\begin{equation}
\lambda_{peak} \simeq 1.24\pm 0.6(E_\gamma [TeV])\; \mu m .
\label{energy}
\end{equation}
The optical depth for a high-energy photon $E_0$ travelling through a cosmic medium
filled of low-energy photons with density $\rho(z)$ from $z_e$ to the present time is
\begin{equation}
\tau(E_0,z_e)  = c\int_0^{z_e} dz {dt \over dz } \int_0^2 dx {x \over 2} \int_0^\infty d\nu
(1+z)^3 {\rho_\nu (z) \over h\nu } \sigma_{\gamma\gamma} (s)
%1.24\pm 0.6 \times (E_\gamma, TeV)\; \mu m .
%
\label{tau}
\end{equation}
$$\sigma_{\gamma\gamma} (s)={3\sigma_T \over 16} (1-\beta^2) [2\beta(\beta^2-2)
+(3-\beta^4) ln({1+\beta \over  1-\beta})]   $$
$$ s \equiv 2 E_0 h\nu x (1+z); \ \ \  \beta \equiv (1-4m_e^2c^4/s)^{1/2} . $$
Coppi \& Aharonian (1999) report the following analytical approximation, good
to better than 40\%, to eq.(\ref{tau}): 
\begin{equation}
\tau(E_0,z_e) \simeq 0.24 {E_\gamma \over TeV } {\rho(z=0) \over 10^{-3}eV/cm^3}
{z_e \over 0.1} h_{60}^{-1}\simeq 0.063 {E_\gamma \over TeV } 
{\nu I_\nu \over nW/m^2/sr} {z_e \over 0.1} h_{60}^{-1}
\label{inu}
\end{equation}

Interesting applications of this concept have been possible when data from
the Compton Gamma Ray Observatory and from hard X-ray space telescopes 
have been combined with observations at TeV energies by the Whipple and other 
Cherenkov observatories on the Earth.

Stanev \& Franceschini (1998) have obtained model--independent upper limits 
on the CIRB with no a-priori guess about the CIRB spectrum, using HEGRA data
for the Blazar MKN 501 (z=0.034) during an outburst in 1997, on the assumption that the 
high-energy source spectrum is the flattest allowed by the data.
These limits (see Fig. \ref{bkg})
get quite close to the CIRB background already resolved by the ISO mid-IR deep surveys 
(see Sect. 9).

More recently, Krawczynski et al. (1999) have combined the observations of MKN501
during the 1997 outburst with X-ray data from RossiXTE and BeppoSAX, providing a simultaneous
high-quality description of the whole high-energy spectrum. These data are very well fitted
by a Synchrotron Self Compton (SSC) model in which the spectrum at $\nu=10^{27} Hz$ is produced
by Inverse Compton of the hard X-ray spectrum at $\nu=10^{18} Hz$: the combination
of the two provides solid constraints on the shape of the "primary" (i.e. before
cosmic attenuation) spectrum at TeV energies. This is used to derive $\tau_{\gamma\gamma}$
as a function of energy and, after eqs. \ref{tau} and \ref{inu}, a constraint on the spectral
intensity of the CIRB.
The result is compatible with the limits by Stanev \& Franceschini (1998)
and allows to get a tentative estimate of the CIRB intensity in the interval from $\lambda=10$ 
to 40 $\mu m$ (see Fig.[\ref{bkg}]), which is formally dependent, however, on the SSC model
adopted for the intrinsic source spectrum.
 
Less model dependent is the constraint set by the observations of purely power-law
Blazar spectra around $E_\gamma \simeq 1\ TeV$, which translates into the upper limit
of about $10\ nano Watt/m^2/sr$ at $\lambda \simeq 1\ \mu m$ shown in Fig. \ref{bkg}.
Substantially exceeding that, as suggested by some authors (Bernstein et al., 
Gorjian et al.), would imply either very "ad hoc" $\gamma-$ray source spectra or new physics
(Harwit, Proteroe \& Bierman 1999).

\subsection{Contribution of cosmic sources to the CIRB: the formalism}

A simple formalism relates background intensity and cell-to-cell anisotropies
to the statistical properties (luminosity functions and number counts) 
of the contributing sources.

\subsubsection{Source contribution to the background intensity}

The differential number counts (sources/unit flux interval/unit solid angle) at a given flux 
$S$ write:
\begin{equation}
{dN \over dS} = 
\int_{z_l}^{z_h}\,dz\,{dV\over dz}\,{d \log L(S;z)\over dS}\, 
\rho[L(S,z),z] \label{dNdS}
\end{equation}
where $\rho[L(S,z),z]$ is the epoch-dependent luminosity function and 
$dV/dz$ is the differential volume element.
Flux $S$ and rest-frame luminosity $L$ are related by 
\begin{equation}
S_{\Delta \nu} = {L_{\Delta\nu} K(L,z) \over 4\pi d_L^2}, \label{S}
\end{equation}
where $d_L$ is the luminosity distance and
$K(L,z)= (1+z) {L[\nu (1+z)]\over L(\nu)} $ the K-correction.  
The contribution of unresolved sources (sources fainter than the 
detection limit $S_d$) to the background intensity is given by: 
\begin{equation}
I = \int_0^{S_d}{dN\over dS} S\, dS =
{1\over 4\pi}\,{c\over H_0} \int_{z(S_d,L_{\rm min})}^{z_{\rm max}}
{dz\over (1+z)^6(1+\Omega z)^{1/2}}
 j_{\rm eff}(z) ,
\label{int}
\end{equation}
having defined the volume emissivity $j_{\rm eff}(z)$ as
\begin{equation}
j_{\rm eff}(z) = \int_{L_{\rm min}}^{\min\left[ L_{\rm max}, L(S_d,z) 
\right] } d\log L 
\ L \ n_c(L,z) K(L,z), \label{eq:22} 
\end{equation}
where $L_{\rm min}$ and $L_{\rm max}$ are the minimum and the maximum source
luminosities. From eq.(\ref{int}) we can note that, when the counts converge
like $dN/dS \propto S^{-2}$ or flatter, the contribution by faint sources to the
background intensity becomes almost insensitive to the source minimum flux 
[$I\propto ln(S_{min})$ or less]. This property has been used by Madau \& Pozzetti (2000)
to estimate the optical background intensity (see Fig. \ref{bkg}) from ultra-deep HST
counts of galaxies, by exploiting the convergence of the optical counts fainter than
$m_{AB}\sim 22$. A similar property of faint IR sources is used in Sect. 9.4 to estimate the 
contribution of IR galaxies to the CIRB.

%we finally have
%
%\begin{equation}
%I={1\over 4\pi}\,{c\over H_0} \int_{z(S_d,L_{\rm min})}
%^{z_{\rm max}}
%{dz\over (1+z)^6(1+\Omega z)^{1/2}} j_{\rm eff}(z) 
%\end{equation}

\subsubsection{Small scale intensity fluctuations}

In addition to the average integrated flux by all sources in a sky area, the background 
radiation contains also spatial information (the cell-to-cell fluctuations) which can 
be used to further constrain the source flux distribution and spatial correlation
properties (e.g. De Zotti et al. 1996).
The usually most important contribution to the cell-to-cell intensity fluctuations 
comes from the stochastic nature of the spatial distribution of sources among
elementary cells with an effective solid angle 
$\omega _{\rm eff, P}$ (Poisson fluctuations). They can be expressed as
\begin{equation}
(\delta I)^2 \equiv C(0) = {\omega_{\rm eff,P}(0)\over 4\pi}\int_0^{S_d} 
S^2\,{dN \over dS}\, dS .
\end{equation}
What is really measured, however, is not 
the flux $S$ but the detector's response $x=f(\vartheta , \varphi )S$, 
$f(\vartheta , \varphi )$ being the angular power pattern of the detector. 
Let $R(x) = \\$ 
$\int dN\left[x/f(\vartheta , \varphi )\right]/ dS \cdot d\omega/f(\vartheta ,\varphi ) $ 
be the {\it mean number of source responses of intensity $x$}.
For a Poisson distribution of the number of sources producing a response $x$,
its variance equals the mean $R(x)dx$. 
Adding the variances of all responses up to the cutoff value $x_c$ (brighter 
sources are considered to be individually detected) gives 
the contribution of unresolved sources to fluctuations:
\begin{equation}
(\delta I)^2 = \int_0^{x_c}x^2\,R(x)\,dx. 
\label{eq:47}
\end{equation}
The cutoff $x_c$ is chosen to be some factor $q$ times $(\delta I)^2$; 
usually $q= 3$--5. The rms background fluctuations ($\delta I$) imply a sky noise
$\sigma_{conf}={\langle(\delta I)^2\rangle}^{1/2}$ for observations with spatial 
resolution $\omega_{eff}$.

The integrated signal $D$ recorded by the detector 
is the sum of the responses $x$ due to all sources in the angular resolution element. 
Its probability distribution function 
$P(D)$ is informative on the amplitude and slope of counts of unresolved sources.
Scheuer (1957) has shown that its Fourier transform, $p(\omega)$, is a simple 
function of the FT $r(\omega)$ of $R(x)$:  $p(\omega) = \exp [r(\omega) - r(0)]$.  
It follows:
$$  %\begin{equation}
P(D)  =  \int _{-\infty}^{\infty} p(\omega) \exp(-2\pi i \omega D)\,d\omega 
= \int _{-\infty}^{\infty} \exp\left[r(\omega)-r(0)-2\pi i \omega D\right]
\,d\omega =
$$%   \end{equation}
\begin{equation}
2 \int _{0}^{\infty} \exp \left\{-\int _{0}^{\infty} R(x) 
\left[1-\cos(2\pi \omega x)\right] dx\right\} \\
     \cdot \cos \left[ \int_{0}^{\infty} R(x) \sin(2\pi \omega x)\,dx - 
2\pi \omega D \right] \, d\omega. \label{P(D)} 
\end{equation}
This synthetic $P(D)$ has to be convolved with the noise distribution to be compared with the 
observations. Assumed that the number count distribution below the detection limit
can be represented as a power-law, $N(>S)=K(S/S_k)^{-\beta}$, then eq. [\ref{P(D)}] 
can be integrated to get (Condon 1974):
\begin{equation}
\sigma_{conf}= \left[{ q^{2-\beta} \over 2-\beta}\right]^{1/\beta} 
(\omega _{eff} \beta K)^{1/\beta}
S_k,   \ \ \ \ \ \ \omega_{eff}=\int f(\vartheta ,\varphi)^\beta d\Omega
\label{sigma} 
\end{equation}
which allows to estimate the slope of the counts ($\beta$) below the detection
limit from a given measured value of the cell-to-cell fluctuations $\sigma_{conf}$. 
This constraint on $N(S)$ applies down to a flux limit corresponding to 
$\sim 1$ source/beam. Assumed that $S_k$ represents the confusion limit ($S_k=q\times
\sigma_{conf}$) of a survey having an areal resolution $\omega_{eff}$, 
then eq. \ref{sigma} further simplifies to a relation
between the number of sources $K$ resolved by the survey (and brighter than
$S_k$) and the parameters $q$ and $\beta$:
\begin{equation}
K = {2-\beta \over \beta q^2} {1 \over \omega_{eff}}:
\label{conf}
\end{equation}
this implies the confusion limit to occur at the flux corresponding to an areal density 
of $(\beta q^2/[2-\beta])^{-1}$ sources per unit beam area $\omega_{eff}$. For euclidean
counts and $q=3$, this corresponds to 1 source/27 beams. Confusion limits based on this
criterion for various IR observatories are indicated in Figs. \ref{c175} and \ref{c850} below.

\section{DEEP SKY SURVEYS WITH THE INFRARED SPACE OBSERVATORY (ISO)}

ISO has been the most important IR astronomical
mission of the 1990s. Launched by ESA, it consisted of a 60 cm 
telescope operative in a highly eccentric 70000 Km orbit. It included two instruments
of cosmological interest (in addition to two spectrographs): a mid-IR 32$\times$32 camera 
(ISOCAM, 4 to 18 $\mu m$), and a far-IR imaging photometer (ISOPHOT, with small
3$\times$3 and 2$\times$2 detector arrays from 60 to 200 $\mu m$).
The whole payload was cooled to 2 K by a $He^3$ cooling system so performant to allow
ISO to operate for 30 months (Nov 1995 to Apr 1998), instead of the nominal 18 months.
An excellent review of the extragalactic results from ISO can be found in Genzel \&
Cesarsky (2000).

\subsection{Motivations for deep ISO surveys}

While designed as an observatory-type mission, the vastly improved sensitivity
offered by ISO with respect to the previous IRAS surveys motivated to spend a 
relevant fraction of the observing time to perform a set of deep sky explorations
at mid- and far-IR wavelengths. The basic argument for this was to parallel
optical searches of the deep sky with complementary observations
at wavelengths where, in particular, the effect of dust is far less effective
in extinguishing optical light. This could have been particularly
relevant for investigations of the distant universe, given the large uncertainties
implied by the (pobably large) extinction corrections in optical spectra of 
high redshift galaxies (e.g. Meurer et al. 1997).

Observations in the mid- and far-IR also sample the portion of the e.m. spectrum 
dominated by dust re-processed light, and are then ideally complementary to optical 
surveys to evaluate the global energy output by stellar populations and active
nuclei.

Organized in parallel with the discovery of the CIRB,
a major intent of the deep ISO surveys was to start to
physically characterize the distant sources of the background and to single
out the fraction contributed by nuclear non-thermal activity in AGNs.

Finally, exploring the sky to unprecedented sensitivity limits
should have provided an obvious potential for discoveries of new unexpected phenomena
from our local environment up to the most distant universe.

\subsection{Overview of the main ISO surveys}

Deep surveys with ISO have been performed in two wide mid-IR
(LW2: 5-8.5\,$\mu$m and LW3: 12-18\,$\mu$m)
 and two far-IR ($\lambda_{eff}=90$ and 170 $\mu m$) bands. 
The diffraction-limited spatial resolutions were $\sim$5 arcsec at 10 $\mu m$ and 
$\sim$50 arcsec at 100 $\mu m$. Mostly because of the better imaging quality,
ISO sensitivity limits in the mid-IR are 1000
times better than at the long wavelengths (0.1 mJy versus 100 mJy).
At some level the confusion problem will remain a fundamental limitation also for   
future space missions (SIRTF, FIRST, ASTRO-F).
A kind of compensation to these different performances as a function
of $\lambda$ derives from the typical FIR 
spectra of galaxies and AGNs, which are almost typically one order of magnitude 
more luminous at 100 $\mu m$ than at 10 $\mu m$.
We detail in the following the most relevant programs of ISO surveys.

\subsubsection{The ISOCAM Guaranteed Time (GT) Extragalactic Surveys }

Five extragalactic surveys with the LW2 and LW3 filters have been performed 
in the ISOCAM GT (GITES, P.I. C. Cesarsky), including large-area shallow surveys and
small-area deep integrations. A total area of 1.5 square degrees in the
Lockman Hole and the "Marano" southern field have been surveyed, where more
than one thousand sources have been detected (Elbaz et al. 1999). These two areas 
were selected for their low zodiacal and cirrus emissions and because of
the existence of data at other wavelengths (optical, radio, X).

%Since the 7\,$\mu$m band catalogues include a large fraction of 
%Galactic stars, we will limit our analysis in the following sections
%to the LW3 15 $\mu m$ data, which are expected to be dominated by dust
%emission.

\subsubsection{The European Large Area ISO Survey (ELAIS)}

ELAIS is the most important program in the ISO Open Time (377
hours, P.I. M. Rowan-Robinson, see Oliver et al. 2000a).
A total of 12 square degrees have been surveyed at 15 $\mu m$ with ISOCAM
and at $90\ \mu m$ with ISOPHOT, 6 and 1 sq. degrees have been
covered with the two instruments at 6.7 and 170 $\mu m$.
To reduce the effects of cosmic variance, ELAIS was split into
4 fields of comparable size, 3 in the north, one in the south, plus six smaller areas.
While data analysis is still in progress, a  source list of over 1000 (mostly $15\ \mu m$)
sources is being published, including starburst galaxies and AGNs (type-1 and type-2), 
typically at z$<$0.5, with several quasars (including various BAL QSOs) 
found up to the highest z.

\subsubsection{The ISOCAM observations of the two Hubble Deep Fields}

Very successful programs by the Hubble Space Telescope have been the two
ultradeep exposures in black fields areas, one in the North and the
other in the South, called the Hubble Deep Fields (HDF). 
These surveys promoted a substantial effort of multi-wavelength studies aimed at
characterizing the SEDs of distant and high-z galaxies. These areas,
including the Flanking Fields for a total of $\sim 50$ sq. arcmin, 
have been observed by ISOCAM (P.I. M. Rowan-Robinson) at 6.7 and 15 $\mu m$,
achieving completness to a limiting flux of 100 $\mu Jy$ at 15 $\mu m$.

These have been among the most sensitive surveys of ISO and have allowed to
discover luminous starburst galaxies over a wide redshift interval up
to $z=1.5$ (Rowan-Robinson et al. 1997; Aussel et al, 1999). 
In the inner 10 sq. arcmin, the exceptional images of HST provided a detail 
morphological information for ISO galaxies at any redshifts (see Figure 4).
Furthermore, these two fields benefit by an almost complete redshift
information (Cohen et al. 2000), allowing a very detailed 
characterization of the faint distant IR sources.
%The outcome of these analyses (Elbaz et al. 1998; Aussel et al. 1999, 2000) 
%was that more than half of sources at $z>0.4$ are classified as Irregulars/Mergers.

\subsubsection{ISOCAM survey of two CFRS fields}

Two fields from the Canada-France Redshift Survey (CFRS) have been
observed with ISOCAM to intermediate depths: the '14+52' field (observed at 6.7 and 
15 $\mu m$) and the '03+00' field (with only 15\,$\mu$m data, but twice as deep). 
The CFRS is, with the HDFs, one of the best studied fields with  multi-wavelength data. 
Studies of the galaxies detected in both fields have provided the first
tentative interpretation of the nature of the galaxies detected in
ISOCAM surveys (Flores {\it et al. } 1999). 

\subsubsection{The ISOPHOT FIRBACK survey program}

FIRBACK is a set of deep cosmological surveys in the far-IR,
specifically aimed at detecting at 170 $\mu m$ the sources of the far-IR background
(P.I. J.L. Puget, see Puget et al. 1999). 
Part of this survey was carried out in the Marano area, and
part in collaboration with the ELAIS team in ELAIS N1 and N2, for a total of 4 sq. degrees.
This survey is limited by extragalactic source confusion in the large ISOPHOT beam 
(90 arcsec) to $S_{170}\geq$ 100 mJy. Some constraints
on the counts below the confusion limit obtained from a fluctuation analysis of
one Marano/FIRBACK field are discussed by Lagache \& Puget (1999) (Sect. 9.4).    
The roughly 300 sources detected are presently
targets of follow-up observations, especially using deep radio exposures
of the same area to help reducing the large ISO errorbox and  to
identify the optical counterparts. Also an effort is being  made
to follow-up these sources with sub-mm telescopes (IRAM, SCUBA): this
can provide significant constraints on the redshift of sources which
would be otherwise very difficult to measure in the optical (Sect 12.2).

\subsubsection{The Lensing Cluster Surveys}

Three lensing galaxy clusters, Abell 2390, Abell 370 and Abell 2218,  
have received very long integrations by ISOCAM (Altieri et al 1999). 
The lensing has been exploited to achieve even better
sensitivities with respect to ultra-deep blank-field surveys (e.g. the HDFs),
and allowed detection of sources between 30 and 100 $\mu Jy$ at 15 $\mu m$.
However this was obviously at the expense of distorting the areal projection
and ultimately making uncertain the source count estimate.

\subsubsection{The Japanese Guaranteed Time surveys}

An ultra-deep survey of the Lockman Hole in the 7\,$\mu$m ISOCAM band was
performed by Taniguchi {\it et~al.} (1997; the survey field is different from
that of the GITES Lockman survey). 
Another field, SSA13, was covered to a similar depth (P.I. Y. Taniguchi). 
The Lockman region was also surveyed with ISOPHOT
by the same team: constraints on the source counts at 90 and 175 $\mu$
were derived by Matsuhara et al. (2000) based on a fluctuation analysis.

%{The PHOT Serendipitous and CAM Parallel Surveys}
%
%These are side-benefits of the ISO mission. The former utilizes the slew
%time between ISO's pointed observations, and consists of scanning strips of
%the sky with the PHOT camera at 170 $\mu m$. In such a way roughly 15\%
%of the sky has been observed with a 50\% completeness for sources
%brighter than 1.5 Jy (Stickel et al. 1998).
%
%The CAM Parallel survey has been possible because
%ISO can have 2 instruments simultaneously working and observing 
%two sky regions 10' to 20' apart (though one should be
%in a non-optimal configuration). The survey 
%is performed (mostly using the 6.7 $\mu m$ broad-band
%filter) when one of the other three instruments were observing the prime target
%(Siebenmorgen et al. 1996).
%In this way a total (unbiased) area of 33 sq.deg. has been observed during
%the ISO mission. However, given the moderate limiting flux, 
%we expect that the vast majority of the detected sources should be galactic stars.
%

\subsection{Data reduction}

ISOCAM data need particular care to remove the effects of glitches induced
by the frequent impacts of cosmic rays on the 
detectors (the 960 pixels registered on average 4.5 events/sec).
This badly conspired with the need to keep them cryogenically cooled to reduce
the instrumental noise, which implied a slow electron reaction time
and longterm memory effects. For the deep surveys this implied a problem to
disentangle faint sources from trace signals by cosmic ray impacts.

To correct for that, tools have been developed by various groups for the
two main instruments (CAM and PHOT),
essentially based on identifying patterns in the time history
of the response of single pixels, which are specific to either astrophysical
sources (a jump above the average background flux when a source
falls on the pixel) or cosmic ray glitches (transient spikes followed by
a slow recovery to the nominal background). The most performant algorithm
for CAM data reduction is PRETI (Stark et al. 1999), a tool exploiting multi-resolution 
wavelet transforms (in the 2D observable plane of the position on the detector 
vs. time sequence). An independent method limited to brighter flux sources, 
developed by D\'esert et al. (1999), has been found to provide consistent 
results with PRETI, in the flux range in common. Other methods have been 
used by Oliver et al. (2000a) and Lari et al. (2000).
These various detection schemes and photometry algorithms have been tested
by means of very sophisticated Monte Carlo simulations, including all possible
artifacts introduced by the analyses.

With simulations it is has been possible to control as a function 
of the flux threshold: the detection reliability, the completeness, the Eddington bias and 
photometric accuracy ($\sim$10\% where enough redundancy was available, as for
CAM HDFs and Ultradeep surveys). 
Also the astrometric accuracy is good (of order of 1-2 arcsec for deep highly-redundant
images), allowing straightforward identification of the sources (Aussel et al. 1999,
see Fig. \ref{hdfs}).
The quality of the results for the CAM surveys is proven by the very good consistency
of the counts from independent surveys (see Fig.[\ref{cdif15}] below).

Longer wavelength ISOPHOT observations also suffered from similar problems.
The $175\mu m$ counts from PHOT C200 surveys are reliable above the confusion
limit $S_{170}\sim$ 100 mJy, and required only relatively standard procedures of 
baseline corrections and "de-glitching".
More severe are the noise problems for the C100 $90\mu m$ channel, which would 
otherwise benefit by a better spatial resolution than C200. 
The C100 PHOT survey dataset is still presently under analysis.

\subsection{Mid-IR and far-IR source counts from ISO surveys}

IR-selected galaxies have typically red colors, because of the dust 
responsible for the excess IR emission. The most distant are also 
quite faint in the optical. For this reason the redshift information is available 
only for very limited subsamples (e.g. in the HDF North and CFRS areas).
In this situation, the source number counts, compared with 
predictions based on the local luminosity function, provide important
constraints on the evolution properties.

\begin{figure}
\caption{ISOCAM LW3 map ($\lambda_{eff}=15\ \mu m$, yellow contours) of the
Hubble Deep Field North by Aussel et al. (1999), overimposed on the HST image.
The (green) circles are the LW2 ($\lambda_{eff}=6.7\ \mu m$) sources.
The figure illustrates the spatial accuracy of the ISO deep images with LW3,
allowing a reliable identification of the IR sources [courtesy of H. Aussel].
}
\label{hdfs}
\end{figure}

Particularly relevant information comes from the mid-IR samples selected from the CAM 
GITES and HDF surveys in the LW3 (12-18 $\mu m$) filter,
because they include the faintest, most distant and most 
numerous ISO-detected sources. They are also easier to identify because of the
small ISO error box for redundant sampling at these wavelengths.

\begin{figure}
\psfig{figure=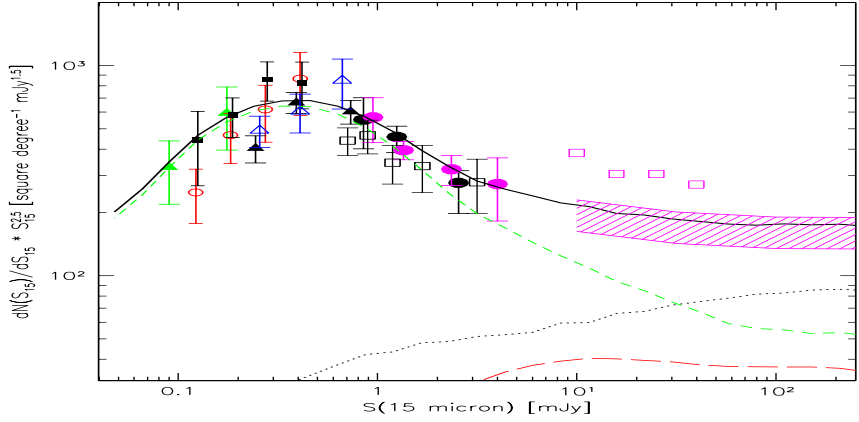,height=60mm,width=120mm}
\caption{Differential counts at $\lambda_{eff}=15\ \mu m$ normalized to the 
Euclidean law ($N[S]\propto S^{-2.5}$; the differential form is preferred here because
all data points are statistically independent). 
The data come from an analysis of the GITES surveys by Elbaz et al. (1999).
The dotted line corresponds to the expected counts for a population of
non-evolving spirals. The dashed line comes from our modelled population
of strongly evolving starburst galaxies, while the long-dashes are AGNs. 
The shaded region at $S_{15}>10\ mJy$ comes from an extrapolation of the 
faint 60 $\mu m$ IRAS counts by Mazzei et al. (2000).
}
\label{cdif15}
\end{figure}

Surveys of different sizes and depths are necessary to cover a wide dynamic
range in flux with enough source statistics, which justified performing a variety of
independent surveys at different flux limits.
The differential counts based on these data, shown in Fig. \ref{cdif15}, 
reveal an impressive agreement between so many independent samples.
Including ELAIS and IRAS survey data, the range in fluxes would reach four orders 
of magnitude. 
The combined 15 $\mu m$  differential counts display various remarkable features
(Elbaz et al. 1999):
a roughly euclidean slope from the brightest IRAS observed fluxes down to 
$S_{15}\sim 5\ mJy$, a sudden upturn at $S_{15}< 3\ mJy$, with the counts increasing as
$dN\propto S^{-3.1}dS$ to $S_{15}\sim 0.4 mJy$, and evidence for a 
flattening below $S_{15}\sim 0.3 mJy$ (where the slope becomes quickly sub-Euclidean,
$N\propto S^{-2}$). 

The areal density of ISOCAM 15\,$\mu$m sources
at the limit of $\sim$50-80 $\mu$Jy is $\sim$5 arcmin$^{-2}$.
This is nominally the ISO confusion limit at 15 $\mu m$, if we consider that the
diffraction-limited size of a point-source is $\sim 50$ arcsec$^2$:
from eq. (\ref{conf}) and for $\beta=-2$, confusion sets in at a source areal density 
of 0.1/resolution element, or $7/arcmin^2$ in our case.
The IR sky is so populated at these wavelengths that ISO was confusion
limited longwards of $\lambda=15\ \mu m$. This will also be the case for NASA's SIRTF
(due to launch in mid 2002), in spite of the moderately larger primary collector (85cm).

\begin{figure}
\psfig{figure=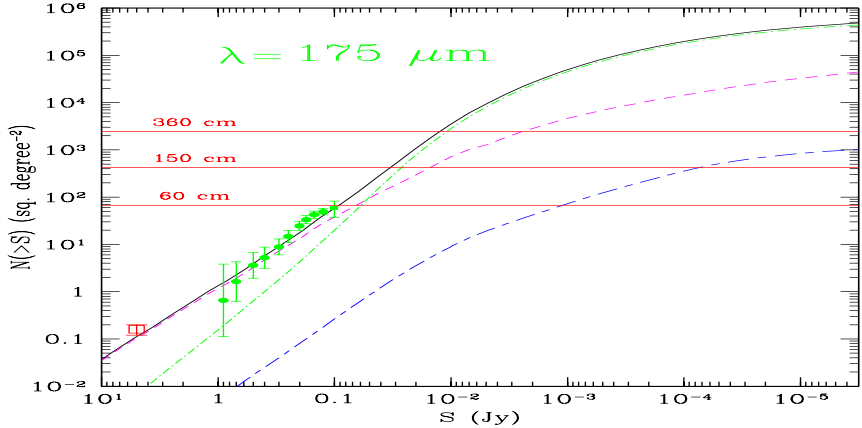,height=60mm,width=120mm}
\caption{Integral counts based on the ISOPHOT FIRBACK survey (Sect.10.2.5)
at $\lambda_{eff}=175\ \mu m$ (filled circles, from Dole et al. 2000) and on the ISOPHOT 
Serendipitous survey.  The dashed and dot-dashed lines correspond to
the non-evolving and the strongly evolving populations as in Fig.\ref{cdif15}.
The lowest curve is the expected (negligible) contribution of AGNs.
The horizontal lines mark the confusion limits for three
telescope sizes (based on eq.\ref{conf}): the lines marked "60cm" and "360cm" 
correspond to the ISO and FIRST limits for faint source detection.
}
\label{c175}
\end{figure}

Obviously, far-IR selected samples are even more seriously affected by confusion.
The datapoints on the 175$\mu m$ integral counts reported in Fig. \ref{c175}
come from the FIRBACK survey. Similarly deep observations at 90, 150 and 175 $\mu m$
are reported by Iuvela, Mattila \& Lemke (2000).
Given the moderate depth of these direct counts,
background fluctuation analyses were used to constrain their continuation below the
survey detection limit. The analysis of small-scale
fluctuations in one FIRBACK field by Lagache \& Puget (1999) produced $\sigma_{conf}
\simeq 0.07\ MJy/sr$ with a beam of size $\omega\simeq 6\ 10^{-4}\ sr$. 
From eq.[\ref{sigma}], this may be used to constrain the continuation of the
counts in Fig. \ref{c175} fainter than 100 mJy.
%The shaded region in Fig.[] comes from application of eq.[\ref{sigma}] inserting the
%measured value for $\sigma_{conf}$.

The 15$\mu m$ counts in Fig.\ref{cdif15} display a remarkable convergence
below $S_{15}\sim 0.2\ mJy$, proven by at least three independent surveys. The
asymptotic slope flatter than -1 in integral count units implies a modest contribution
to the integrated CIRB flux by sources fainter than this limit, unless a sharp
upturn of the counts would happen at much fainter fluxes
with very steep number count distributions, a rather unplausible situation.
 A meaningful estimate of the CIRB flux can then be obtained from direct integration 
of the observed mid-IR counts (the two datapoints at 15 and 7 $\mu m$ in Fig.\ref{bkg}).
If we further consider how close these are to the upper limits set by the observed TeV 
cosmic opacity 
(Fig. \ref{bkg}), the ISOCAM surveys appear to have resolved a significant (50-70\%)
fraction of the CIRB in the mid-IR.
On the other hand, the depth of the ISO far-IR surveys (FIRBACK) is not enough
to resolve more than ten percent of the CIRB at its peak wavelenth.

\section{EXPLORATIONS OF THE DEEP UNIVERSE BY LARGE MILLIMETRIC TELESCOPES}

Galaxy surveys in the sub-millimeter waveband offer a unique advantage 
for the exploration of the distant universe: the capability to naturally
generate volume-limited samples from a flux-limited survey.
This property is due to the peculiar shape of galaxy spectra in the
sub-mm, with an extremely steep slope from 1 mm to 100 $\mu$m,
as illustrated in Figure \ref{M82} for the prototype dusty starburst galaxy M82.

While above a few mm the luminosity is dominated by synchrotron and free-free
radio emission, from 100 $\mu$m to 1 mm dust continuum emission dominates,
with slopes as steep as $L(\nu)\propto \nu ^{3.5}$ (see Sect. 3). 
Then, as we observe at sub-mm wavelengths galaxies
at larger and larger redshifts, the rest-frame flux density
moves to higher and higher frequencies along a steeply increasing
spectrum, and the corresponding K-correction almost completely counter-balances 
the cosmic dimming of the observed flux, for a source of given luminosity at $z \geq 1$. 
The source flux keeps roughly constant with redshift up to $z\sim 10$, 
assuming cosmic sources were already present and dusty so early.

A further related advantage of sub-mm surveys is that local
galaxies emit very modestly at these wavelengths. Together with the
very favorable K-correction, this implies that a sensitive sub-mm survey
will avoid local objects (stars and nearby galaxies) and will select
preferentially sources at high and very high redshifts: a kind of
direct picture of the high-redshift universe, impossible
to obtain at other frequencies, where surveys are dominated by galaxies at 
modest redshifts if not by galactic stars.
Finally, and similarly to the ISO surveys, observing in the sub-mm has the advantage 
of producing samples completely unaffected by intergalactic opacity and dust extinction.

The third breakthrough event after 1996 for IR/sub-mm cosmology has come from
operation of a powerful array of bolometers (SCUBA) at the focal plane of the sub-mm
telescope JCMT on Mauna Kea. 
The success of SCUBA on JCMT was due to a combination of three crucial factors: 
a sensitive detector array with good multiplexing capability
(37 bolometers on a field of 2 arcmin diameter, with a diffraction-limited 
spatial resolution of 15 arcsec), 
put at the focal plane of a powerful sub-mm telescope (15m dish),
on a site allowing to operate at short enough wavelengths (850 $\mu m$) to exploit 
the very steep shape of sub-mm SED's of galaxies. 
For comparison, in spite of the larger collecting area, the competing bolometer array camera
on the IRAM 30m telescope at Pico Veleta (Spain) is limited to work at wavelengths $>1.2$ mm
by the poorer, lower-altitude site, which means by itself a factor 5 penalty 
in the detectable source flux with respect to SCUBA/JCMT.

The latter had a long development phase (almost like a space project!), 
partly because of the difficulty to keep the microphonic noise within acceptable limits.
But eventually, its long-sought results have come, and
the instrument is providing new very exciting facts to observational cosmology.

Basically, SCUBA/JCMT has allowed to partly resolve the long-$\lambda$ (850 $\mu m$) 
CIRB background into a population of faint distant, mostly high-z sources,
as discussed in Sect. 12.3 below. During three years of activity, largely dedicated to
deep surveys, SCUBA has discovered several tens of sub-millimetric sources,
mostly at 850 $\mu m$.

\begin{figure}
\psfig{figure=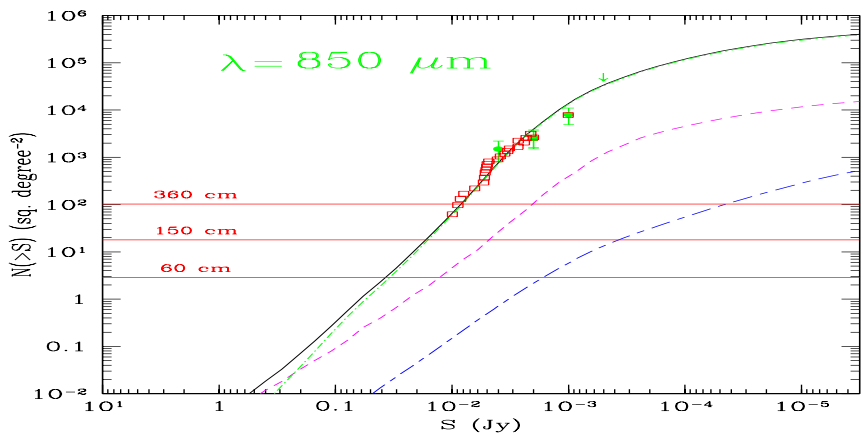,height=60mm,width=120mm}
\caption{Integral counts at $\lambda_{eff}=850\ \mu m$ (see also caption to Fig. \ref{c175}).
}
\label{c850}
\end{figure}

Four main groups have used SCUBA for a variety of deep integrations. 
Smail et al. (1997, 1999) have undertaken an ingenious program exploiting
distant galaxy clusters as cosmic lenses to amplify the flux of background sub-mm
sources and to improve the spatial resolution at the source. Their 
sample includes now 17 sources brighter than $S_{850}=6\ mJy$.
Hughes et al. (1998) published a single very deep image of the HDF North
containing 5 sources at $S_{850}(4\sigma) \geq 2\ mJy$.

Barger et al. (1998), while detecting only 2 sources down to 3 mJy, have 
carried out a very successful program of follow-up of SCUBA sources with optical
telescopes on Mauna Kea. 
Eales et al. (1999) and Lilly et al. (1999) have published 12 sources to 3 mJy
[a richer sample of 20 more sources is being published].

All these deep integrations are requiring many tens of hours each of
especially good weather, which meant a substantial fraction of the JCMT observatory time.
In spite of this effort, the surveyed areas (few tens of $arcmin^2$) and number of detected
sources are quite modest, which illustrates the difficulty to work from
ground at these wavelengths.

The extragalactic source counts at 850 $\mu$m, reported in Figure \ref{c850}, 
show a dramatic departure from the Euclidean law [$N(>S) \propto S^{-2}$
in the crucial flux-density interval from 1 to 10 mJy], a clear signature of the strong 
evolution and high redshift of SCUBA-selected sources.
Only 4 of them have been detected also at 450 $\mu m$, the sky transmission at Mauna 
Kea in this atmospheric channel is usually poor.

More recently, a new powerful bolometer array (MAMBO) has been put in operation
on IRAM. Bertoldi et al. (2000) report the first results of observations at
$\lambda_{eff}=1.2\ mm$ from a survey of 3 fields with a total area of over 300
$arcmin^2$ to a flux limit of few mJy.

%{The general outcome of these surveys is a, perhaps unexpected, 
%large areal density
%of faint sources, much more than predicted by estimates based on local luminosity
%functions of galaxies at the mm} (Franceschini, Andreani, Danese, 1998, 
%MNRAS)
%
%The frequency of source detection (better measure than the \# of sources which
%is small...) then requires strong evolution with cosmic time of the population emissivity

\section{INTERPRETATIONS OF FAINT IR/MM GALAXY COUNTS}

\subsection{Predictions for non-evolving source populations in the mid-IR}

A zero-th order approach to interprete the deep count observations is to
compare them with the expectations of
models assuming no-evolution for cosmic sources. Any such calculations
have to account for the effects of the very complex spectrum of
galaxies in the mid-IR (including strong PAH emission and
silicate absorption features, see Fig.\ref{source8}) in the K-correction factor appearing in
eq.(\ref{S}), which in terms of the system transmission function $T(\lambda)$ 
is more appropriately written as:
$$ K(L,z)={(1+z)\  \int_{\lambda_1}^{\lambda_2} d\lambda\ \left( \lambda_0 \over \lambda \right)\ 
T(\lambda)\ L[\nu(1+z)]            \over
\int_{\lambda_1}^{\lambda_2} d\lambda\ \left( \lambda_0 \over \lambda \right)\ 
T(\lambda)\ L[\nu,z=0]} .  $$
The effect on the source flux and on the counts [eq. \ref{dNdS}]
may be particularly important in the wide LW3 (12-18 $\mu m$) filter.
The prominent mid-IR features imply a complication when interpreting the counts, 
but at the same time they imply an enhanced sensitivity
of the LW3 source selection to the details of the evolution of sources in the
redshift interval $0.5<z<1.3$, which is known to be so critical for the formation 
of structures in the universe.

\begin{figure}
\psfig{figure=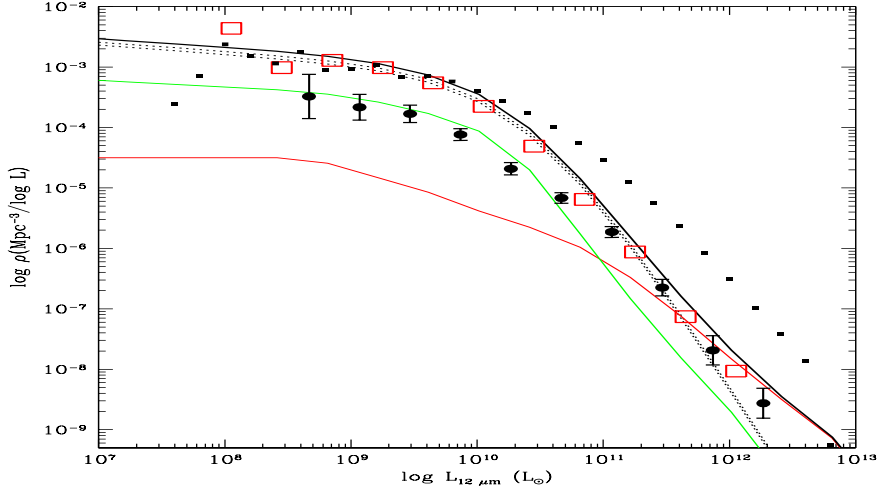,height=70mm,width=120mm}
\caption{Galaxy LLF's at $12\ \mu m$ from Xu et al. (1998, red open squares) compared 
with the IRAS $60\ \mu m$ LLF by Saunders et al. (1990, small filled squares).
Black ellipses are an estimate of the $12\ \mu m$ LLF of active galaxies (including
type-I [red line] and type-II AGNs plus starbursts [green line]) based on the (revised)
catalogue 
by Rush et al. (1993). Active galaxies clearly dominate the LLF at high luminosities.
}
\label{llf12}
\end{figure}

Local mid-IR luminosity functions have been published by Rush et al.
(1993), Xu et al. (1998) and Fang et al. (1998) based on the 12 $\mu m$ 
all-sky IRAS survey, see Figure \ref{llf12}. Unfortunately, in spite of the proximity of the 
CAM LW3 and IRAS 12 micron bands, at the moment we do not have a reliable
LLF at 15 $\mu m$ because of: a) uncertainties in the IRAS 12~$\mu$m photometry,
b) the effects of local inhomogeneities, particularly the
local Virgo super--cluster; and c) the flux conversion between the IRAS and CAM-LW3
bands (Elbaz et al. 1999).

The dotted line in Fig. \ref{cdif15} corresponds to the present
best estimate of the contribution from a non-evolving population
with a luminosity function consistent with that in the IRAS 12 $\mu m$ band derived by 
Xu et al. and Fang et al. The correction to the CAM LW3 band
is made assuming a 12 to 15 $\mu m$ flux ratio which is a function 
of the 12 $\mu m$ luminosity: for the less luminous objects the ratio
is based on the observed mid-IR spectrum of quiescent spirals, while for
the highest luminosity galaxies the ratio is the one expected for ultraluminous
IR galaxies, and for intermediate objects it is close to a typical starburst spectrum
like the one of M82 (see continuous line in Fig. \ref{source8}).
The 15 to 12 $\mu m$ flux ratio increases continuously with luminosity, the
flux at long-wavelength being increasingly dominated by the starburst emission.

It is clear that the no-evolution prediction, even taking into account the
effects of the PAH features on the K-corrections, falls very short of the
observed counts at fluxes fainter than a few mJy. Also the observed slope
 in the 0.4 to 4 mJy flux range ($N[S]\propto S^{-3\pm0.1}$)
is very significantly different from the no-evolution predicted dependence
$N(S)\propto S^{-2}$. 
The extrapolation to the bright fluxes is instead consistent, within the uncertainties,
with the IRAS 12 $\mu m$ counts with a slope close to Eclidean.

\subsection{Evidence for a strongly evolving population of mid-IR galaxies}

The shape of the differential counts shown in Fig. \ref{cdif15} contains relevant
indications about the properties of the contributing source populations. 
In particular the almost flat (Euclidean) normalized
counts extending from the bright IRAS fluxes down to a few mJy, followed by the sudden 
upturn below, suggests that is not likely the whole population of IR galaxies that evolve:
in this case and for the observed IR galaxy LLF, the super-Euclidean 
increase in the counts would appear at brighter fluxes and not be as abrupt. 
This behaviour is better consistent with a locally
small fraction of IR galaxies to evolve.

The IR counts in Fig. \ref{cdif15} are reproduced with the contribution of two 
source populations, one evolving, the other with constant properties as a function of
time.     The local fraction of the evolving starburst
population is only several percent of the total, consistent with the observed fraction 
of interacting galaxies ($\sim 5\%$ locally), the quick upturn in the counts then 
requiring quite a strong evolution to match the peak in the normalized
counts around $S_{15}\simeq 0.5$ mJy.
The details of the fit depend on the assumed values for the
geometrical parameters of the universe. For a zero-$\Lambda$ open universe (in our
case $H_0=50\ Km/sec/Mpc,\ \ \Omega_m=0.3$),
a physically credible solution would require a redshift increase of the comoving
density of the starburst sub-population and at the same time an increase of the luminosities 
respectively as
 \begin{equation}
 n(L[z],z) = n_0(L_0)\times (1+z)^6 \ \  \ \ \ 
% \end{equation}
% \begin{equation}
 L(z) = L_0\times (1+z)^3 .
 \end{equation}
These are quite extreme evolution rates, if compared with those observed in optical samples 
for the merging and interacting galaxies (e.g. Le Fevre et al. 2000).
The inclusion of a non-zero cosmological constant, and the corresponding increase 
of the cosmic
timescale from z$=0$ to 1, tend to make the best-fitting evolution rates less extreme.
For $H_0=50\ Km/sec/Mpc, \ \Omega_m=0.2, \ \Omega_\Lambda=0.8,$ a best-fit to the counts
requires:
 \begin{equation}
 n(L[z],z) = n_0(L_0,z)\times (1+z)^{4.5}, \ \ \ \  L(z) = L_0\times (1+z)^2
\label{solu}
 \end{equation}

To be consistent with data on the $z-$distributions from the ISO source samples
in the HDF (Aussel et al. 1999, 2000, see Fig. \ref{zdist}) and with the observed CIRB intensity, 
this fast evolution should turn over
at $z\simeq 1$ and the IR emissivity should keep roughly constant at higher $z$.
An accurate probe, however, of hidden SF in the interval $1\leq z \leq 2$ will only
be possible with the longer-wavelength broad-band channel of SIRTF at $\lambda_{eff}=24\ 
\mu m$. 

In our scheme, {\sl any single galaxy would be expected to spend most of its life in the
quiescent (non-evolving) phase, being occasionally put by interactions in a short-lived
(few $10^7\ yrs$) starbursting state. The evolution for the latter may simply be due
to an increased probability in the past to find a galaxy in such an excited mode.
Then the density evolution in eq. (\ref{solu}) scales with redshift as the rate of 
interactions due to a simple geometric effect following the increased source volume density. 
The luminosity evolution may be interpreted as an effect of the larger
gas mass available to the starbursts at higher z.}

Note, however, that the above evolutionary scheme is by no means the only one able to fit
the data, other solutions may be devised (e.g. the one by Xu [2000] allowing the whole
local population to evolve with cosmic time).

%
%We have tried to reproduce this behaviour by playing with various evolutionary 
%recipes (luminosity or density evolution with various rates) applied to the
%whole galaxy population, or to a sub-population significantly present locally
%(see models by AF88 anf PRR96):
%
%{\sl the observed counts are in no way reproduced }(excess number of galaxies expected at
%$S_{15}\sim 10 mJy$ and lack of at 0.5 mJy).
%
%The LF of the evolving SB population combines L and \# evolution in such a way
%that, at z=1, the total LF is roughly $\times 10$ more numerous at 
%$L_{15}=10^9 L_\odot$ than the global local LF of galaxies and $\times 5$ higher 
%in L at $L_{15}=10^{11} L_\odot$
%
% altogether, at z=1 the total luminosity density at 15 $\mu$ is 40 times
%higher than locally
%
%
%Note that nor pure \# density nor pure L evolution can explain the counts, but
%only a combination:
%
%\# evol. is not enough, unless unplausibly high rates ($n\propto [1+z]^{13}$ !) 
%are assumed;
%
%simple L evol. would imply unphysical rates ($L(z)\propto [1+z]^{6}$ !),
%and could'nt explain the LF at z=0.5 to 1 in the critical L range 
%$L_{15}=10^9$ to $10^{11} L_\odot$.

\subsection{A panchromatic view of IR galaxy evolution}

Deep surveys at various IR/sub-mm wavelengths can be exploited to simultaneously 
constrain the evolution properties and broad-band spectra of faint IR sources.
Franceschini et al. (2000) have compared the 15 $\mu m$ survey data with those coming from 
the IRAS 60 $\mu m$, the FIRBACK 175 $\mu m$, the ELAIS 90 $\mu m$, and the SCUBA 850 $\mu m$
surveys, which are the deepest, most reliable available at the moment.
Information on both number counts and the source redshift distributions were used in these 
comparisons. 

Further essential constraints, providing the local boundary conditions on the 
evolution histories, are given by the multi-wavelength local luminosity functions. 
In addition to the 12 and 15 $\mu m$ LLF's, as discussed in Section 11.1, 
the galaxy LLF is particularly well known at 60 $\mu m$ after the IRAS all-sky 
surveys and their extensive spectroscopic follow-up (Saunders et al. 1990). 
Dunne et al. (2000, see also Franceschini, Andreani, Danese 1998) attempted to constrain 
the galaxy LLF in the millimeter, based on $mm$ observations of complete samples of 
IRAS 60 $\mu m$ galaxies. 

As previously mentioned, the properties of LLF's observed at various IR/sub-mm wavelengths 
can be explained only assuming that the galaxy IR SED's depend on bolometric luminosity. 
Fig. \ref{llf12} shows that the 60$\mu m$ LLF has a flatter 
(power-law) shape at high-L compared with the mid-IR LLF's (a fact explained in 
Sect. 6.6 as an effect of spectra for luminous active galaxies showing excess 60 $\mu m$
emission compared to inactive galaxies [see also the L-dependence of the IRAS colours]).

%A first robust result of comparing the various IR survey data with typical galaxy SED's
%and multi-wavelength LLF's is the assessment of the contribution by low-redshift
%sources to the observed counts and CIRB intensity: both are far from being
%explained by a local source population without evolution (see Figs. \ref{bkg},
%\ref{cdif15},\ref{c850}), and all require a substantial amount of increase of
%the volume emissivity with redshift.

\begin{figure}
\psfig{figure=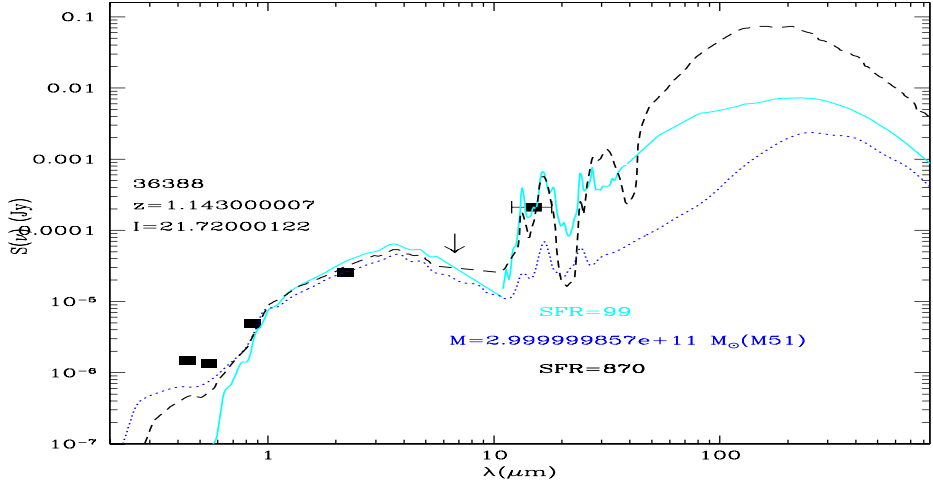,height=70mm,width=130mm}
\caption{Broad-band spectrum of a mid-IR source selected by ISOCAM LW3 
in the Hubble Deep Field North (Aussel et al. 1999), compared with
the SED's of M82 (thick continuous line), Arp 220 (dashed line), and M51
(dotted line). Estimates of the SF rate [based on the M82 and Arp 220
templates] and of the stellar mass [based on the M51 template] are indicated.
}
\label{source8}
\end{figure}

Franceschini et al. (2000) have modelled in some detail the redshift-dependent 
multi-wavelength LLF's of galaxies by assuming for both non-evolving
spirals and active starburst galaxies spectral energy distributions dependent
on luminosity, with spectra ranging from those typical of inactive spirals 
for low-luminosities,
to the 60$\mu m$--peaked spectra of luminous and ultra-luminous IR galaxies
as previously described. 
For the SED's of intermediate luminosity objects, linear 
interpolations between the two as a function of bolometric luminosity were assumed. 
This allows to simultaneously fit the LLF's at the various wavelengths.
For comparison, solutions with single spectral energy distributions for the 
evolving populations were also tried.

Altogether, the observed long-wavelength counts and CIRB intensity, when compared with typical 
galaxy SED's and the multi-wavelength LLFs, 
require a substantial increase of the IR volume emissivity of galaxies with redshift
(see Figs. \ref{bkg}, \ref{cdif15}, \ref{c850}).

Should one assume that the IR SED of the ultra-luminous galaxy Arp 220 is 
representative of the average spectrum of the evolving population detected
by ISOCAM LW3, then the consequence would be that the observed far-IR counts and the 
CIRB intensity are far exceeded.
On the contrary, if we assume for the IR evolving sources a more typical starburst
spectrum (like the one of M82, by all means similar to those of other luminous
starbursts observed by ISO), then most of the observed properties
of far-IR galaxy samples (number counts, redshift distributions, luminosity functions)
are appropriately reproduced. Best-fits to the counts based on the M82 template
are given in Figs. \ref{c175} and \ref{c850}.

The good match to the multi-wavelength counts obtained by assuming a typical starburst 
spectrum for the evolving population already indicates that 
the faint IR-selected source population is likely dominated by processes of star-formation 
in distant galaxies more than by AGN emissions.
This seems indeed the result of the first spectroscopic studies of faint ISO sources
(Sect. 12.1), although a more substantial effort is required to confirm it.
Considering the different shapes of the IR SEDs for SBs and AGNs, this would imply 
that the population detected by ISO in the mid-IR not only contributes a major
fraction of CIRB at 15$\mu$, but is also responsible for a majority contribution
of the CIRB at any wavelengths.

\section{NATURE OF THE FAST EVOLVING SOURCE POPULATION}

\subsection{Tests of the evolving IR population in the HDFs and CFRS fields}

The ISO observatory has deeply surveyed with CAM LW3 some of the 
best investigated sky areas, in particular the two Hubble Deep Fields (North \& South, 
Rowan-Robinson et al. 1997, Oliver et al. 2000b) and the area CFRS 1415+52 (Flores et al. 1999).
%achieving completeness down to $S_{15}\sim 100 \mu Jy$ and $S_{15}\sim 300 \mu Jy$ respectively. 
Given the variety of multi-wavelength data
and the almost complete spectroscopic follow-up, the surveys in these areas have
allowed to achieve important tests of
the evolving population responsible for the upturn of the ISO mid-IR counts and for
a substantial fraction of the CIRB.

\begin{figure}
\psfig{figure=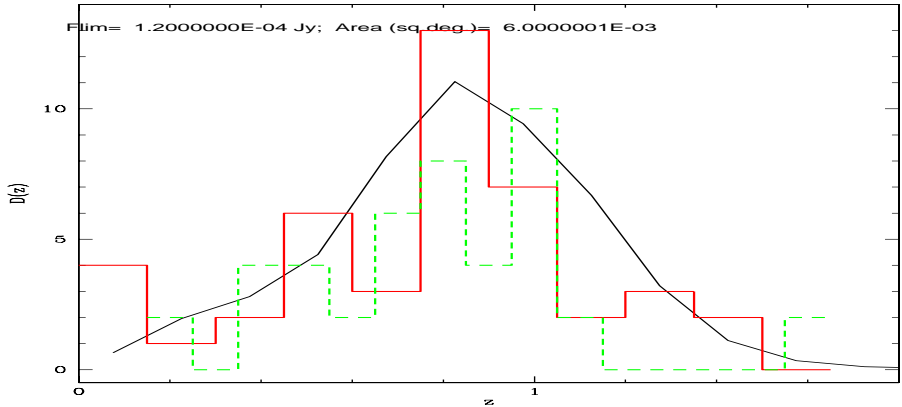,height=60mm,width=130mm}   %@  rifare figura
\caption{Redshift distributions from the HDFN (Aussel et al. 2000) and CFRS 1415+52
ISOCAM LW3 samples, compared with model predictions.
}
\label{zdist}
\end{figure}

Aussel et al. (1999 and 2000) report reliably tested (see Sect. 9.3) complete samples of 
49 and 63 sources to $S_{15}\geq 100 \mu Jy$ in the HDF North and South respectively, 
covering similar areas of 25 sq. arcmin each. Flores et al. (1999) analyse a sample of 
41 sources brighter than $S_{15}\sim 300 \mu Jy$ ($S/N>4$) over an area of 10'x10' 
in CFRS 1415+52.
The vast majority (90\%) of the ISO sources in the HDF surveys have spectroscopic
redshifts, and for the remaining objects photometric redshifts are easily estimated.
The redshift distributions $d(z)$
for the HDF and CFR1415 surveys are reported in Figure \ref{zdist}, and compared with the model
fitting the multi-wavelength counts mentioned in Sect. 11.2.
Although the two surveys cover individually small sky areas, the fair match between
them gives some confidence about the overall reliability of the result.
These data set a stringent limit on the rate of cosmological evolution for
IR galaxies above $z\sim 1$, which needs to level off to avoid exceeding
the observed $d(z)$ on the high-$z$ tail. Note however that the observed high-$z$ convergence
of $d(z)$
is also partly an effect of the strong K-correction in the LW3 flux for dust-rich galaxies
(see an example in Fig. \ref{source8}): disentangling K- from evolutionary-corrections
at $z>1$ will require SIRTF and FIRST.

HST imaging data on these fields provide detailed morphological information on ISO
sources. Elbaz et al. (1999) and Aussel et al. (1999) find that 30 to 50\% of them
show clear evidence of peculiarities and multiple structures, in keeping with the local 
evidence that galaxy interactions are the primary trigger of luminous IR starbursts.
From their Caltech redshift survey in the HDF North, 
Cohen et al. (2000) report that over 90\% of the faint LW3 ISO sources are members 
of galaxy concentrations and groups, which they identify as peaks in their redshift
distributions.  Indeed, it is in these dense galaxy environments with low velocity dispersion 
that interactions produce resonant perturbation effects on galaxy dynamics.

\subsubsection{Optical and NIR spectral properties: nature of the IR sources}

Flores et al. (1999) report a preliminary analysis of the spectra of IR sources
in CFRS 1415+52, noting that a majority fraction of these display both weak emission
(OII 3787) and absorption ($H_\delta$) lines, as typical of the {\sl e(a)} galaxy spectral
class: the latter is mentioned in the literature as a post-starbursting population,
one in which a vast population of A-type absorption-line stars from a $\sim 1$ Gyr old
massive starburst combine with a small residual of ongoing SF evidenciated by
the weak OII emission.
Given the far-IR selection of the faint ISO sources, which is expected to preferentially
detect dusty star-forming galaxies, this result would be difficult to understand,
as it lets open the question of "why the ongoing active starbursts are
not detected".

\begin{figure}
\vspace*{-1.cm}
\psfig{figure=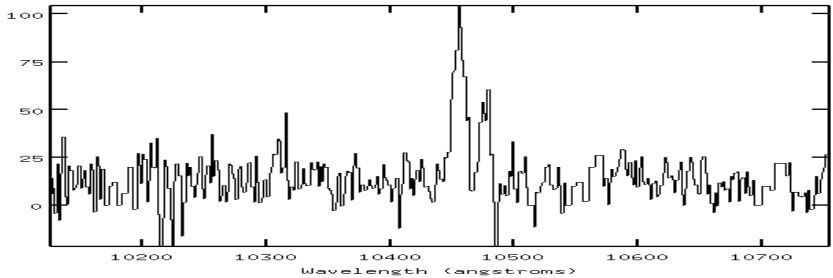,height=50mm,width=120mm}
\vspace*{1.cm}
\caption{
ISAAC/VLT spectrum of HDFS source \# 53 at z=0.58. The $H\alpha$ and NII redshifted lines
are clearly visible [from Rigopoulou et al. 2000].
}
\label{s53}
\end{figure}

Rigopoulou et al. (2000) and Franceschini et al. (2000b) have observed with ISAAC on VLT 
a sample of 13 high-z (0.2$<z<$1.4) galaxies selected in the HDF South to 
$S_{15}>100\ \mu Jy$:
{\sl the $H\alpha$ line is detected in virtually all of the sources, and found quite
prominent (EW$>50\ \AA$), indicating substantial rates of SF 
after de-reddening corrections, and demonstrating that these optically faint but IR luminous 
sources are indeed powered by an ongoing massive dusty starburst}.

The {\sl e(a)} spectral appearence is interpreted by Poggianti \& Wu (1999) and
Poggianti, Bressan, Franceschini (2000) as due to selective dust attenuation,
extinguishing more the newly-formed stars than the older ones which have already
disrupted their parent molecular cloud.

{\sl These papers independently found that $\sim 70 - 80\%$ of the energy emitted by young
stars and re-processed in the far-IR leaves no traces in the optical spectrum, hence 
can only be accounted for with long-wavelength observations.}

\subsubsection{Evaluating baryonic masses and the SFR of the IR population}

\begin{figure}
\psfig{figure=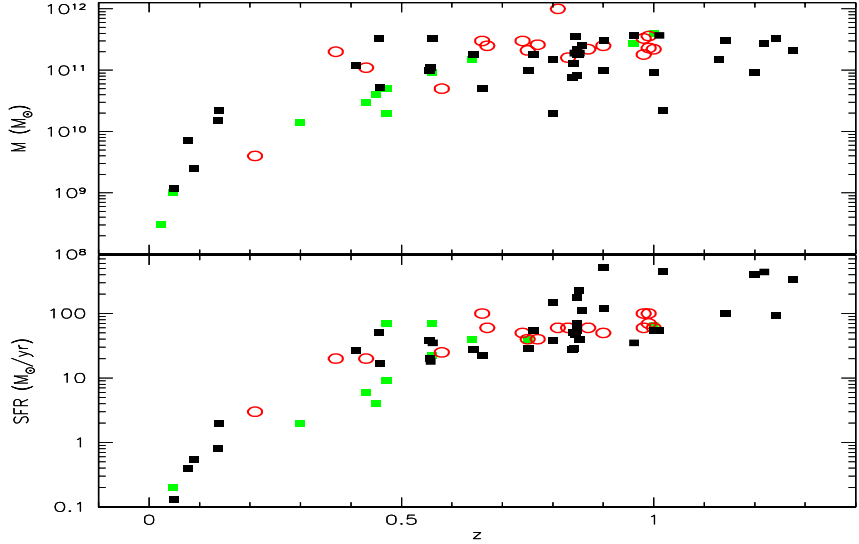,height=80mm,width=120mm}
%\vspace*{1.cm}
\caption{Star formation rates and baryonic masses as a function of redshift
for galaxies selected by ISOCAM LW3 at 15$\mu m$ in the HDFN and CFRS 1415+52.
}
\label{sfr_z}
\end{figure}

Further efforts of optical-NIR spectroscopic follow-up of faint
IR sources are planned for the next years, including attempts to address the 
source kinematics and dynamics based on line studies with the next-generation of
IR spectrographs (e.g. SINFONI on VLT). The latter would be particularly relevant in
consideration of the typically complex dynamical structure of luminous IR starbursts.
At the moment, for an evaluation of the main properties of the IR population
we have to rely on indirect estimates exploiting the near-IR and far-IR fluxes.
One important parameter is the baryonic mass in stars, for measure of which 
fits of local template SEDs to the near-IR broad-band spectrum can be used.
Our estimated values of the baryonic mass ($\sim 10^{11}\ M_\odot$, with 1 dex typical
spread, see Figure \ref{sfr_z}) indicate that already evolved and massive
galaxies host the powerful starbursts.

As a measure of the rate of star-formation (SFR), the other fundamental parameter
describing the physical and 
evolutionary status of the sources, we have exploited the mid-IR flux as an alternative to 
the (heavily extinguished) optical emissions, since it is much more directly 
related to the bolometric (mostly far-IR) flux, which is the most robust indicator of 
the number of massive reddened newly-formed stars.  Vigroux et al. (1998) find that the
ISOCAM mid-IR fluxes (from both LW3 and LW2 ISOCAM observations) 
are tightly and linearly related with the bolometric
emission in local galaxies, evidence contradicted only in very extinguished peculiar 
sources (e.g. Arp 220), for which the mid-IR spectrum is self-absorbed.
Using several HDF North sources having both the mid-IR and radio flux, Aussel et al. (2000)
find that the two SFR estimators, both largely unaffected by
dust extinction, provide consistent results on the SFR.
However, the mid-IR flux has the advantage over the radio to be less affected by AGN
emission, providing a more reliable SF measurer (Cohen et al. 2000; Aussel et al. 2000;
Franceschini et al. 2000b). Also the fact that only 7 of the 49 IR SBs in the HDFN
are detected in radio to a flux limit of few tens of $\mu Jy$ tells that the mid-IR flux
is a more sensitive indicator of SF.
This untill dedicated space missions (in particular the 3.6m FIRST observatory)
will measure the peak of dust emission at $\lambda\sim 100\ \mu m$ in high-redshift
galaxies with high accuracy.

Altogether, the galaxy population dominating the faint mid-IR counts and substantially
contributing to the bolometric CIRB intensity (assumed typical SB SEDs)
appears to be composed of luminous ($L_{bol}\sim 10^{11}-10^{12}\ L_\odot$) 
starbursts in massive ($M\sim 10^{11}\ M_\odot$) galaxies at $z\sim 0.5-1$,
observed during a phase of active stellar formation.
The typically red colors of these systems suggest that they are mostly unrelated to
the faint blue galaxy population dominating the optical counts (Ellis 1997),
and should be considered as an independent manifestation of (optically hidden)
star formation (Elbaz 1999; Aussel 1998).

%
%We warn that the alternative of using the radio emission as a measure may not be 
%reliable:
%the analysis of Schmitt et al. has shown that, particularly in SBs, the radio 
%emission may be substantially depressed (+ very large spread in radio)
%
%Among local templates, apparently Arp220 has a remarkably variant spectrum, with
%excess far-IR emission, while M82, NGC 6090, and others have a more typical IR
%spectrum:
%as a result, estimating the SFR from mid-IR luminosities using Arp220 as a template
%would imply SFR values larger by factors 3-5 than using other more typical spectra.
%From our previous analysis, we use M82 (or the similar N6090) to estimate SFR
%
%At $z>0.5$, typical values of SFR are around $100\ M_\odot$ (with 5-6 out of 17
%galaxies showing SFR=200: these are typically 2 mag higher than implied by
%optical obs (as inferred from the $L_{2800}$ or the OII line EW)
%
%For the ISO-HDF the results are similar, except that the SFR shows a wider range
%to lower values, as expected due to the fainter fluxes
%
%Note a correlation of SFR and $S_{15}$ and (an obvious one) with z
%
%the fraction of inactive (SFR $<10\ M_\odot/yr$) to actively star-forming
%galaxies at $S_{15}>100\ \mu Jy$ (i.e. 10\% of the population) is consistent with
%expectation

\subsection{What are the FIRBACK 175 $\mu m$ sources? }

The nature of the 175 $\mu m$ sources discovered by FIRBACK/ISO, and contributing
$\sim 10\%$ of the CIRB intensity, is presently the target of intense observational and 
modellistic investigations, although no conclusions are possible at the moment.
Because of the missing knowledge of the LLF, the interpretation of the 175$\mu m$ counts 
themselves is subject to some uncertainties: 
is there strong or marginal evidence for evolution at the survey limit of
100 mJy (Fig.\ref{c175})? Dole et al. (2000) argue in favour of the former,
while Fig. \ref{c175} reports a solution in which a moderate-redshift ($z\sim 0.5$)
population still dominates there.

The basic limitation comes from the difficulty to identify the optical counterparts,
due to the large (40 arcsec) ISOPHOT error-box. 
Progress is being achieved by cross-correlating with deep radio
surveys available in the FIRBACK fields (exploiting the good radio/FIR correlation,
eq. \ref{FIRrad}) and by means of some limited SCUBA follow-up.
Scott et al. (2000) have obtained data at 450 and 850 $\mu m$ for 10
FIRBACK sources: the FIR-mm SEDs tentatively indicate, for plausible far-IR spectra,
redshifts in the range from 0 to 0.4 for the majority of the sources, while
a few may be at $z>1$.

Mid-IR 15 $\mu m$ fluxes from an ISOCAM map are available in the {\sl "FIRBACK Marano"}
area, which indicate that the 15$\mu m$ counterparts of the 175$\mu m$ sources
are rather faint (Elbaz, 1999).
Three interpretations have been suggested: (a) FIRBACK sources are typically very 
high-luminosity Arp220-like at low redshift (z$\sim$0.1-0.4); 
(b) they are more standard starbursts at $z>1$;
(c) they are low-activity spirals at moderate $z$ with significant 
amounts of cold-dust and excess emission at $\lambda>100\mu m$.

Although the results of the SCUBA observations might indicate that the last interpretation 
could be more probable, the nature of the FIRBACK source population is far from proven, 
further multi-wavelength data being required to address it.
Deeper far-IR observations will be possible with SIRTF, but a more final solution will 
probably require the  FIRST's better spatial resolution.

\subsection{The nature of the high-z galaxies detected in the millimeter}

Thanks to the unique advantage for deep sub-mm observations offered by the very 
peculiar $K-$correction, sub-mm surveys with sensitivities of few mJy at 850$\mu m$, 
have been able to detect high-redshift (very luminous) sources in flux-limited samples. 
The observed 850 $\mu m$ counts, far in excess of the no-evolution prediction,
already tell incontrovertibly about the cosmological distance and evolutionary status 
of the SCUBA-selected source population.

\begin{figure}
%\vspace*{-1.5cm}
\hspace*{1.0cm}
\psfig{figure=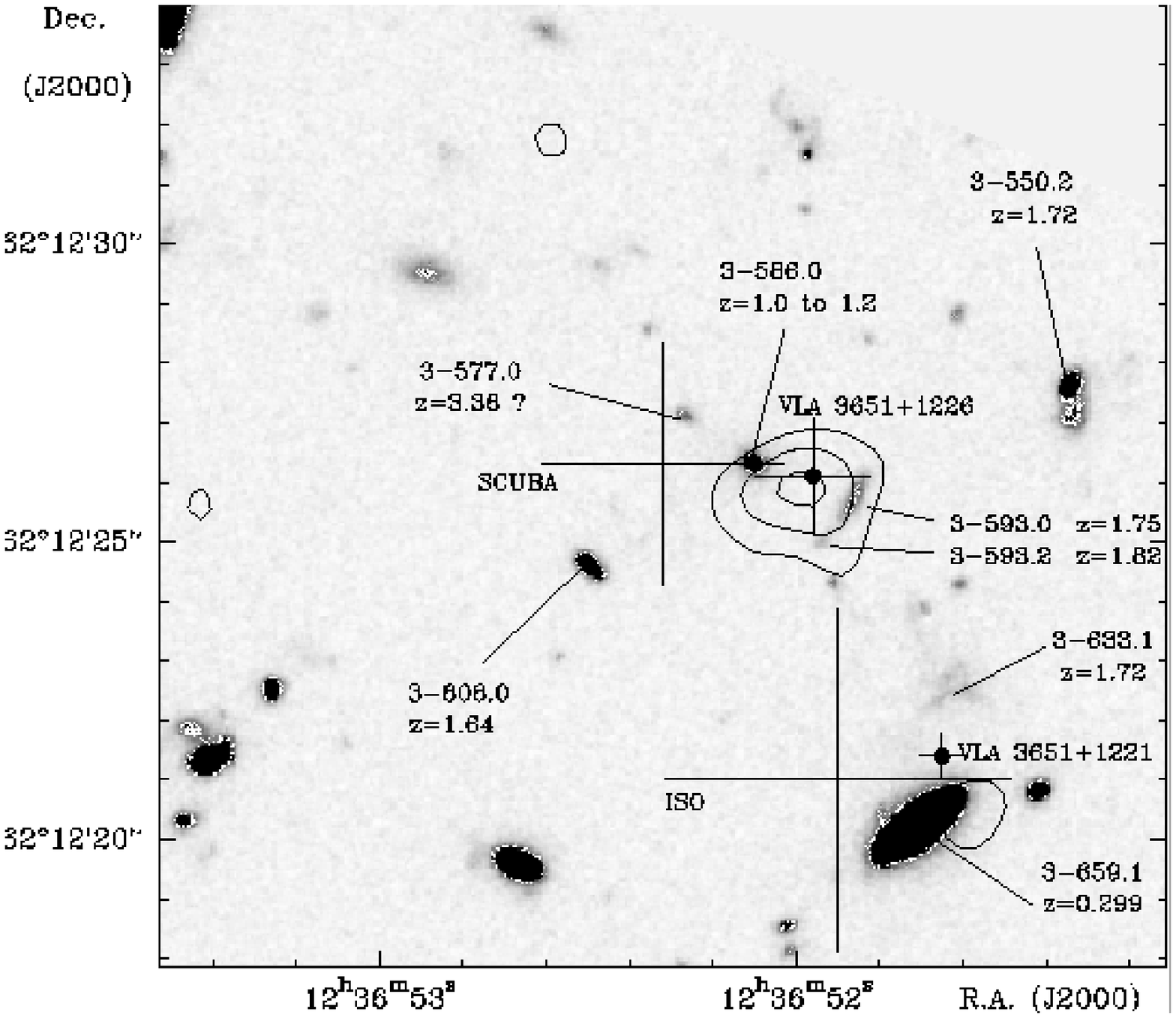,height=100mm,width=110mm}
%\vspace*{-3.5cm}
\caption{Map of the 1.3 mm continuum obtained with the IRAM interferometer in the field
of the source HDF 850.1 by Downes et al. (1999). HDF 850.1 is the brightest source discovered
at 850 $\mu m$ by SCUBA (Hughes et al. 1998), and has a flux density of 2.2 mJy at 1.3 mm. 
The field center coincides with the
center position of the SCUBA error-box, whose size is however comparable to the whole
image area. The colour image is a 
composite of BVI data from HDF. Positions of VLA and ISO sources,
as well as photometric redshift data, are also indicated. IRAM and VLA position clearly
point to a faint optical counterpart of HDF 850.1 (3-593.0), possibly influenced by gravitational
lensing by the elliptical 3-586.0, in a similar configuration to the prototypical
primeval galaxy IRAS F10214 [courtesy of D. Downes].
}
\label{downes}
\end{figure}

Unfortunately, probing directly the nature of these objects via optical
identification and spectroscopic follow-up turned out to be very difficult,
in spite of the substantial efforts dedicated. The SCUBA diffraction-limited 
HPBW at 850 $\mu m$ is large, $\sim 15$ arcsec FWHM, and the
difficulty of the identification is further exacerbated by the usual extreme
faintness of the optical counterparts, as demonstrated in the (few) cases in which
the identification has been possible (see e.g. Figure \ref{downes}).

The reliability of the identification has been evaluated
by computing the probability that the nearest member of a population of
candidate identifications with surface density $n$ falls by chance within a distance 
$d$ from the SCUBA source: $ P= e^{-\pi n d^2}$.
For a sample of size $N$ of SCUBA detections, the product $NP$ gives the number of spurious 
identifications (Lilly et al. 1999). 
This analysis has shown that the situation is not quite comfortable for the SCUBA 
surveys, essentially because of the
faintness of the optical counterparts: roughly 50\% of all identifications may be 
spurious.

Two approaches have been followed to improve the identification and try to characterize the 
population.
One was to systematically survey spectroscopically all optical sources
falling in the SCUBA beam, the other was to exploit cross-identifications with ultra-deep 
radio catalogues.
Particularly well studied are the fields in the Cluster Lens Survey 
(Smail et al. 1997), exploiting the flux-amplification by massive foreground galaxy
clusters. The current situation about redshift measurements in this survey is: the 
16 SCUBA sources have 24 possible counterparts with spectroscopic redshifts,
 6 reliable $z$ estimates (a $z=2.8$ combined AGN/starburst, a $z=2.6$ galaxy pair,
2 galaxies with AGN signatures at $z=1.16$ and $z=1.06$, and finally 2 foreground
cD cluster members [Barger et al. 1999]). Note that the identification with the 
galaxy pair has been later confirmed by CO mm observations (Frayer et al. 1999).

An interesting case is illustrated in Fig. \ref{downes}, showing the brightest object HDF-850.1 
in the Hughes et al. (1998) survey, confirmed by IRAM interferometry as a probable 
ultra-luminous lensed 
starburst with $L_{bol}\sim 2\ 10^{12} L_\odot$ at $z_{photom}\simeq 1.7$

The difficulty of the identification process is illustrated by the recent finding
(Smail et al. 1999) of the presence of two Extremely Red Objects (ERO's) as probable 
counterparts of two SCUBA sources.
Given the faintness of optical counterparts and the extreme difficulty to get the
redshift from optical spectroscopy, some millimetric estimators of the redshift 
have been devised to override optical measurements.
Hughes et al. (1998) use the $S_{450}/S_{850}$ flux ratio as a measure of $z$.
However, given the rather wide temperature-distribution of cosmic dust (see e.g. the 
three quite different spectral templates, for Arp220, M82, and M51 in 
Fig. \ref{source8}), this test proved to be very uncertain.
Much more reliable the technique proposed by Carilli \& Youn (1999) to exploit the
$S_{850\mu}/S_{20cm}$ flux ratio, which has the advantage to rely on very robust
mm spectral shapes at 850 $\mu$ ($S_\nu \propto \nu^{3.5}$, see Sect. 3)
and in the radio (typical power-law synchrotron spectra), with opposing spectral 
slopes. Assuming an Arp 220 spectral template they got:
$$ S_{850\mu}/S_{20cm} = 1.1 \times (1+z)^{3.8}  $$ 
whose small scatter mostly reflects the tight FIR to radio correlation.

Population constraints on the $z$-distributions have been derived in this way,
and the basic result (still tentative and requiring confirmation) is that
{\sl faint SCUBA sources are mostly ultra-luminous galaxies at typical 
$z\sim 1$ to $\sim 3$} (e.g. Barger et al. 1999).
Clearly, the details of the z-distribution cannot yet be quantified with precision,
this will likely require new instrumentation (mm interferometers -- e.g. ALMA -- 
are particulalry needed, in addition to space FIR observatories).

As suggested by many authors, the similarity in properties between this
high-z population and local ultra-luminous IR galaxies argues in favour
of the idea that these represent the long-sought "primeval galaxies", those in particular
originating the local massive elliptical and S0 galaxies.
This is also supported by estimates of the volume density of these
objects in the field $\sim 2-4\times 10^{-4}\ Mpc^{-3}$, high enough to allow most
of the field E/S0 to be formed in this way (Lilly et al. 1999).
As for the E/S0 galaxies in clusters, a very interesting result was the recent 
discovery by SCUBA of a 
significant excess of very luminous ($L\sim 10^{13}L_\odot$) sources at $850 \ \mu m$
close to the z=3.8 radiogalaxy 4C41.17 (Ivison et al. 2000), which parallels
the evidence of a similar excess of EROs and Lyman-break galaxies in this area.
It is tentalizing to interprete these data as indicative of the presence of a
forming cluster surrounding the radiogalaxy, where the SCUBA sources
would represent the very luminous ongoing starbursts.

By continuity, the less extreme starbursts ($L\sim 10^{11}-10^{12}\ L_\odot$) 
discovered by ISOCAM at lower redshifts can possibly originate the spheroidal
components in later morphological type galaxies (see more in Sect. 13.2.4 below).

\subsection{AGN contribution to the energetics of the faint IR sources}

Within this interpretative scheme, a margin of uncertainty still exists about the 
possible contribution 
by gravitational accretion from a nuclear quasar to the energy budget in these high-z 
IR-mm sources.
While stellar energy production provides a modest overall efficiency for baryon
transformations of quite less 
than a percent at most, the theory of gravitational accretion predicts values 
in the range $\epsilon \sim 5 - 40\%$. A natural question then arises as of how much
of the bolometric flux in these sources is contributed by an AGN. 
 Unfortunately, the optical--UV--soft-X ray primary source spectrum
in the high-redshift IR-mm sources is almost completely re-processed by dust into 
an IR spectrum largely insensitive to the properties of the primary incident one. 

As for SCUBA sources, there have been indications for AGN activity for at least
a fraction (20-30\%) of them. Indeed, since SCUBA selects the top 
luminosity end of the IR population, and considering
the local evidence of a larger incidence of AGNs among ULIRGs, an important AGN 
contribution to the SCUBA sources would be expected (potentially biasing our conclusions 
about their contribution to the SFR history).
Risaliti et al. (2000) and Bassani et al. (2000) claim
evidence for a significant AGN contribution in the large majority ($>60\%$) of the local 
ULIRGs based on hard X-ray data, something confirmed also by high spatial resolution IR 
imaging by Soifer et al. (2000). 

Since its launch the last year, the CHANDRA X-ray observatory (the ultimate imager
in hard X-rays) has allowed to probe very deeply into the nature of the high-z SCUBA
sources, using the hard X-ray flux as diagnostic tool (SB are weaker
X-ray emitters than any kind of AGNs). 
Among several tens of hard X-ray and 850 $\mu m$ sources detected in various 
independent survey areas,  (Fabian et al. 2000, Hornschemeier et al.
2000, Barger et al. 2000), only very few are in common, the two samples
being essentially orthogonal. Unless all these are Compton-thick and any 
hard X-ray scattered photons are also photoelectrically absorbed,
the conclusion is that the bulk of the emission by high-luminosity SCUBA sources is due 
to star formation (in agreement with a dominant stellar emission 
in local ULIRGs found by Genzel et al. 1998).

While the detailed interplay between starburst and AGN 
is still an open issue even for local sources, the estimated fraction of the CIRB at 
850 $\mu m$ due to AGNs is not larger than 10\% (Barger et al. 2000).
Preliminary results of spectroscopic studies of the $H_\alpha$ line properties
in faint ISO mid-IR sources
(D. Rigopoulou, private communication) seem also to indicate a modest incidence of AGN,
which would imply that the overall AGN contribution to the bolometric CIRB is likely 
around 10\% or so.

\subsection{Discussion}

ISO and SCUBA surveys have proven nicely complementary capabilities to explore,
within the limitations of the current instrumentation, long-wavelength emission
by galaxies over most of the Hubble time, up to $z$ of several.
Unfortunately, this has been possible only at the short- and long-wavelength tails
of the CIRB background spectrum: 
a bad coincidence makes the wavelength interval including peak emission by distant 
dusty galaxies ($\lambda\sim 30$ to 300 $\mu m$) hardly accessible at present. 

All mentioned exploratory surveys of the distant universe have indicated
that the overall volume
emissivity of galaxies at long wavelengths drastically increases as a function of 
redshift, to explain the very steep observed  multi-wavelength counts and the
redshift distributions showing substantial high-z tails.
This evolution, however, should level off by $z\sim 1$ (see Fig. \ref{sfr} below) 
to allow consistency with the observed z-distributions (Franceschini et al. 2000)
and the CIRB spectral shape.

A spectacular finding by the deep SCUBA surveys was the discovery of ultra-luminous
galaxies at high-redshifts, mostly emitting in the far-IR and possibly at the origin
of present-day galaxy spheroids. 
However, the most precise quantification of the cosmic history of the IR population 
comes at the moment from the ISO
deep and ultra-deep surveys, which provide very detailed constraints on the counts
(Fig. \ref{cdif15}) and also allow to unambiguously identify in the optical the 
faint IR sources (Fig. \ref{hdfs}).
The outcome of our spectroscopic observations is that the faint population 
making up the CIRB in the mid-IR is dominated by actively star-forming galaxies
with substantial $H\alpha$ emission (Sect. 12.1.1). Preliminary inspection
of $H\alpha$ line profiles and constraints set by the 15 to 7 micron flux ratio 
indicate that the majority of sources are powered by a SB rather than an AGN.

Mid-IR ISO counts and the redshift distributions of the sources require extremely
high rates of evolution of the 15$\mu m$ luminosity function up to $z\sim 1$.
Taking into account all effects due to the detector spectral response function
to the complex mid-IR spectral features, the observable statistics may be explained
in terms of a strong evolution for a population of IR starbursts
contributing little to the local LF. Consequently, a plausible evolution 
pattern should involve both the source luminosities and spatial densities.

A natural way to account for this very high dependence on redshift
of the IR starburst population is to assume that it consists of otherwise normal
galaxies, but observed during a dust-extinguished luminous starburst phase, and that
its extreme evolution is due to
{\sl an increased probability with z to observe a galaxy during a starburst event.}

The common wisdom that SBs are triggered by interactions and merging suggest that
the inferred strong number density evolution 
may be interpreted as an increased probability of interaction
with z. Assuming that the phenomenon is dominated by interactions in the field and a 
velocity field constant with $z$, than this probability would scale roughly as
$\propto n(z)^2 \propto (1+z)^6$, $n$ being the number density in the proper (physical) 
volume. A more complex situation is likely to occur, as the velocity field evolves with $z$
in realistic cosmological scenarios and if we consider that the most favourable
environment for interactions are galaxy groups, which indeed are observed to include the
majority of ISOCAM distant sources (Cohen et al. 1999).
The increased luminosity with z of the typical starburst is due, qualitatively, 
to the larger amount of gas available in the past to make stars.

To note is that closed or zero-$\Lambda$ world models require evolution 
rates quite in excess of those inferred from deep optical imaging (Le Fevre et al. 2000),
whereas our best-fit solution for $\Omega_\Lambda=0.8$ and $\Omega_m=0.2$
(eq. [\ref{solu}]) is closer to the optical results.

%Note that n(z) exceeds at $z\sim 1$ the local galaxy density because of the 
%intervening effects of merging

How this picture of a 2-phase evolution of faint IR sources compares with results
of optical and near-IR deep galaxy surveys is matter of debate.
Since, because of dust, most of the bolometric emission during a starburst comes out in 
the far-IR, we would not expect the optical surveys to see much of this violent IR
starbursting phase. Indeed, B-band counts of galaxies and spectroscopic surveys 
are interpreted in
terms of number-density evolution, consequence of merging, and essentially no evolution 
in luminosity.
The Faint Blue Object population found in optical surveys may be interpreted as
the "post-starburst" population, objects either observed mostly after the major event of SF,
or more likely ones in which the moderately extinguished intermediate age ($\sim 10^7\ yrs$)
stars in a prolonged starburst (several $10^7\ yrs$) dominate the optical
spectrum.
In this sense optical and far-IR selections trace different phases of the evolution of 
galaxies, and provide independent sampling of the cosmic star formation.

A lively debate is currently taking place about the capabilities of UV-optical
observations to map accurately by themselves the past and present
star-formation, based on suitable corrections for dust extinction in distant
galaxies.
%(as expected, answers to this question are perhaps not completely free from biases introduced by 
%personal attitude and involvement in long-term programs).
Adelberger et al. (2000) suggest that the observed 850 $\mu m$ galaxy counts and the 
background could possibly be explained with the optical Lyman drop-out high-z
population by applying a proportionality correction to the optical flux and by taking into 
account the locally observed distribution of mm-to-optical flux ratios.

On the other hand, a variety of facts indicate that optically-selected and 
IR/mm-selected faint high-redshift sources form almost completely disjoint samples.
Chapman et al. (2000) observed with SCUBA a subset of z=3 Lyman-break galaxies having the
highest estimeted rates of SF as inferred from the optical spectrum, but detected only one
object out of ten. For this single detected source the predicted SFR based on the
extinction-corrected optical spectrum was 5 times lower than found by SCUBA.
A similar behaviour is also shared by local luminous IR galaxies, whose bolometric
flux is unrelated to the optical spectrum (Sanders \& Mirabel 1996).

Finally, our previously mentioned observational results by Rigopoulou et al. 
(2000) and the theoretical ones by Poggianti \& Wu (2000) and Poggianti et al. (2000) 
report {\sl independent evidence from both local and high-z luminous 
starbursts that typically 70\% to 80\% of the bolometric flux from young stars
leaves no traces in the UV-optical spectrum, because it is completely obscured
by dust}. As there seems to be no "a priory" way to correct for this missing energy,
we conclude that only long-wavelength observations, with the appropriate instrumentation,
can eventually {\sl measure} SF in galaxies at any redshifts.

%On the contrary, selection in the K-band is closer related to the lower-mass and older 
%stellar population emission, a sort of integral of the SF much more slowly evolving
%with redshift.
%Also the modest evolution in K-selected galaxies is due to the old stellar populations 
%slowly evolving in galaxies.

%this is in keeping with the evidence that the FBOs are mostly normal-sized
%galaxies, and not e.g. small compact objects to z=1 (Ellis 1997). 
%The evolution in size becomes evident only at $z>2$.
%
%Masses estimated for IR/Starbursts ($\sim 10^{11}\ M_\odot$)
%in faint ISO samples also clearly indicate them to be mostly normal (rather than
%low-mass compact) galaxies;
%
%this may also be consistent with the observed (2D and 3D) correlation properties 
%of faint opt. galaxies, which are consistent
%with the emergence of a population slightly less clustered than optical gals,
%and rather clustered as IRAS galaxies (which sample the later type - Sb/Sd -
%population), or with a single population slightly
%evolving with z (as expected in the linear gravitational instability theory,
%Peacock 1997)

\section{GLOBAL PROPERTIES: THE SFR DENSITY AND CONTRIBUTIONS TO THE CIRB}

\subsection{Evolution of the comoving luminosity density and SFR}

As illustrated in Fig. \ref{bkg}, the CIRB intensity and spectral distribution 
are in clear support of models for evolving starbursts discussed above.

Unfortunately, we are not yet in the position to derive an independent assessment 
of the evolutionary SFR density based on the available complete samples of faint IR
sources:
although a substantial effort to follow them up in the optical 
has started (particularly good chances are offered by ongoing spectroscopic follow-up
of the statistically rich faint ISOCAM samples like the GITES and HDFs), 
the process is far from complete.
As a consequence, no detailed conclusions can yet be drawn about 
the contribution of IR sources to the global comoving luminosity and SFR densities 
(Madau et al. 1996). 

\begin{figure}
\psfig{figure=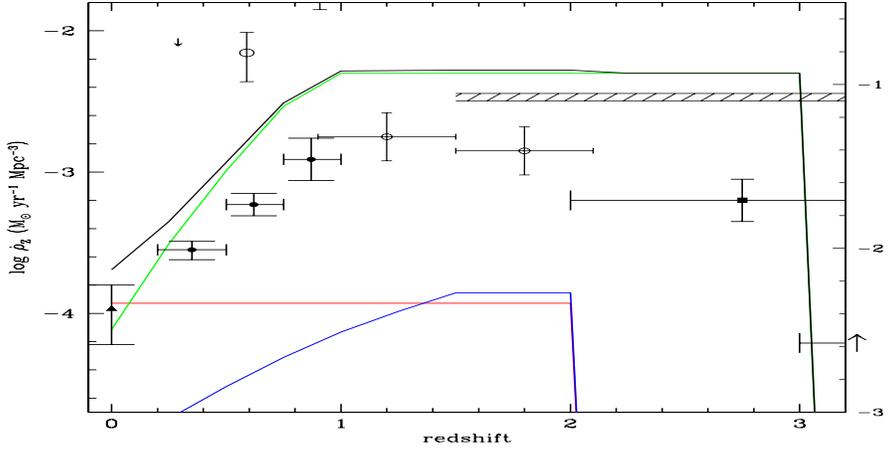,height=65mm,width=120mm}   %@ controllare il right-hand side
%\vspace*{1.cm}
\caption{Evolution of the metal production rates (left axis) and of the star formation rates 
(right axis) based on the modelisation of IR counts and z-distributions in Sect 11.1
(case $\Omega_m=0.3, \ \Omega_\Lambda=0$). Data points come from optical observations.
Very premilinary evaluations based on SCUBA results indicate values of the comoving
densities close to or even higher than the horizontal line from z=1 to 3 (Smail et al. 
1999). The shaded horizontal line is an evaluation of the average SFR in spheroidal 
galaxies by Mushotzky \& Loewenstein (1997).
}
\label{sfr}
\end{figure}

Only rather model-dependent estimates are possible at the moment, based for example 
on the evolution scheme described in Sect. 11 and whose predictions 
are summarized in Figure \ref{sfr}.
There is a clear indication here that the contribution of 
IR-selected sources to the luminosity density at high-z
should significantly exceed those based on optically selected sources, and that the
excess may be progressive with redshift up to $z\sim 1$.

This evolution should however level off at higher $z$, to allow consistency with the 
observed z-distributions for faint ISOCAM sources (Franceschini et al. 2000)
and with the estimates of the average time-dependent emissivity $j_{eff}(z)$ 
(eq. [\ref{eq:22}]) based on deconvolution of the CIRB spectrum (Gisper et al. 2000).

Altogether {\sl these results indicate that the history of galaxy long-wavelength emission
does probably follow a path similar to that revealed by optical-UV observations, 
by showing a similar peak activity around $z\sim 1$, rather than being
confined to the very high-z,} as sometimes was suggested.
This confirms that the bulk of the galaxy activity, and particularly the bulk of the
energy released in the CIRB background, is to 
be placed around $z=1$, which is obvious from Fig. \ref{sfr} if the dependence of the 
cosmological timescale on redshift is considered (Harwit 1998;
Haarsma \& Partridge 1998).

These results can only be preliminary untill we will have more substantial identifications
of existing IR-selected source samples, or, better,
after the fleet of IR/mm facilities planned for this and the next 
decade will have eventually provided data of enough quality to allow a full long-wavelength
complement to the optical-UV high-z observations.

%
%Note the very pronounced luminosity (and SFR) density peak at z=0.8 to z=2 implied 
%by the model, still not inconsistent with predictions of hierarchical clustering
%CDM models

\subsection{Energy constraints from background observations}

In the present situation, the most robust constraints on the high-redshift 
far-IR/sub-mm population come from observations of the global energetics residing in the
CIRB and optical background radiations. 
The latter imply a very substantal demand on contributing sources,
as detailed below in schematic terms.

%Finally, all {CIRB observations (including the upper limits in the MIR range
%from high-energy opacity evaluations) are consistent with negligible additional
%contributions to the universal energy budget with respect to those
%coming from SF and AGN activity}:
%all hypothesized emissions at high-z from e.g. decaying particles, primeval
%BH accretion, supermassive Pop.III stars, etc. should produce minor contributions.
%

Let us assume that a fraction $f_\ast$ of the universal mass density in baryons
\begin{equation} 
\rho_b = {3H_0^2 (1+z)^3 \over 8 \pi G } \Omega_b
%\simeq 5\ 10^{-30}(1+z)^3{H_0\over 50\ Km/s/Mpc}^2 \Omega_B [gr/cm^3] 
\simeq  7\ 10^{10}(1+z)^3 \left(H_0\over 50\ Km/s/Mpc\right)^2 \Omega_b\ [M_\odot/Mpc^3]
\label{rho}
\end{equation} 
undergoes a transformation (either processed in stars or by gravitational fields) 
with radiative efficiency $\epsilon$, then the locally observed energy density
of the remnant photons is
\begin{equation} 
\rho_\gamma = \rho_b {c^2\epsilon f_\ast \over (1+z)^4} \simeq  5\ 10^{-30}
\left(H_0\over 50\ Km/s/Mpc\right)^2 {\Omega_b \over 0.05} {f_\ast \over 0.1} 
{2.5 \over (1+z_\ast)} {\epsilon \over 0.001} [gr/cm^3] .
\label{energy}
\end{equation} 
For stellar processes, $\epsilon$ is 
essentially determined, within the moderate uncertainties of stellar models,
by the IMF: $\epsilon = 0.001$ for a Salpeter IMF and a low-mass cutoff 
$M_{min}=0.1 \ M_\odot$ (see eq. \ref{salp}), $\epsilon = 0.002$ and $\epsilon = 0.003$
for $M_{min}=2$ and $M_{min}=3$, while $\epsilon$ gets the usually quoted value of
$\epsilon = 0.007$ only for $M_{min} > 10\ M_\odot$ [A. Bressan, private communication].

Note how the contribution to the photon background energy by very high redshifts
is penalized in eq. (\ref{energy}) by the $(1+z)^{-1}$ factor: measurements
of the photon background preferentially constrain source emission at moderate $z$,
whereas estimates of the local average metal abundance (obviously much more difficult
and indirect!) would in principle provide a less biased integral over the total stellar 
yield in the past.

\subsubsection{Constraints from the integrated optical background}

As already noticed (Sect. 8.3.1), the converging galaxy counts at faint magnitudes
observed in the optical
and near-IR allow to estimate with fair accuracy the total diffuse flux at these 
wavelengths (Fig. \ref{bkg}, see Madau \& Pozzetti 2000). The  
bolometric emission from 0.1 to 7 $\mu m$ by distant galaxies turns out to be
\begin{equation} 
\nu I(\nu)|_{opt} \simeq (17 \pm 3) \ 10^{-9}\ Watt/m^2/sr ,
\label{optical}
\end{equation} 
which in fact is a lower limit if we give credit to claims of a ($\times 2-3$) larger
optical/NIR background, see Sect. 8.1 (but see also a counter-argument in Sect. 8.2).

We discussed evidence that for the most luminous starbursts the optical spectra are only 
moderately contributed by starburst emission, which is mostly hidden in the far-IR.
Accordingly, let us assume 
that the optical/NIR BKG mostly originates by quiescent SF in spiral disks and by
intermediate and low-mass stars. As observed in the Solar Neighborhood, a good
approximation to the IMF in such relatively quiescent environments is the
Salpeter law with standard low-mass cutoff, corresponding to a mass--energy
conversion efficiency  $\epsilon \sim 0.001$. With these parameter values, we
can reproduce the whole optical BKG intensity of eq.(\ref{optical}) by
transforming a fraction $f_\ast \simeq 10 \%$ of all nucleosynthetic baryons into
low-mass stars, assumed the bulk of this process happened at $z_\ast \sim 1.5$
and 5\% of the closure value in baryons (for our adopted $H_0=50\ Km/s/Mpc$, or
$\Omega_b h^2=0.012$, consistent with the theory of primordial nucleosynthesis):
$$\nu I(\nu)|_{opt} \simeq 20 \ 10^{-9}\left(H_0\over 50\ Km/s/Mpc\right)^2 
{\Omega_b \over 0.05} 
{f_\ast \over 0.1} \left(2.5 \over 1+z_\ast\right) {\epsilon \over 0.001}  \ Watt/m^2/sr .
$$
It is generated in this way a local density in low-mass stars consistent with the observations
(based on photometric surveys, Ellis et al. 1996, and assuming standard mass to light ratios):
\begin{equation} 
\rho_b(stars) \simeq 7\ 10^{10} f_\ast \Omega_b \simeq
3.4\ 10^8\ M_\odot/Mpc^3 ,
\label{star}
\end{equation} 
which, assuming typical solar metallicities, corresponds to a local density in metals of
\begin{equation} 
\rho_Z(stars) \simeq 1.6\ 10^{9} f_\ast {Z\over Z_\odot} \Omega_b\ M_\odot/Mpc^3\simeq 
7.7\ 10^6  M_\odot/Mpc^3 .
\label{Z}
\end{equation} 
Note that a factor 2-3 larger optical/NIR background than in eq. (\ref{optical}) could 
still be consistent with the present scheme if a similar scaling factor would also apply to 
eqs. (\ref{star}) and (\ref{Z}): that is, if both the excess background and low-mass stars
and stellar metals would be due to extended low-brightness halos, unaccounted for
by deep HST imaging as well as by local photometric surveys.

\subsubsection{Explaining the CIRB background}

The total energy density between 7 and 1000 $\mu m$ contained in the CIRB,
including modellistic extrapolations as in Fig. \ref{bkg} consistent with the constraints set
by the cosmic opacity observations, amounts to 
\begin{equation}
\nu I(\nu)|_{FIR} \simeq 40 \ 10^{-9}\ Watt/m^2/sr .
\label{CIRB}\end{equation} 

Following our previous assumption that luminous starbursting galaxies
emit negligible energy in the optical-UV and most of it in the far-IR,
we coherently assume that the energy resident in the CIRB background originates from 
star-forming galaxies at median $z_\ast \simeq 1.5$. The amount of baryons 
processed in this phase and the conversion efficiency $\epsilon$
have to account for the combined constraint set by eqs.(\ref{star}) and 
(\ref{CIRB}), that is to provide a huge amount of energy with essentially 
no much stellar remnant in the local populations. The only plausible solution
is then to change the assumptions about the stellar IMF characterizing
the starburst phase, for example to a Salpeter distribution cutoff below 
$M_{min}=2\ M_\odot$, with a correspondingly higher efficiency $\epsilon=0.002$
(see discussion in Sect. 13.2). This may explain the energy density in the CIRB:
$$\nu I(\nu)|_{FIR} \simeq 40 \ 10^{-9}\left(H_0\over 50\ Km/s/Mpc\right)^2 
{\Omega_b \over 0.05} 
{f_\ast \over 0.1} \left(2.5 \over 1+z_\ast\right) {\epsilon \over 0.002}  \ Watt/m^2/sr ,
$$
assumed that a similar amount of baryons, $f_\ast\simeq 10\%$, as processed with low 
efficiency during the ``inactive'' secular evolution, are processed with higher 
efficiency during the starbursting phases, producing a two times larger amount of metals:
$\rho(metals)\sim 1.4\ 10^7\ M_\odot/Mpc^3$.
Note that by decreasing $M_{min}$ during the SB phase would decrease the efficiency 
$\epsilon$ and increase
the amount of processed baryons $f_\ast$, hence would bring to exceed the locally
observed mass in stellar remnants (eq.[\ref{star}]).

The above scheme is made intentionally extreme, to illustrate the point. The reality
is obviously more complex than this, e.g. by including a flattening at low mass 
values in the 
Salpeter law (see Zoccali et al 1999) for the solar-neighborhood SF and, likewise, a more
gentle convergence of the starburst IMF than a simple low-mass cutoff.

\subsubsection{Galactic winds and metal pollution of the inter-cluster medium}

A direct prediction of our scheme above is that most of the metals produced
during the starburst phase have to be removed by the galaxies to avoid largely exceeding
the locally observed metals in galaxies. As discussed in Sect. 6.4,
there is clear evidence in local starbursts, based on optical and 
X-ray observations, for large-scale super-winds out-gassing high-temperature
enriched plasmas from the galaxy. Our expectation would be that a substantial
amount of metals, those originating from the same SF processes producing the CIRB
background, are hidden in the hot inter-cluster medium.

But where all these metals are?

While densities and temperatures of the polluted plasmas in the diffuse (mostly
primordial and un-processed) inter-cluster medium are such to hide easily these products of the
ancient SB phase, an interesting support to the above scheme comes from consideration 
of the metal-enriched intra-cluster plasma (ICP) in clusters of galaxies. 
Rich clusters are considered to constitute a representative sample of the universe,
while at the same time -- given their deep gravitational potential -- they are to be
considered from a chemical point of view as closed boxes (all metals produced
by cluster galaxies are kept inside the cluster itself).

The mass of metals in the ICP plasma is easily evaluated from the total amount of 
ICP baryons (measured to be $\sim$5 times larger than the mass in galactic stars)
and from their average metallicity, $\sim$40\% solar. The mass of ICP metals
is $M_{metals, ICP} \simeq 5\times 0.4\ (Z/Z_\odot)\ M_{stars}$, which is
two times larger than the mass of the metals present in galactic stars
and consistent with the mass in metals produced during the SB phase.

Then the same starbursts producing the ICP metals are also likely responsible
for the origin of the CIRB. As mentioned, the starburst enrichment process could have 
been pictured in a deep SCUBA image of the candidate proto-cluster surrounding
the z=3.8 radio-galaxy 4C41.17.
In a similar fashion,  Mushotzky \& Loewenstein (1997) used their metallicity measurements
in clusters to estimate the contribution of spheroidal galaxies to the SFR
density (see Fig. \ref{sfr}).

\subsubsection{A two-phase star-formation: origin of galactic disks and spheroids}

The above scheme, best-fitting the available IR data as discussed in Sect. 11, 
implies that {\sl star formation in galaxies has proceeded in two phases: 
a quiescent one taking place during most of the Hubble time, slowly building stars with 
standard IMF 
from the regular flow of gas in rotational supported disks; and a transient actively
starbursting phase, recurrently triggered by galaxy mergers and interactions. }
During the merger,
{\it violent relaxation} redistributes old stars, producing de Vaucouleur profiles typical 
of galaxy spheroids, while young stars are generated following a top-heavy IMF.

Because of the geometric (thin disk) configuration of the diffuse ISM and the modest
incidence of dusty molecular clouds, the quiescent phase is only moderately
affected by dust extinction, and naturally originates most of the optical/NIR background 
(included early-type galaxies completely deprived of an ISM).

The merger-triggered active starburst phase is instead characterized by 
a large-scale redistribution
of the dusty ISM, with bar-modes and shocks, compressing a large fraction of
the gas into the inner galactic regions and triggering formation of molecular clouds.
As a consequence, this phase is expected to be heavily extinguished
and the bulk of the emission to happen at long wavelengths, naturally originating
the cosmic CIRB background.
Based on dynamical considerations, we expect that during this violent SB phase 
the elliptical and S0 galaxies are formed in the most luminous IR SBs at higher-z 
(corresponding to the SCUBA source population), while galactic bulges in later-type galaxies
likely originate in lower IR luminosity, lower-z SBs (the ISO mid-IR population). 

The presently available IR data cannot assess if the different luminosity ranks of
SCUBA and ISO selected sources are characterized also by
different formation timescales (SF activities
being confined to the higher-z for the former and to lower-z for the latter), since
the present samples are far dominated by K-correction and selection effects. Assumed however
this is indeed the case, this could still be reconciled with the expectations 
of hierarchical clustering models if we consider that SCUBA sources  likely trace the
very high-density (galaxy clusters) environment with an accelerated merging rate at
high-z, while ISO sources are likely related with lower-density environments 
(galaxy groups or the field) entering the non-linear collapse phase at later cosmic
epochs (e.g. Franceschini et al. 1999).

Finally, {\sl if indeed the IMF characteristic of the SB phase is deprived of low-mass 
stars, as suggested in
the previous paragraphs, a consequence would be that the excess blue stars
formed during the SB would quickly disappear, leaving the colors of the emerging 
remnant as typically observed for early-type galaxies and keeping consistent with
the evidence that the stellar mass content in galaxies does not change much for $z<1$.}

\subsection{Contribution by gravitational accretion to the global energetics}

The remarkable similarities between the cosmic evolution of galaxy
and AGN emissivities have been taken as evidence that the same processes triggering
SF also make a fraction of the gas to accrete and fuel the AGN (Hasinger 
Franceschini et al. 1999).
Furthermore, detailed studies of local high-luminosity IR galaxies are showing that 
SF and AGN activities happen very often concomitantly in the same object (Genzel et al. 
1998; Risaliti et al. 2000; Bassani et al. 2000). After all, this is a 
natural outcome of the scheme discussed in previous Sections, 
the violent radial inflow of gas following the merger/interaction should likely 
fuel not only nuclear star-clusters, but the BH itself at some stage.

Waiting for forthcoming and future powerful instrumentation (X-ray observatories
CHANDRA and XMM, Constellation-X and XEUS in the future, and 
large space IR observatories like NGST and FIRST) to have a detailed quantification of 
the relative merits of the two fundamental baryon drivers, 
some order-of-magnitude estimates may be useful as a guideline.
From a combined analysis of the AGN and starburst average 
bolometric emissivities as a function
of redshift, Franceschini et al. (1999) infer a relationship between the mass $M_{BH}$
of the local remnant super-massive BH after the AGN phase to the mass $M_\ast$
in galactic stars from the SB phase:
\begin{equation}
 M_{BH} \simeq 0.001 \left({\epsilon \over 0.001}\right) 
\left({0.1 \over \eta}\right) \left({n[type\ II]/n[type\ I] \over 5}\right) \ M_\ast,
\label{BH}
\end{equation}
where $\eta$ is the radiative efficiency by BH accretion and $n[type\ II]/n[type\ I]$
is the ratio of the absorbed to unabsorbed AGNs (which should be close to
3-5 to explain the local AGN statistics and the observed intensity of the XRB).
On the other hand, observations of supermassive BH's in local spheroidal
galaxies (Magorrian et al. 
1998, Faber et al. 1997) indicate a quite higher mass in the BH accreted material
with respect to that in stars: $M_{BH} \simeq (0.002-0.006)\ M_{\ast}$.  
Assumed that $\eta$ should not be lower than 
0.1, this may require a stellar mass-energy conversion efficiency $\epsilon >> 0.001$, 
which is further independent support to the idea of a top-heavy IMF during the SB phase.

\section{CONCLUSIONS}

During the last few years a variety of observational campaigns, in particular 
by ISO from space in the far-IR and by large mm telescopes from ground,
have started to provide a complementary view of the distant universe at long
wavelengths with respect to that offered by standard optical-UV-NIR deep explorations. 
Also of crucial importance in this context was the discovery of an 
intense diffuse background radiation in the far-IR/sub-mm of extragalactic
origin, the CIRB. These results are challenging those obtained
from optical-UV observations only, by revealing luminous to very luminous phases
in galaxy evolution at substantial redshifts, likely corresponding to violent
events of star-formation in massive systems. In the most extreme of these sources, 
however, a quasar contribution cannot be excluded, and sometimes has indeed been proven.

Whereas the process of optical identification and spectroscopic characterization 
of the long-wavelength selected high-redshift sources is only at the beginning
(and will keep being a challenging task for the next several years because of the
faintness of the optical counterparts), some interesting constraints on the 
cosmic evolution can already been inferred from observations of the CIRB spectral
intensity and the multi-wavelength source counts.
The most robust conclusions at the moment appear to be those of a very rapid
increase of galaxy long-wavelength emissivity with redshift, paralleled by an
increased incidence in high-redshift sources of dust extinction and thermal dust
reprocessing with respect to locally observed sources.

A way to interprete these results is to consider as a crucial cosmogonic ingredient
the role of galaxy interactions and merging. The strong increases with redshift
of the {\sl probability} of interactions (as partly due to a plain geometrical effect
in the expanding universe) and of the {\sl effects} of interactions (due to the 
more abundant fuel avaliable in the past), likely explain the observed rapid
evolution. 

Altogether, the large energy content of the CIRB is not easily explained, unless
the powerful infrared starburst phase is characterized by a stellar IMF somewhat 
deprived in low-mass stars.

Although the subject is presently subject to some controversies, we think we have provided
enough evidence, based on pioneering efforts of deep sky surveys in the IR and mm,
that only such long wavelengths contain the clue to an exhaustive description
of the star formation phenomenon, now and in the past. It seems clear that there are no 
alternatives, neither in X-rays, optical nor radio, to the IR/mm flux measurement for a 
reliable determinantion of the rate of SF in galaxies, simply because it is
there that a dominant fraction of photons from young very luminous stars emerges, 
and no ways are available to determine "a priory" what precisely this fraction is. 
Fundamental aspects of galaxy formation and evolution (e.g. the origin of
galaxy spheroids, and the onset of quasar activity) can effectively be observed at 
long wavelengths.
In this sense the variety of ground-based and space projects in this field planned for
the present decade promises extremely rewarding benefits for observational cosmology.

\begin{acknowledgments}{} 

This paper has benefited by a large collaboration, in particular
concerning items discussed in the last chapters, including some yet unpublished
results. I want to mention the people who have particularly 
contributed:
H. Aussel, S. Bressan, C. Cesarsky, D. Clements, FX. Desert,
D. Elbaz, D. Fadda, R. Genzel, G.L. Granato, M. Harwit, 
S. Oliver, B. Poggianti, J.L. Puget, D. Rigopoulou, M. Rowan-Robinson, L. Silva.
I am also glad to thank L. Danese, G. De Zotti for a long-standing collaboration in this
field, A. Cavaliere and C. Chiosi for many 
fruitful discussions.  Finally, I want to warmly thank the organizers of the
Canary Islands Winter School on "High-Redshift Galaxies" for their kind invitation.

\end{acknowledgments}

\end{document}